\documentclass[prx,twocolumn,superscriptaddress]{revtex4-1}

\usepackage{graphicx}
\usepackage{color}
\usepackage[dvipsnames]{xcolor}
\usepackage{amsmath,amssymb,mathrsfs}
\usepackage{braket}
\usepackage{mathtools}
\usepackage{lmodern}

\usepackage{bbm}

\definecolor{mildblue}{rgb}{.2,0.7,.9}
\definecolor{warmred}{rgb}{.68,0.19,.14}

\definecolor{dante}{rgb}{.2,0.7,.9}

\newcommand{\hs}[1]{{\color{black}#1}}

\definecolor{daniel}{rgb}{.0,1,0.5}

\date{\today}                  

\usepackage{array, multirow, bigdelim, makecell} 

\begin{document}



\title{Second-order topology and supersymmetry in two-dimensional topological insulators}

\author{Clara S. Weber}
\affiliation{Institut f\"ur Theorie der Statistischen Physik, RWTH Aachen and JARA - Fundamentals of Future Information Technology, D-52056 Aachen, Germany}
\affiliation{Department of Physics and Astronomy, University of Pennsylvania, Philadelphia, Pennsylvania 19104, USA}
\author{Mikhail Pletyukhov}
\affiliation{Institut f\"ur Theorie der Statistischen Physik, RWTH Aachen and JARA - Fundamentals of Future Information Technology, D-52056 Aachen, Germany}
\author{Zhe Hou}
\affiliation{Department of Physics, University of Basel, Klingelbergstrasse 82, 
CH-4056 Basel, Switzerland}
\author{Dante M. Kennes}
\affiliation{Institut f\"ur Theorie der Statistischen Physik, RWTH Aachen and JARA - Fundamentals of Future Information Technology, D-52056 Aachen, Germany}
\affiliation{Max Planck Institute for the Structure and Dynamics of Matter, Center for Free Electron Laser Science, 22761 Hamburg, Germany}
\author{Jelena Klinovaja}
\affiliation{Department of Physics, University of Basel, Klingelbergstrasse 82, 
CH-4056 Basel, Switzerland}
\author{Daniel Loss}
\affiliation{Department of Physics, University of Basel, Klingelbergstrasse 82, 
CH-4056 Basel, Switzerland}
\author{Herbert Schoeller}
\email[Email: ]{schoeller@physik.rwth-aachen.de}
\affiliation{Institut f\"ur Theorie der Statistischen Physik, RWTH Aachen and JARA - Fundamentals of Future Information Technology, D-52056 Aachen, Germany}

\begin{abstract}

We unravel a fundamental connection between supersymmetry (SUSY) and a wide class of two dimensional (2D) second-order topological insulators (SOTI). This particular supersymmetry is induced by applying a half-integer Aharonov-Bohm flux $f=\Phi/\Phi_0=1/2$ through a hole in the system. Here, three symmetries are essential to establish this fundamental link: chiral symmetry, inversion symmetry, and mirror symmetry. At such a  flux of half-integer value the mirror symmetry anticommutes with the inversion symmetry leading to a nontrivial $n=1$-SUSY representation for the absolute value of the Hamiltonian in each chiral sector, separately. This implies that a unique zero-energy state and an exact twofold degeneracy of all eigenstates with non-zero energy is found even at finite system size. For arbitrary smooth surfaces the link between 2D-SOTI and SUSY can be described within a universal low-energy theory in terms of an effective surface Hamiltonian which encompasses the whole class of supersymmetric periodic Witten models. Applying this general link to the prototypical example of a Bernevig-Hughes-Zhang(BHZ)-model with an in-plane Zeeman field, we analyze the entire phase diagram and identify a gapless Weyl phase separating the topological from the non-topological gapped phase. Surprisingly, we find that topological states localized at the outer surface remain in the Weyl phase, whereas topological hole states move to the outer surface and change their spatial symmetry upon approaching the Weyl phase. Therefore, the topological hole states can be tuned in a versatile manner opening up a route towards magnetic-field-induced topological engineering in multi-hole systems. Finally, we demonstrate the stability of localized states against deviation from half-integer flux, flux penetration into the sample, surface distortions, and random impurities for impurity strengths up to the order of the surface gap.    

\end{abstract}


\maketitle

\section{Introduction} 
\label{sec:introduction}

Supersymmetry (SUSY) in nonrelativistic quantum mechanics \cite{junker_book_19,cooper_etal_book_01,bagchi_book_01} is a special type of symmetry 
allowing one to classify system's eigenstates into the so-called ``bosonic'' and ``fermionic'' subspaces as well as to establish mappings between these subspaces by the so called SUSY transformations. SUSY deepens our 
understanding of the level structure and the states, and in certain cases the SUSY algebra generators facilitate an exact solution of the eigenvalue problem by purely algebraic means. One of the central models in nonrelativistic quantum mechanics exhibiting SUSY is the Witten's model \cite{witten_npb_81}, which serves as a prototypical example for the explicit demonstration of the SUSY properties and their application. The SUSY structure of the Dirac equation \cite{thaller_book_92} also paves the way for the application of SUSY in solid state systems, particularly in their low-energy description and draws an important bridge between this field and the field of high-energy physics where SUSY remains a central topic to this date. Thus, the occurrence of SUSY in the description of heterojunctions with band-inverting contact has been highlighted in Ref.~\cite{pankratov_etal_ssc_87} and the SUSY algebra has been applied for the description of the quantum Hall effect in graphene \cite{ezawa_pla_07}. A SUSY formulation of the two-dimensional electron gas with Rashba and Dresselhaus spin-orbit coupling is also feasible \cite{tomka_etal_scr_15}. The emergence of the (space-time) SUSY (which is a generalization of the quantum mechanical SUSY) in topological insulators and superconductors has been unveiled in Refs.~\cite{grover_etal_science_14,ponte_lee_njp_14}. Most recently, it has been proposed \cite{queralto_etal_cph_20} to exploit the SUSY transformations for topological state engineering.

The focus of the present work is to establish an important link between SUSY and the field of second-order topological insulators (SOTI), a field of tremendous recent interest in condensed matter physics \cite{benalcazar_etal_science_17,benalcazar_etal_prb_17,song_etal_prl_17,langbehn_etal_prl_17,geier_etal_prb_18,imhof_etal_ntp_18,schindler_etal_sca_18,trifunovic_brouwer_prx_19}. In particular, we establish that a wide subclass of 2D-SOTI are close to a supersymmetric point stabilizing zero-dimensional bound states. We find that SUSY is an exact symmetry if one applies a half-integer Aharonov-Bohm flux through a hole in the 2D system. The corresponding effective 1D-surface Hamiltonian describing the second-order topological phase transition in a low energy description turns out to be a realization of the whole class of supersymmetric Witten models playing a central role in the discussion of SUSY models \cite{junker_book_19,cooper_etal_book_01,bagchi_book_01}, see also Refs.~\cite{khare_sukhatme_jpa_04,arenas_master_20} for the discussion of periodic Witten models relevant for this work, together with the special case of the double-sine potential in Refs.~\cite{razavy_pla_81,ulrich_etal_prb_14,venn_master_18}. The important subclass of 2D-SOTI with SUSY properties consists of those models where the topological zero-energy states emerge as interface bound states at those positions of the surface where the mass term of the effective surface Hamiltonian changes its sign \cite{langbehn_etal_prl_17}. Such models have been classified within the general classification scheme of higher-order TIs \cite{geier_etal_prb_18} in terms of specific Shiozaki-Sato symmetry classes \cite{shiozaki_sato_prb_14}. A prototypical example is the combination of a standard Bernevig-Hughes-Zhang (BHZ) model \cite{bernevig_etal_science_06} with an in-plane Zeeman field inducing a mass gap between the counterpropagating helical edge modes, see, e.g., Refs.~\cite{khalaf_prb_18} and \cite{ren_etal_prl_20}. For pedagogical reasons this model will be the backbone of this work, although our conclusions hold for an extended subclass of such 2D-SOTI. We note that Zeeman fields play a very important role in controlling higher-order topological states and, besides the combination with the BHZ-model, have also been used in superconducting systems to realize and control Majorana states  \cite{laubscher_etal_prr_19,plekhanov_etal_prr_19,volpez_etal_prl_19,laubscher_etal_prb_20,laubscher_etal_prr_20,plekhanov_etal_prr_20,plekhanov_etal_prb_21}. Similiarly, it will also turn out in this work that the Zeeman term is a very flexible tool to control the shape of the topological states, implying versatile possibilities of topological engineering with magnetic-field-only control.

\begin{figure*}[t!]
    \centering
    \includegraphics[width =\textwidth]{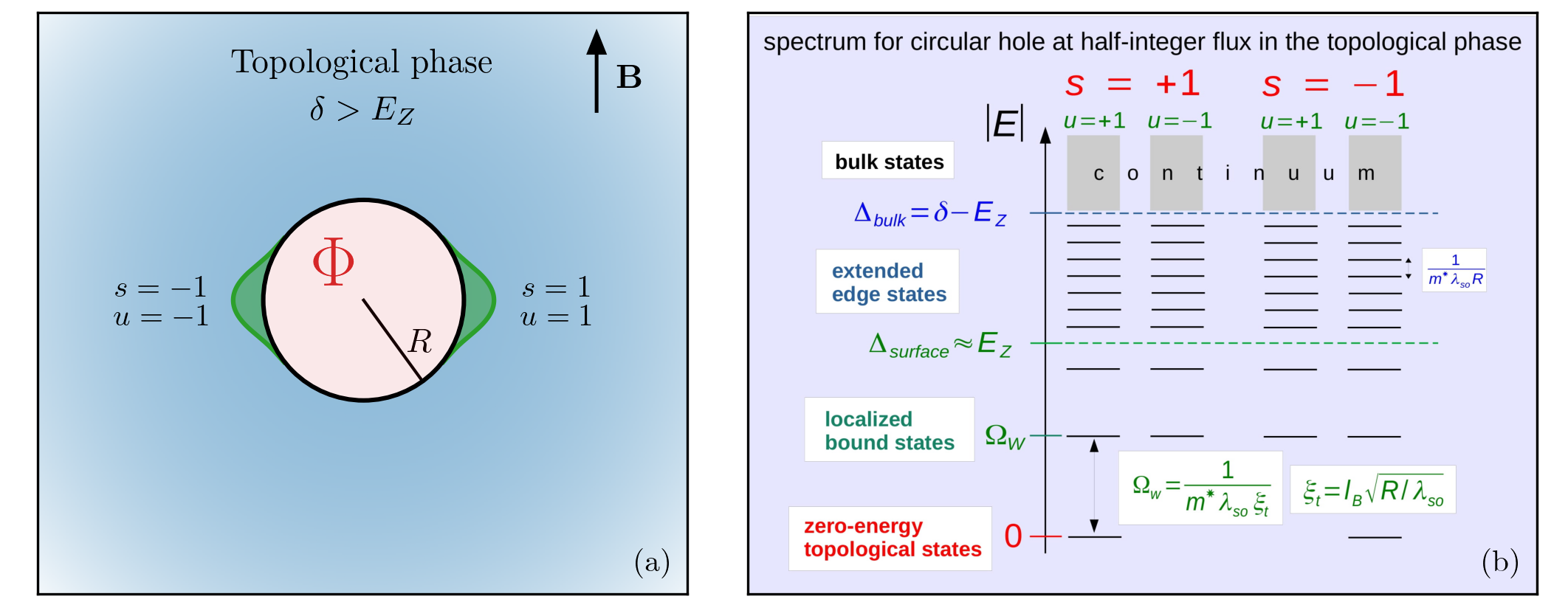}
	  \caption{(a) Sketch of a circular hole with radius $R$ in an infinite system. The hole is threaded by a magnetic flux $\Phi$ and the system is subject to a homogeneous in-plane Zeeman field ${\bf B}$. In the topological phase where the band inversion parameter $\delta$ is larger than the Zeeman energy $E_Z$, one finds two topological states (indicated by the green regions) localized at the hole's surface where the normal component with respect to the circle's surface of the Zeeman field changes sign. At half-integer flux $f=\Phi/\Phi_0=1/2$, the topological states are exactly at zero energy due to chiral symmetry and SUSY. The eigenvalues of chiral symmetry and SUSY are labeled by $s=\pm 1$ and $u=\pm 1$, respectively. (b) Sketch of the spectrum of the absolute value of the eigenenergies $|E|$ in the topological phase at half-integer flux. The spectrum decomposes into two chiral sectors $s=\pm 1$, each of them revealing an unbroken SUSY spectrum with one single zero-energy state and a set of twofold degenerate states at positive energies labeled by the SUSY $u=\pm 1$. Below the surface gap $\Delta_{\rm surface}\approx E_Z$, a set of discrete localized bound states appears due to a second-order topological mechanism, with a typical energy scale set by the Witten frequency $\Omega_W = 1/(m^* \lambda_{\rm so} \xi_t)$, where $m^*$ is the effective mass, $\lambda_{\rm so}$ the spin-orbit length, and $\xi_t=l_B\sqrt{R/\lambda_{\rm so}}$ the tangential localization length of the bound states. Here, $l_B$ is the magnetic length characterizing the Zeeman energy $E_Z=1/(2m^* l_B^2)$. Between the surface gap $\Delta_{\rm surface}$ and the bulk gap $\Delta_{\rm bulk}=\delta-E_Z$, pairs of degenerate edge states appear in each chiral sector which are extended along the hole's surface and localized in radial direction with normal localization length $\xi_n\sim\lambda_{\rm so}$. Due to finite size quantization in angular direction, their spacing is given by the order $\sim 1/(m^* \lambda_{\rm so} R)$ which is much smaller than the Witten frequency $\Omega_W$ if the magnetic length fulfils the condition $l_B\ll \sqrt{R\lambda_{\rm so}}$ such that the topological states are well localized $\xi_t \ll R$. Beyond the bulk gap a continuum of states appears which is twofold degenerate in each chiral sector such that the SUSY structure of the spectrum applies exactly to all states.  
	  } 
    \label{fig:hole_system}
\end{figure*}
The fact that inversion and/or mirror symmetries can stabilize higher-order topological states in 2D systems has been emphasized in previous works \cite{langbehn_etal_prl_17,geier_etal_prb_18,khalaf_prb_18,ren_etal_prl_20}. However, what we add here is the insight that the simultaneous presence of both inversion and mirror symmetry commuting with each other can be tuned at half-integer Aharonov-Bohm flux to two {\it anti-commuting} unitary symmetries by multiplying the mirror symmetry with an exponential factor $e^{-i\varphi}$, where $\varphi$ denotes the polar angle with respect to the mirror symmetry axis. This exponential factor respects periodic boundary conditions, removes the half-integer flux, and enforces the anti-commutation of inversion and mirror symmetry. As a result, one can prove that there is an exact twofold degeneracy of all eigenstates of the model, quite analog to a Kramer's degeneracy but here realized via two anti-commuting {\it unitary} symmetries with one of them being an involution. If, in addition, the model fulfils chiral symmetry, this twofold degeneracy leads to a protection of a pair of zero-energy topological states. Importantly, even in the absence of chiral symmetry, it turns out that the mirror symmetry is the involution of an exact $n=1$ SUSY representation \cite{combescure_etal_jpa_04} with the Hermitian supercharge operator given by the product of the Hamiltonian and the inversion symmetry. These properties show that a wide subclass of 2D-SOTI has a supersymmetric spectrum and, if zero-energy states are present, those are topologically protected by SUSY. We note that this protection is exact at half-integer flux even for a finite system with an exact degeneracy of the two zero-energy states, irrespective of whether they have a significant orbital overlap or not. When tuning the flux away from half-filling an approximate protection up to exponentially small splittings is found if the two topological states have an exponentially small orbital overlap (which is realized for a sufficiently large system). At the SUSY point, the topological index playing the role of the topological invariant is the Witten index distinguishing broken from unbroken SUSY in the absence/presence of zero-energy states, see e.g. Refs.~\cite{durand_vinet_pla_90,beckers_debergh_mpl_89,beckers_debergh_jmp_90}.

\begin{figure}
	 \includegraphics[width =1.0\columnwidth]{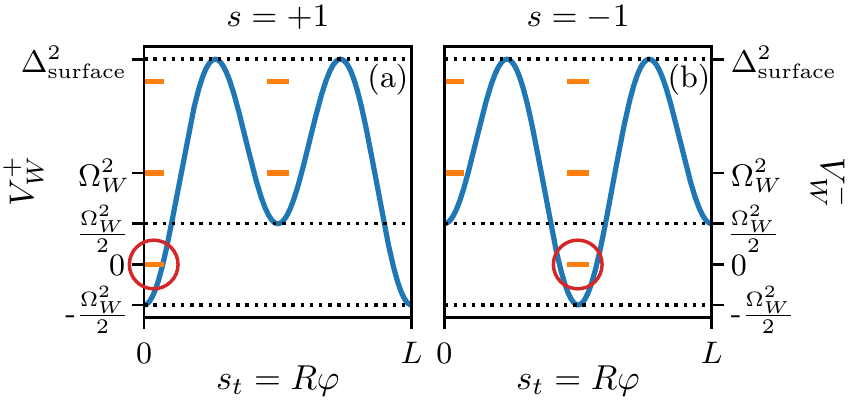}
	  \caption{The two effective surface Witten potentials for a circular hole with radius $R=1000 \lambda_{\rm so}$ in the two chiral sectors (a) $s=1$ and (b) $s=-1$ as function of the line element $s_t=R \varphi$ along the surface ($0<\varphi<2\pi$ denotes the polar angle and $L=2\pi R$ is the circumference of the surface). The Zeeman energy is chosen as $E_Z=E_{\rm so}/25$, where $E_{\rm so}$ is the spin-orbit energy. For both chiral sectors localized states (shown in orange) are trapped in the two potential minima which appear to be exactly twofold degenerate due to SUSY except for one single topological bound state with zero eigenvalue (highlighted by the red circle) which is localized around $\varphi= 0$ for $s=1$ and $\varphi=\pi$ for $s=-1$. The potential maximum is given by the square of the surface gap $\Delta_{\rm surface}^2=E_Z^2 + (1/(2m^* R^2))^2\approx E_Z^2$, and the two potential minima appear at $\pm \Omega_W^2/2$ with the Witten frequency defined in Fig.~\ref{fig:hole_system}(b). As a consequence, the lowest state has zero eigenvalue and all the excited states are twofold degenerate with eigenvalue $n\,\Omega_W^2$, $n=1,2,\dots$ (in a semiclassical picture), consistent with the sketch of the spectrum shown in Fig.~\ref{fig:hole_system}(b).     
	  } 
    \label{fig:double_sine}
\end{figure}

The main results of our work are summarized in Figs.~\ref{fig:hole_system}(a,b) and Fig.~\ref{fig:double_sine}. In Fig.~\ref{fig:hole_system}(a) we show a prototypical example of a circular hole in an infinite system, where the SUSY structure of the spectrum applies to all states and is exact. In the topological phase, where the band inversion parameter is larger than the Zeeman energy, one finds topological states localized at the positions of the hole's surface where the normal component  of the Zeeman field with respect to the surface of the hole changes sign (a generic rule  for any shape of the surface). The spectrum of the absolute value of the Hamiltonian applying an additional half-integer flux through the hole is sketched in Fig.~\ref{fig:hole_system}(b), which demonstrates  the close relationship of the typical spectrum of a second-order TI with the spectrum of an unbroken SUSY in each chiral sector. The later manifests itself by an exact twofold degeneracy of SUSY partners (labeled by the SUSY eigenvalue $u=\pm 1$) at all positive eigenvalues, and a unique zero-energy topological state with fixed SUSY value $u=s$ in each chiral sector $s=\pm 1$. As typical for second-order TIs with a smooth surface, the spectrum reveals a set of localized bound states below the surface gap set by the Zeeman energy. Besides the two zero-energy topological states, their energy is characterized by a new emerging energy scale, the Witten frequency $\Omega_W$ which scales inversely proportional to the tangential localization length $\xi_t$. Importantly, it turns out that $\xi_t$ scales with the square root of the hole radius, leading to well-localized bound states in tangential direction for sufficiently large hole radius. In between the surface and bulk gap, we find a set of helical edge states which are extended over the whole surface. Since the circumference of the surface is much larger than $\xi_t$ for a large hole radius, their finite-size spacing is much smaller than the Witten frequency. 

Fig.~\ref{fig:double_sine} shows the effective surface potentials in each chiral sector for the case of a circular hole from which all boundary states can be analyzed analytically. It results from squaring the effective surface Hamiltonian $H^{\rm eff}_{\rm surface}$ which can be written in the generic form of a periodic $2$-band Dirac model (here, $\hbar=1$, and the Pauli matrices $\sigma_i$ result from a convenient spinor transformation to be specified later) 
\begin{align}
    \label{eq:effective_surface_H_with_dimension}
    H^{\rm eff}_{\rm surface} = \alpha \sigma_x (-i\partial_{s_t}) + \sigma_y E_{Z,n}(s_t)\,, 
\end{align}
where $\alpha$ is the spin-orbit interaction, $s_t$ the line element along the surface, and $E_{Z,n}(s_t)$ the normal component of the Zeeman field along the surface. From this Hamiltonian one can calculate the non-trivial tangential part of the wave function along the surface, whereas the normal part is described by an exponentially decaying wave function with a normal localization length of the order of the spin-orbit length. The effective surface Hamiltonian brings the relationship of second-order topology and SUSY to a universal low-energy form. On the one hand side, it contains the two important basic ingredients to generate second-order topology: the spin-orbit interaction $\alpha$ generating two counter-progagating helical edge modes along the surface (with helicity $\sigma_x=\pm 1$) as familiar from the BHZ model \cite{bernevig_etal_science_06}, and the normal component of the Zeeman term acting as a mass term generating a surface gap in which topological states are trapped at the positions where the mass term changes sign \cite{langbehn_etal_prl_17,ren_etal_prl_20}. On the other side, by squaring the effective surface Hamiltonian, one can demonstrate the SUSY structure of the spectrum shown in Fig.~\ref{fig:hole_system}(b) for all boundary states below the bulk gap. In the two chiral sectors $s=-\sigma_z=\pm 1$ one obtains two periodic Witten models describing a particle in an effective surface potential
\begin{align}
    \label{eq:witten_model_with_dimension}
    {\cal H}_W^\pm&= (H^{\rm eff}_{\rm surface})^2|_{\sigma_z=\mp}= -\alpha^2 \partial_{s_t}^2 + V_W^{\pm}(s_t)\,,\\
    \label{eq:witten_potential_with_dimension}
    V_W^\pm(s_t) &= E_{Z,n}(s_t)^2 \mp \alpha\partial_{s_t} E_{Z,n}(s_t)\,.
\end{align}
Here, $V_W^\pm(s_t)$ are the two partner Witten potentials shown in Fig.~\ref{fig:double_sine}, which are given by a double-sine potential for the special case of a circular hole but can be tuned to any generic form depending on the choice for the shape of the surface. Most importantly, for any mirror-symmetric surface around the two axis parallel and perpendicular to the Zeeman field, the Witten model has supersymmetric properties in each chiral sector, consistent with the spectrum of the boundary states shown in Fig.~\ref{fig:hole_system}(b) and Fig.~\ref{fig:double_sine}. Obviously, all bound states below the surface gap can be described by states localized in the potential minima of the Witten potentials, with harmonic oscillator form in a semiclassical approximation. 

We note that the topological protection of zero-energy states does neither require inversion nor mirror symmetry, consistent with Ref.~\cite{langbehn_etal_prl_17}. The effective surface Hamiltonian (\ref{eq:effective_surface_H_with_dimension}) has always two zero-energy solutions irrespective of the symmetry of the surface. However, regarding the exact twofold degeneracy of all states induced by SUSY, both inversion and mirror symmetry are essential.

\hs{Furthermore, we note that the SUSY properties obtained here via the realization of periodic Witten models is a nontrivial SUSY essentially related to the SOTI physics. In the absence of the surface gap (i.e., for zero Zeeman field), the Witten model (\ref{eq:witten_model_with_dimension}) turns into a model of a free particle on a ring with a trivial SUSY spectrum. Only the presence of the Zeeman field gives rise to a potential with non-trivial SUSY properties. In addition, the SUSY spectrum appears here for each chiral sector separately, and is not a consequence of the trivial SUSY spectrum of two partner potentials as usually discussed within Witten models for extended systems (where one of the zero-energy states is absent due to the asymptotic conditions). In contrast, for periodic Witten models, it is essential to have additional {\it nonlocal} symmetries to realize non-trivial SUSY spectra for each of the two partner potentials separately.}

%

\begin{figure}
	 \includegraphics[width =1.0\columnwidth]{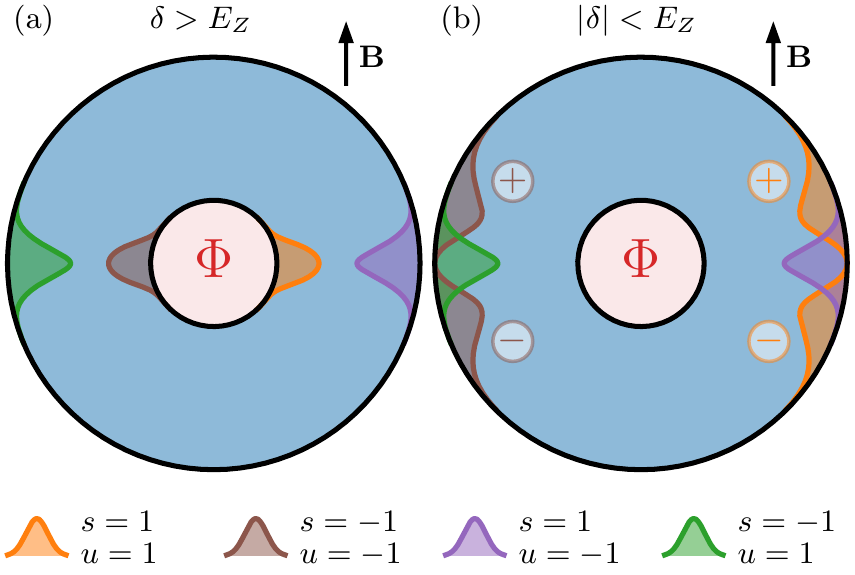}
	  \caption{Topological bound states for a Corbino disk for (a) the topological gapped phase $\delta > E_Z$ and (b) the Weyl phase $|\delta|<E_Z$. The two topological states at the outer surface persist in the Weyl phase, whereas the two topological hole states disappear at the phase transition and are replaced by two additional anti-symmetric states (indicated by the plus and minus sign symbol) at the outer surface. The indices $s=\pm 1$ and $u=\pm 1$ indicate the eigenvalues of the chiral symmetry and the SUSY.       
	  } 
    \label{fig:states_corbino}
\end{figure}

The fundamental connection between SUSY and higher-order topological phenomena is the center result of this work. Additionally, we apply and relate this insight to  aspects including the analysis of the phase diagram of the prototypical Bernevig-Hughes-Zhang (BHZ) model, the possibilities for topological engineering and the stability of topological states against various perturbations. First, for a finite system in the form of a Corbino disk (see Fig.~\ref{fig:states_corbino}), we discuss the topological  phase diagram both analytically and numerically. 
It turns out that the normal component of the Zeeman term controls the localization of the bound states along the surface, whereas the tangential component determines the normal localization length and drives the phase transition. Of particular interest is the gapless Weyl phase separating the topological from the non-topological gapped phase. At strong Zeeman field  the two topological states at the outer surface persist in the Weyl phase, whereas the two topological hole states disappear and are replaced by two anti-symmetric topological states at the outer surface. We will calculate all topological states in the two phases analytically and find excellent agreement with numerical results. Furthermore, by studying the lowest and next-lowest absolute value of the energy numerically in the whole phase diagram, we identify all phase boundaries and find perfect qualitative agreement with the analytical considerations. 

Secondly, we propose the hole states to be of particular interest for topological engineering. Taking a $2D$ system with several holes, one can control the topological states of each hole independently by local Zeeman fields and Aharonov-Bohm fluxes. With these two magnetic-field-only control elements, we show that one-hole and two-hole operations can be realized by tuning the shape of the topological states in tangential and normal direction via local Zeeman fields and by inducing a controlled coupling between the states of the same hole via tuning the flux away from half-integer value. 

Finally, we analyze the stability of the topological states against deviations from half-integer flux, flux penetration, surface distortions, and random impurities. For well-localized topological states we find rather robust stability up to impurity strengths of the order of the surface gap (or even beyond for particular spinor dependencies). Together with the fact that the BHZ model is a standard model discussed in topology with various realizations proposed in density functional theory \cite{marrazzo_etal_nlt_19} and experiments, such as in quantum wells of HgTe/CdTe and InAs/GaSb \cite{koenig_etal_science_07,knez_etal_prl_11}, $\rm ZrTe_5$ single crystal \cite{wu_etal_prx_16}, and in cold atom systems \cite{goldman_etal_prl_10,beri_cooper_prl_11,beeler_etal_nature_13,peng_etal_prl_14}, we expect that our proposal for the generic model involving only very basic ingredients can be realized in various material systems with sufficient stability. 

Our work is organized as follows. In Section~\ref{sec:model} we set up the basic model and discuss the phase diagram of the bulk spectrum. The central subject of the SUSY will be described in Section~\ref{sec:SUSY}, where we show that the squared Hamiltonian has an exact $n=1$ SUSY representation. Section~\ref{sec:corbino} is devoted to the full analytical theory for a Corbino disc, a prototypical example for an outer and inner surface, which contains the essential physics for any smooth surface. The general setup of the differential equations needed to solve for the topological states is outlined in Section~\ref{sec:topological_states_corbino}. Subsequently we will discuss the spectrum for zero Zeeman field in Section~\ref{sec:zero_B}, present the derivation of the effective surface Hamiltonian in the topological phase for a weak Zeeman field in Section~\ref{sec:corbino_weak_fields}, and study the topological states in the topological and Weyl phase for a strong Zeeman field in Section~\ref{sec:corbino_strong_fields}. In Section~\ref{sec:states_summary_numerics} we summarize the results for the topological states and compare to numerics. The general validity range of the effective surface Hamiltonian for any ratio of Zeeman and spin-orbit energy is presented in Section~\ref{sec:validity}. The whole phase diagram for a Corbino disk is presented numerically and compared to the analytical predictions in Section~\ref{sec:phase_diagram_corbino} by analysing the lowest and next-lowest absolute value of the energy eigenvalues as function of the model parameters. Section~\ref{sec:witten_model} is devoted to the derivation of the effective surface Hamiltonian for any smooth surface by using orthogonal coordinates. In Section~\ref{sec:witten_derivation}, we show that the square of the surface Hamiltonian is given by a generic periodic Witten model. The supersymmetric properties of the Witten model and the low-energy spectrum of the surface Hamiltonian is analyzed in Sections~\ref{sec:witten_SUSY} and \ref{sec:spectrum_surface_H}, respectively. An example for an area of peanut shape is discussed numerically and analytically in Section~\ref{sec:peanut}. In the final Section~\ref{sec:stability_tunability}, we discuss the stability of the topological states against various perturbations in Section~\ref{sec:stability}, and the possibilities for topological engineering in Section~\ref{sec:tunability}. We close with a summary and an outlook in Section~\ref{sec:summary}.

\section{Model}
\label{sec:model}

The 2D-SOTI with SUSY properties considered in this work are an important subclass of the 2D-SOTI listed within the general classification scheme developped in Refs.~\cite{langbehn_etal_prl_17,geier_etal_prb_18,trifunovic_brouwer_prx_19} (see, e.g., Section VI in Ref.~\cite{geier_etal_prb_18}). At zero Aharonov-Bohm flux the continuum version reads
\begin{align}
    \label{eq:H_zero_flux_general}
    H_0 = \Gamma_0 \left(\frac{p^2}{2m^*} - \delta \right) 
    \,+\,\alpha \,{\bf p}\cdot{\bf \Gamma} \,+\, E_Z \,{\bf b}\cdot{\bf \Gamma}\,\gamma,
\end{align}
where ${\bf p}=(p_x,p_y)$ is the momentum, $m^*$ denotes the effective mass, and ${\bf b}=(b_x,b_y)$ is a two-dimensional real unit vector in the plane of the system playing the role of the direction of a generalized in-plane Zeeman field. The generalized band inversion parameter, spin-orbit coupling, and Zeeman energy are denoted by the real numbers $\delta$, $\alpha$ and $E_Z$, respectively. In the remainder of this work we use units $\hbar=1$. The Hermitian and unitary spinor matrices $\Gamma_0$, ${\bf \Gamma}=(\Gamma_x,\Gamma_y)$ and $\gamma$ fulfill a certain algebra characteristic for a certain Shiozaki-Sato class \cite{shiozaki_sato_prb_14}. Specifically, we need the properties
\begin{align}
    \label{eq:Gamma_properties}
    \Gamma_i \Gamma_j &= - \Gamma_j \Gamma_i \,,\, \Gamma_i^2 = 1 \quad \text{for all}\quad i,j=0,x,y \,,\\
    \label{eq:gamma_properties}
    \Gamma_0 \gamma &= - \gamma \Gamma_0 \,,\, \Gamma_{x,y} \,\gamma = \gamma \,\Gamma_{x,y} \,,\, \gamma^2 = 1\,.
\end{align}
Due to rotational invariance we choose ${\bf b}={\bf e}_y$ by convention in the direction of the $y$-axis. Furthermore, without loss of generality, we assume $\alpha>0$ since a sign change of $\alpha$ is equivalent to changing ${\bf x}\rightarrow -{\bf x}$.

\begin{figure}
	 \includegraphics[width =0.95\columnwidth]{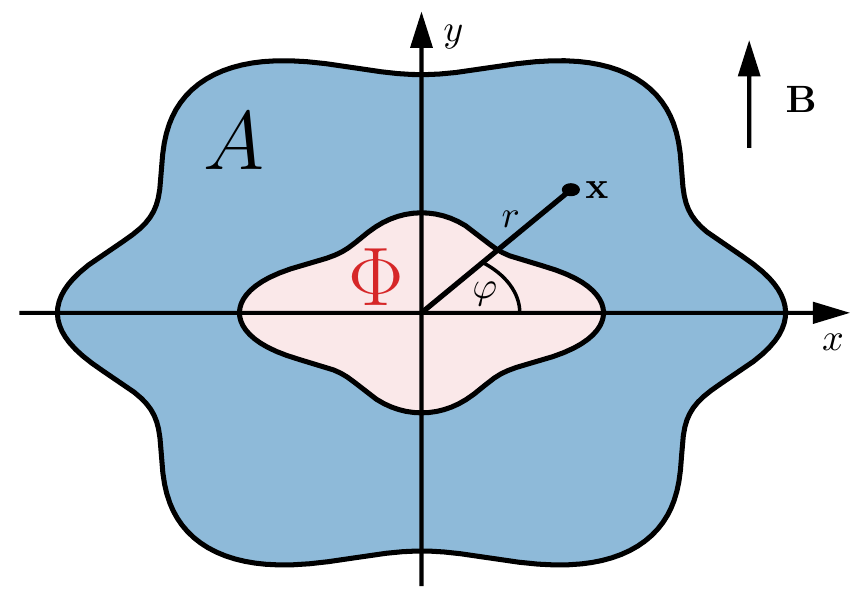}
	  \caption{Sketch of the system (shaded area $A$) with two boundaries formed by the inner and outer surface. A flux $\Phi$ is threaded through the hole formed by the inner surface. The $y$-coordinate is chosen in the direction of the Zeeman field $\bf{B}$. Polar coordinates are denoted by $(r,\varphi)$. If the area $A$ is chosen symmetrically to the $x$- and $y$-axis (as shown), the SUSY properties are exact and apply to all states.    
	  } 
    \label{fig:system}
\end{figure}

Taking a hole in the system around ${\bf x}=(x,y)=0$ and applying a perpendicular magnetic flux $\Phi$ through this hole, we have to shift the momentum ${\bf p}\rightarrow {\bf p} + \frac{e}{c}{\bf A}$ via the vector potential ${\bf A}=(A_x(x,y),A_y(x,y))$ (the $z$-component $A_z=0$ vanishes) with
\begin{align}
    \label{eq:vector_potential}
    \frac{e}{c}{\bf A}({\bf x})=\frac{f}{r^2}(-y,x) = f\,{\bf \nabla}\varphi\quad,\quad f=\frac{\Phi}{\Phi_0}\,, 
\end{align}
where $\Phi_0=\frac{hc}{e}$ is the flux quantum, $r=\sqrt{x^2 + y^2}$ denotes the radial coordinate and $0<\varphi<2\pi$ is the polar angle, which we choose by convention relative to the axis perpendicular to the Zeeman field, see Fig.~\ref{fig:system}. Our final Hamiltonian then reads
\begin{align}
    \label{eq:H_trafo_flux_general}
    H_f &= e^{-if\varphi}\,H_0\,e^{if\varphi} \\
    \label{eq:H_flux_general}
     &= \Gamma_0\,(\frac{1}{2m^*}{\bf p}_K^2 - \delta) 
    + \alpha\,{\bf p}_K\cdot{\bf \Gamma} + E_Z \Gamma_y\gamma\,.
\end{align}
where 
\begin{align}
    \label{eq:kinetic_momentum}
    {\bf p}_K = {\bf p} + \frac{e}{c}{\bf A} = e^{-if\varphi}\,{\bf p}\,e^{if\varphi}
\end{align}
denotes the kinetic momentum. Alternatively, using the transformation (\ref{eq:H_trafo_flux_general}), we note that the external flux can also be treated via the Hamiltonian (\ref{eq:H_zero_flux_general}) at zero flux but with twisted boundary conditions for the wave functions
\begin{align}
    \psi_0({\varphi+2\pi}) = e^{i2\pi f} \psi_0({\varphi})\,.
\end{align}

Although all our conclusions hold for the general Hamiltonian (\ref{eq:H_flux_general}), for pedagogical reasons we mostly consider in this work the very instructive case of a 2D-BHZ model with an in-plane Zeeman field, realized by the special choice
\begin{align}
    \label{eq:BHZ_algebra}
    \Gamma_0 = \sigma_z \quad,\quad {\bf \Gamma} = \sigma_x \,{\bf s} \quad,\quad \gamma=\sigma_x \,,
\end{align}
leading to the Hamiltonian
\begin{align}
    \label{eq:H_trafo_flux}
    H_f &= e^{-if\varphi}\,H_0\,e^{if\varphi} \\
    \label{eq:H_flux}
     &= \sigma_z\,(\frac{1}{2m^*}{\bf p}_K^2 - \delta) 
    + \alpha\,\sigma_x {\bf p}_K\cdot{\bf s} + E_Z s_y\,.
\end{align}
where ${\bf s}=(s_x,s_y)$ contains the Pauli matrices of the physical spin-$\frac{1}{2}$ operator ${\bf S}=\frac{1}{2}{\bf s}$, and $\sigma_i$ are the Pauli matrices describing the conduction and valence band. Due to the band inversion induced by the energy shift $\delta$ and the spin-orbit interaction $\alpha$ a bulk gap is opened hosting two counter-propagating helical edge modes at the boundary of a finite system. The edge modes are gapless for large system sizes but, as we will show below, acquire a finite gap induced by the Zeeman field realizing a second order topological insulator. We note that the spin-orbit interaction is equivalent to a Rashba-type $\sim (p_x s_y - p_y s_x)$ perpendicular to the system, since a spin rotation $(s_x,s_y,s_z)\rightarrow (-s_y,s_x,s_z)$ around the $z$-axis brings the Rashba interaction into the more convenient rotationally invariant form ${\bf p}\cdot{\bf s}$.   

By convention, we denote energies and length scales with a tilde symbol when they are measured with respect to the spin-orbit energy and spin-orbit length, respectively, defined by
\begin{align}
    \label{eq:E_so}
    E_{\rm{so}}&=\frac{k_{\rm so}^2}{2m^*}=\frac{1}{2}m^* \alpha^2 \,,\\
    \label{eq:so_length}
    \lambda_{\rm so}&=\frac{1}{k_{\rm so}}=\frac{1}{\alpha \,m^*}\,.
\end{align}
For Example, for the Zeeman energy $E_Z$ and the energy shift $\delta$ we define the dimensionless quantities
\begin{align}
    \label{eq:Zeeman_delta_tilde}
    \tilde{E}_Z = \frac{E_Z}{E_{\rm so}} =\frac{1}{\tilde{l}_B^2} \quad,\quad   
    \tilde{\delta} = \frac{\delta}{E_{\rm so}}\,.
\end{align}
In addition, we introduced the dimensionless length scale $\tilde{l}_B=l_B/\lambda_{\rm so}$, which characterizes the Zeeman field and will be called magnetic length in the following (note that it is not related to any orbital magnetic field). 

For an infinite system in the thermodynamic limit, i.e., when the outer surface goes to infinity, one can study in the asymptotic region the energy dispersion of the bulk states, see Appendix~\ref{app:bulk_spectrum}. Since the magnetic flux through the inner hole does not play any role in this regime, one obtains a flux-independent bulk gap given by 
\begin{align}
    \nonumber
    \tilde{\Delta}_{\rm bulk} &\equiv \Delta_{\rm bulk}/E_{\rm so}   \\
    \label{eq:bulk_gap}
    &\hspace{-1cm}
    =  \begin{cases}
        2\sqrt{\tilde{\delta}-1}-1/\tilde{l}_B^2  & \text{for}\quad \tilde{\delta} > {\rm max}\{2,1+1/(4\tilde{l}_B^4)\}\\
        |\tilde{\delta}|-1/\tilde{l}_B^2  & \text{for}\quad \tilde{\delta} < 2 \,\,{\rm and} \,\,|\tilde{\delta}|>1/\tilde{l}_B^2 \\
        0 & \text{otherwise}
        \end{cases}    \,.
\end{align}
Thus, the bulk gap closes at 
\begin{align}
    \label{eq:bulk_gap_closing}
    \tilde{\delta} \,=\,
    \begin{cases} 
        1 \,+\,\frac{1}{4\tilde{l}_B^4} & \text{for} \,\,\tilde{\delta} > 2\\
             \pm\frac{1}{\tilde{l}_B^2} &\text{for} \,\,\tilde{\delta} < 2
    \end{cases}\,.
\end{align}

\begin{figure}
	 \includegraphics[width =1.0\columnwidth]{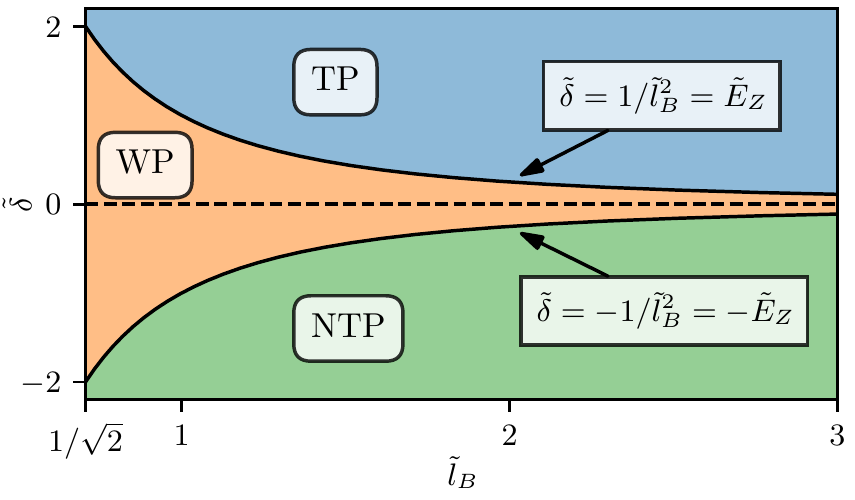}
	  \caption{The three different phases of the model as function of the band inversion shift $\tilde{\delta}=\delta/E_{\rm so}$ and the magnetic length $\tilde{l}_B=l_B/\lambda_{\rm so}$. Two lines at $\tilde{\delta}=\pm 1/\tilde{l}_B^2 = \pm \tilde{E}_Z = E_Z/E_{\rm so}$ separate the topological gapped phase (TP) $\tilde{\delta}>1/\tilde{l}_B^2$ from the Weyl phase (WP) $|\tilde{\delta}|<1/\tilde{l}_B^2$ and the non-topological gapped phase (NTP) $\tilde{\delta}<-1/\tilde{l}_B^2$.
	  } 
    \label{fig:bulk_gap}
\end{figure}

In this work we will restrict ourselves mostly to the regime of strong spin-orbit $\tilde{\delta}<2$ and $\tilde{l}_B > 1/\sqrt{2}$, or, equivalently, to
\begin{align}
    \label{eq:strong_spin_orbit}
    \delta , E_Z < 2 E_{\rm so}    \,.
\end{align} 
In this case, the spin-orbit length $\lambda_{\rm so}$ is the smallest length scale and is used for the lattice spacing within a discrete tight-binding formulation of the model, see Appendix~\ref{app:tb_numerics}. Therefore, only the gap closing lines at $\tilde{\delta}=\pm 1/\tilde{l}_B^2$ are of relevance here, see the black lines in Fig.~\ref{fig:bulk_gap}. They separate two gapped phases with a gapless regime in between. For the numerical implementation of the model we use both a continuum version in terms of a basis set of spherical Bessel functions (for an area of disk shape, see Appendix~\ref{app:disc_continuum_numerics}) as well as a tight-binding version described in Appendix~\ref{app:tb_numerics}. The latter approach has the advantage that it can deal with any shape of the area and the stability of topological states against disorder can be studied. For the special case of a disk we have compared the two different numerical methods and checked for quantitative agreement of the low-energy spectrum.  

The typical band structure in the gapless and gapped phase is shown in Figs.~\ref{fig:spectrum}(a,b). In the gapless case, we show in Appendix~\ref{app:bulk_spectrum} for $\tilde{\delta}<2$ that two Weyl points appear at ${\bf k}=\pm{\bf k}_W$ with ${\bf k}_W=(0,k_W)$ and 
\begin{align}
    \label{eq:weyl_point}
    \tilde{k}_W = k_W/k_{\rm so}=\sqrt{\tilde{\delta}-2+\sqrt{\tilde{E}_Z^2 + 4(1-\tilde{\delta})}}\,,
\end{align}
see also Fig.~\ref{fig:spectrum}(a). As a consequence we will denote the gapless phase by the Weyl phase (WP) in the following and will see later that it has also interesting topological properties (although less stable against disorder due to the absence of a gap). The two Weyl points are characterized by an anisotropic derivative of the dispersion in $k_x$ and $k_y$ direction 
\begin{align}
    \label{eq:weyl_derivative_kx}
    \frac{\partial\tilde{\epsilon}_{\bf k}}{\partial \tilde{k}_x}|_{{\bf k}={\bf k}_W} &= 2\,, \\
    \label{eq:weyl_derivative_ky}
    \frac{\partial\tilde{\epsilon}_{\bf k}}{\partial \tilde{k}_y}|_{{\bf k}={\bf k}_W} &= 
    \frac{2\tilde{k}_W}{\tilde{E}_Z}\sqrt{\tilde{E}_Z^2 + 4(1-\tilde{\delta})}\,.
\end{align}
On the gap closing line $\delta=\pm E_Z$, with $E_Z < 2E_{\rm so}$, the two Weyl points merge together to a single point at ${\bf k}=0$, with a topological gapped phase (TP) for $\delta > E_Z$ and a non-topological gapped phase (NTP) for $\delta < - E_Z$, see the detailed discussion of the phase diagram in Section~\ref{sec:phase_diagram_corbino}. Since we restrict ourselves to the regime of strong spin-orbit $\delta, E_Z < 2 E_{\rm so}$ in this work, we note that the minimum of the dispersion is always at $k=0$ in the gapped phase, see Appendix~\ref{app:bulk_spectrum}.   

\begin{figure}
	 \includegraphics[width =1.0\columnwidth]{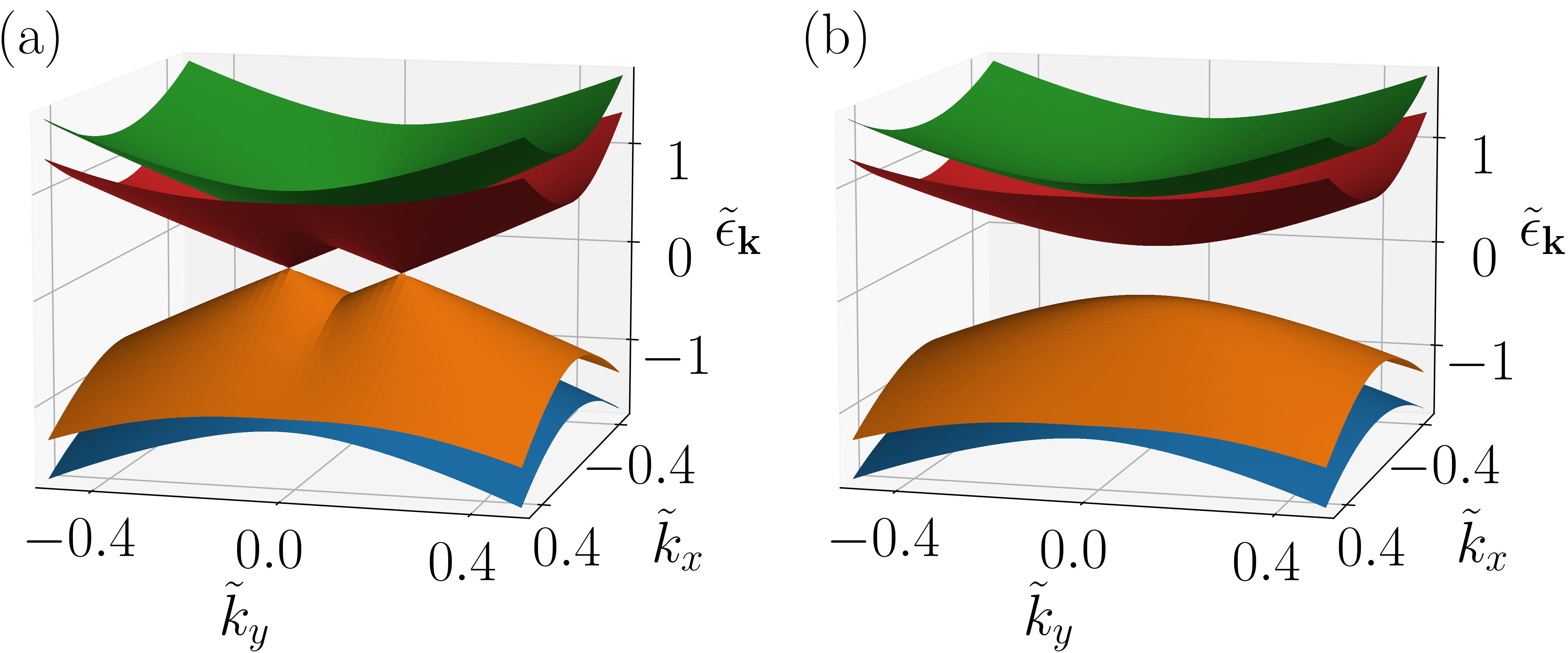}
	  \caption{The spectrum $\tilde{\epsilon}_{\bf k}=\epsilon_{\bf k}/E_{\rm so}$ of the four bands as function of $\tilde{k}_{x,y}=k_{x,y}/k_{\rm so}$ for $\tilde{l}_B=2$, with (a) $\tilde{\delta}=0$ in the gapless phase and (b) $\tilde{\delta}=0.5$ in the gapped topological phase. In the gapless phase (a) one finds two Weyl points at $\pm(0,\tilde{k}_W)$, with $\tilde{k}_W$ given by Eq.~(\ref{eq:weyl_point}). 
	  } 
    \label{fig:spectrum}
\end{figure}

\section{Supersymmetry}
\label{sec:SUSY}

In this section we will analyze the supersymmetric structure of our model. In Section~\ref{sec:symmetries} we state the exact symmetries of the general model Hamiltonian (\ref{eq:H_flux_general}) and find besides chiral and inversion symmetry another important symmetry, a mirror symmetry with respect to the axis perpendicular to the Zeeman field. At the particular value of half-integer flux $f=\frac{1}{2}$ it is shown that the mirror symmetry anti-commutes with the inversion symmetry leading to a nontrivial realization of SUSY. For this reason, the mirror symmetry is called SUSY in the following. We show that the SUSY protects a twofold degeneracy of all eigenstates and, as a consequence, will protect a pair of topological bound states at zero energy (if present). The topological invariant is then given by the Witten index which distinguishes between unbroken and broken SUSY, depending on whether states with zero (or exponentially small) energies are present or not, respectively. In Section~\ref{sec:SUSY_general} we present the formal representation of SUSY in terms of the supercharge operator and show that it relies only on the presence of inversion and mirror symmetry, independent of the special form (\ref{eq:H_flux_general}) of the model.

\subsection{Symmetries of the model}
\label{sec:symmetries}

Starting from the general form of the Hamiltonian (\ref{eq:H_flux_general}) we find chiral and inversion symmetry given by
\begin{align}
    \label{eq:chiral}
    S H_f S = - H_f \quad &,\quad S = S^\dagger = \Gamma_x \Gamma_y \gamma \,,\\
    \Pi H_f \Pi = H_f \quad &,\quad \Pi = \Pi^\dagger = P_{{\bf x}} \,\Gamma_0\,,
\end{align}
where $P_{{\bf x}}$ denotes the parity transformation ${\bf x}\rightarrow -{\bf x}$. The Hamiltonian does not fulfill time-reversal symmetry but, for the special BHZ-type realization (\ref{eq:BHZ_algebra}), we can relate $H_f$ and $H_{-f}$ via the anti-unitary transformation
\begin{align}
    \label{eq:time_reversal}
    \sigma_z s_x H_f^* \sigma_z s_x = H_{-f}\,.
\end{align}
As a consequence, the spectra of $H_f$ and $H_{-f}$ are the same. 

Furthermore, using (\ref{eq:kinetic_momentum}), we find another important mirror symmetry 
\begin{align}
    U_f H_f U_f &= H_f \,,\\ 
    \label{eq:U_f_1}
    U_f = U_f^\dagger &= e^{-if\varphi}\,P_\varphi\,\Gamma_0 \Gamma_y \gamma\,e^{if\varphi} \\
    \label{eq:U_f}
    &= e^{-i2f\varphi}\,P_\varphi\,\Gamma_0 \Gamma_y \gamma \,,
\end{align}
where $P_\varphi$ denotes a sign change of the polar angle $\varphi \rightarrow - \varphi$, which is equivalent to changing the sign of the $y$-coordinate, i.e., the sign of the coordinate along the Zeeman field. The exponentials gauge away the flux such that the two components of the kinetic momentum are transformed with different signs
\begin{align}
    \label{eq:kinetic_momentum_x_trafo}
    U_f \, (p_x + A_x) \, U_f &= p_x + A_x \,,\\
    \label{eq:kinetic_momentum_y_trafo}
    U_f \, (p_y + A_y) \, U_f &= - (p_y + A_y) \,.
\end{align}
However, to respect periodic boundary conditions under $\varphi\rightarrow\varphi + 2\pi$, the transformation $U_f$ is only an allowed symmetry for integer and half-integer fluxes $f=0,1/2\mod(1)$ in units of the flux quantum. As we will show in the following the interesting case is a half-integer flux where $U_{1/2}$ turns out to be a SUSY leading to a typical SUSY spectrum with an exact twofold degeneracy of all eigenstates except for a single state at zero energy (due to chiral symmetry this SUSY spectrum turns out to occur twice for the absolute value of the Hamiltonian, see below).  

The crucial property of the SUSY operator $U_{1/2}$ at half-integer flux is its anticommutation with inversion symmetry
\begin{align}
    \label{eq:U_Pi_anticommutation}
    U_{1/2}\,\Pi = - \Pi\,U_{1/2} \,,
\end{align}
whereas, for integer flux, one gets the commutation $U_0\Pi=\Pi U_0$. Since we can choose all eigenstates $|\psi\rangle$ of the Hamiltonian simultaneously as eigenfunctions of the inversion symmetry $\Pi |\psi\rangle = \pm |\psi\rangle$, we find for half-integer fluxes that $|\psi\rangle$ and its SUSY partner $U_{1/2} |\psi\rangle$ must be orthogonal
\begin{align}
    \nonumber
    &\langle \psi | U_{1/2} |\psi\rangle = \langle \psi|\Pi \,U_{1/2} \,\Pi |\psi\rangle \\
    \label{eq:SUSY_proof}
    &= -\langle \psi|\,U_{1/2} \,\Pi\,\Pi |\psi\rangle = -\langle \psi|\,U_{1/2} |\psi\rangle\,,
\end{align}
where we used (\ref{eq:U_Pi_anticommutation}) and $\Pi^2=1$ in the last two steps. Since both $|\psi\rangle$ and $U_{1/2} |\psi\rangle$ are eigenstates of the Hamiltonian with the same energy, this leads necessarily to an at least twofold degenerate spectrum of the Hamiltonian. This is similiar to Kramers degeneracy but the orthogonality of time-reversed partners is replaced by orthogonality of SUSY partners. For our model it turns out that no further degeneracies are present, i.e., the degeneracy is given precisely by two. 

Due to chiral symmetry all eigenstates $|\psi\rangle$ at positive energy have a counterpart $S|\psi\rangle$ with negative energy. All eigenstates are twofold degenerate due to SUSY and, therefore, also a possible zero energy state must be twofold degenerate. 
Since both the bulk and edge state spectrum is gapped in the presence of spin-orbit and Zeeman interaction, the zero energy states correspond to topological bound states generated by second order topology. If they exist, they are topologically protected by SUSY since a splitting would break the twofold degeneracy (note that chiral symmetry alone would allow for such a splitting). Furthermore, due to chiral symmetry, a pair of zero-energy states can not shift away from zero-energy. As a result we find here a topological protection via the combination of chiral symmetry with SUSY, quite similiar to topological protection induced by chiral symmetry and time-reversal symmetry with $T^2=-1$ (leading to Kramers degeneracy). 

To reveal the typical SUSY structure of the spectrum, it is most convenient to start with a hole in an infinite system, as sketched in Fig.~\ref{fig:hole_system}(a). In this case one obtains in the TP two topological states exactly at zero energy localized at two opposite points of the hole surface with different chirality $s$ and different value $u$ for the SUSY. This happens not only for a circular hole but for all mirror-symmetric hole surfaces where several pairs very close to zero energy can appear at the positions where the normal component of the Zeeman term changes sign. However, as explained later in more detail, one of these pairs will lie exactly at zero energy at the SUSY point for half-integer flux whereas the other ones are at exponentially small energy for a large hole radius. Instead of considering the Hamiltonian $H$ it is then more convenient to identify the SUSY structure of the spectrum by considering the squared Hamiltonian 
\begin{align}
    \label{eq:H_squared}
    {\cal H}_W\equiv H_{1/2}^2 \,,
\end{align}
which has only positive or zero energies. This model (with dimension of energy squared) is called a Witten model for supersymmetric systems. Since ${\cal H}_W$ commutes with the chiral symmetry one obtains a SUSY spectrum in each chiral sector separately, with a twofold degeneracy of all states with positive energy (labeled by the supersymmetry $u=\pm 1$) and a unique zero energy state in the TP, see Fig.~\ref{fig:hole_system}(b) where we show the spectrum of the absolute value of the Hamiltonian $|H|=\sqrt{{\cal H}_W}$. In this case one obtains unbroken SUSY, whereas in the WP and the NTP the zero energy states do not exist, denoted by a broken SUSY. We note that the SUSY structure of the spectrum applies to all states of the system, i.e., to the states localized at the boundary below the bulk gap $\Delta_{\rm bulk}$ and to the bulk states above the bulk gap (where the continuum has a fourfold degeneracy with respect to $s=\pm 1$ and $u=\pm 1$). As already explained in the introduction the boundary states below the bulk gap consist of three different kinds: (1) edge states extended along the surface for energies between the surface gap $\Delta_{\rm surface}$ and the bulk gap, (2) localized bound states at finite energy below the surface gap, and (3) topological bound states exactly at zero energy. Whether the energy scale of the first bound state at non-zero energy is given by the Witten frequency $\Omega_W$ or not, depends on the shape of the hole. If $2n$ points are present on the hole surface where the normal component of the Zeeman term changes sign, one obtains one pair of topological states at zero energy and $n-1$ pairs of localized bound states with an exponentially small energy (i.e., for a disk with $n=1$, the first pair of bound states at non-zero energy starts with the Witten frequency as shown in Fig.~\ref{fig:hole_system}(b)). 

For a finite system with both an inner and an outer surface, we note that there are never states strictly at zero energy, due to an exponentially small splitting induced by a hybridization between zero-energy states localized at the inner and outer surface, see also the more detailed discussion below. Therefore, in a strict mathematical sense the SUSY is always broken for a finite system. However, when neglecting the experimentally unmeasurable and exponentially small splitting between the zero-energy states localized at different positions in a finite system (as is standardly done for all topological systems), it is reasonable from a physical point of view to use the nomenclature of unbroken SUSY also for this case. Nevertheless, one should keep in mind that the SUSY structure of the spectrum of the Witten model ${\cal H}_W$ applies only to the states localized either at the inner or the outer surface but not to the bulk states for a finite system. The discrete bulk states are not related to the inner or outer surface and just have a fourfold degeneracy due to chiral symmetry and SUSY. It is then unclear how to associate a given SUSY pair of bulk states at fixed chirality to the two SUSY spectra of the boundary states at the inner and outer surface with the same chirality, and any choice would be ambigious and very unphysical. Therefore, for a finite system, the SUSY structure of the spectrum applies only to the effective surface Hamiltonian to be introduced later and is closely related to the second-order mechanism of inducing zero-energy topological surface states.         

We note that the symmetry considerations in this section apply to all states of the system, irrespective of whether they are two-dimensional bulk states, one-dimensional edge states along the boundary of the system, or zero-dimensional bound states generated by 2nd-order topology. However, in order for our symmetry arguments to apply for a system with a boundary living only in a finite region $\bf{x}\in A$, we have to require that the corresponding confinement potential  
\begin{align}
    \label{eq:conf_potential}
    V(\bf{x}) = \sigma_z\,\begin{cases} 0 & \text{for}\quad \bf{x}\,\in\, A \\ \infty & \text{for} \quad \bf{x}\,\not\in\, A  \end{cases}
\end{align}
fulfils the same symmetries. Obviously, this is only the case when the area $A$ is both symmetric under reflection of the $x$- or $y$-coordinate, i.e., if $(x,y)\in A$, then also $(-x,y),(x,-y),(-x,-y)\in A$ must be fulfilled, see Fig.~\ref{fig:system}.

Zero energy topological states are bound states localized at the boundary of the system, i.e., either at the inner or outer surface. If present, we will choose them by convenience as eigenfunctions of the two commuting symmetries $S$ and $U_{1/2}$ with eigenvalues $s,u=\pm 1$
\begin{align}
    \label{eq:topological_states_H}
    H_{1/2}|\psi_{su}\rangle &= 0 \,,\\
    \label{eq:toplogical_states_SU}
    S|\psi_{su}\rangle = s|\psi_{su}\rangle\,&,\,U_{1/2}|\psi_{su}\rangle = u|\psi_{su}\rangle\,.
\end{align}
Since $S$ and $U_{1/2}$ anticommute with $\Pi$ we can furthermore choose the eigenstates such that the application of the inversion symmetry changes both the sign of $S$ and $U_{1/2}$
\begin{align}
    \label{eq:change_us}
    |\psi_{-s,-u}\rangle = -\Pi |\psi_{su}\rangle \,,
\end{align}
where, for convenience (see later), we introduced a sign factor here. As a consequence, topological states appear for mirror symmetric areas always in pairs of two bound states localized at two points of the boundary at oppposite positions with different signs for $s$ and $u$. These two wave functions can not hybridize since they have different eigenvalues of the supersymmetry. For a finite system several pairs can occur (either at different or on the same surface), where the wave functions from different pairs have the same value of $u$ and different values of $s$. In this case, the wave functions from two different pairs can hybridize via the Hamiltonian, such that the exact eigenstates are no longer eigenstates of the chiral symmetry but appear in two pairs at non-zero energy $\pm\epsilon$. This is typical for all finite topological systems where localized bound states can appear at two different ends of the system with an exponentially small orbital overlap. This overlap leads to an exponentially small splitting of the two states which can be neglected for a large system. Therefore, when neglecting the exponentially small orbital overlap of wave functions from different pairs, we can still use the states $|\psi_{su}\rangle$ as the topological bound states which are localized at a certain position and are eigenstates of $S$ and $U_{1/2}$. As a consequence, (\ref{eq:topological_states_H}) has to be changed to $H_{1/2}|\psi_{su}\rangle \approx 0$ up to exponentially small terms but (\ref{eq:toplogical_states_SU}) remains the same. Numerically, this is achieved by first determining the two pairs of states with energy closest to zero energy and, subsequently, diagonalizing $S$ and $U_{1/2}$ in this four-dimensional subspace such that (\ref{eq:toplogical_states_SU}) and (\ref{eq:change_us}) are fulfilled. 

We will see that the topological bound states appear always at the positions of the surface where the normal component of the Zeeman field changes sign. This has already been discussed in other works \cite{khalaf_prb_18,ren_etal_prl_20,laubscher_etal_prr_19,plekhanov_etal_prr_19,volpez_etal_prl_19,laubscher_etal_prb_20,laubscher_etal_prr_20,plekhanov_etal_prr_20,plekhanov_etal_prb_21} for sharp corners in $2D$ systems, where the emergence of topological bound states via second-order topology is induced by the application of an in-plane Zeeman field breaking rotational invariance around the $z$-axis. It is related to the occurrence of bound states at the interface of two effective edge state Hamiltonians with the Zeeman term being the mass term and changing sign. Similarly, we will show below that the same mechanism happens here but, instead of considering sharp corners at the boundary as in previous works, we will analyze arbitrary smooth surfaces where the curvature radius is much larger than the localization length of the bound states. This will allow us to derive effective surface Hamiltonians for a given surface in the form of generic periodic Witten models with supersymmetric properties. Moreover, within this formalism, we will find a new viewpoint for the occurrence of topological states being trapped in minima of effective surface potentials. In particular, this allows for a full analytical theory to determine the wave functions of the topological states, together with the analysis of other bound states at higher energy. Since the Corbino disk contains already many of the possible scenarios, we will consider a Corbino disk in section~\ref{sec:corbino} and leave the discussion of arbitrary smooth surfaces to Section~\ref{sec:witten_model}.

\subsection{SUSY representation in terms of supercharge operator} 
\label{sec:SUSY_general}

To write the Witten Hamiltonian ${\cal H}_W$ in the formal framework of SUSY Hamiltonians we define the Hermitian supercharge operator $Q=Q^\dagger$ and the involution $K$ by
\begin{align}
    \label{eq:supercharge}
    Q = H_{1/2} \, \Pi \quad,\quad K = U_{1/2} \,,
\end{align}
and find the so-called $n=1$ SUSY representation \cite{combescure_etal_jpa_04}
\begin{align}
    \label{eq:SUSY_n=1}
    {\cal H}_W = Q^2 \quad,\quad QK=-KQ \quad,\quad K^2=1\,.
\end{align}
Using 
\begin{align}
    \label{eq:S_U_Pi_relations}
    S\,U_{1/2} = U_{1/2}\,S \quad,\quad  S\,\Pi = - \Pi\,S \,,
\end{align}
we find that the chiral symmetry commutes with both $Q$ and $K$. Therefore, the representation (\ref{eq:SUSY_n=1}) is valid in both chiral sectors separately. Within each chiral sector, the twofold degeneracy follows only for all states with positive eigenvalue of ${\cal H}_W$, but not for a possible state with zero eigenvalue. This follows by taking the eigenstates $|\psi\rangle$ of ${\cal H}_W$ simultaneously as eigenstates of the involution $K$. One then gets from $QK=-KQ$ and $K^2=1$ analog to (\ref{eq:SUSY_proof}) that $|\psi\rangle$ and $Q|\psi\rangle$ are orthogonal to each other and are both eigenstates of ${\cal H}_W=Q^2$ with the same eigenvalue $E$. For $E>0$, we get $Q|\psi\rangle\ne 0$, leading to a twofold degeneracy. However, since $Q$ is not unitary, it is also possible that $Q|\psi\rangle=0$, which must be obviously the case for the state with $E=0$. Therefore, one gets a non-degenerate eigenstate of ${\cal H}_W$ with zero eigenvalue in each chiral sector (at least if it exists).    

Equivalently, one can also find a so-called $M=1$ SUSY representation (without an involution and a non-Hermitian supercharge operator) or a $m=2$ realization (with two Hermitian supercharge operators and no involution) \cite{com_1} via the definitions
\begin{align}
    \label{eq:bar_Q}
    \bar{Q} &= \frac{1}{2} Q(1+K) \,,\\
    \label{eq:Q_1}
    Q_1 &= \frac{1}{\sqrt{2}}(\bar{Q} + \bar{Q}^\dagger) = Q_1^\dagger \,,\\
    \label{eq:Q_2}
    Q_2 &= -i\frac{1}{\sqrt{2}}(\bar{Q} - \bar{Q}^\dagger) = Q_2^\dagger \,.
\end{align}
It is then straightforward to show that one gets the $M=1$ form
\begin{align}
    \label{eq:SUSY_M=1}
    {\cal H}_W = \bar{Q}\bar{Q}^\dagger + \bar{Q}^\dagger\bar{Q}  \quad,\quad \bar{Q}^2=0 \,,
\end{align}
and the $m=2$ form
\begin{align}
    \label{eq:SUSY_m=2}
    {\cal H}_W = Q_1^2 = Q_2^2 \quad,\quad Q_1 Q_2 = - Q_2 Q_1 \,.
\end{align}

We note that the general SUSY representation does not rely on chiral symmetry and is possible for any system with inversion and mirror symmetry, independent of the special form (\ref{eq:H_flux_general}) of the Hamiltonian. Let us assume that the Hamiltonian has inversion symmetry $\Pi = P_{\bf x} \Gamma_\Pi$ and mirror symmetry $U_0 = P_\varphi \Gamma_{U}$ at zero flux, where $\Gamma_\Pi$ and $\Gamma_U$ are any spinor matrices which either commute or anti-commute with each other (which is always the case when they consist of any product of Pauli matrices from different spinor degrees of freedom). If they anti-commute already at zero flux, we can use the above construction with $U_{1/2}\rightarrow U_0$ and get a SUSY realization at zero flux. If they commute we can apply a half-integer flux and define the new mirror symmetry
\begin{align}
    \label{eq:mirror_half_integer}
    U_{1/2} = e^{-i\varphi} U_0 = e^{-i\varphi/2}\,U_0\,e^{i\varphi/2}\,.
\end{align}
By construction, $U_{1/2}$ is a mirror symmetry of the Hamiltonian at half-integer flux and anti-commutes with $\Pi$. As a consequence, we obtain a SUSY realization at half-integer flux.

\section{Corbino disc}
\label{sec:corbino}

To discuss the phase diagram of the Hamiltonian in terms of the number of topological states at exponentially small energies it is most convenient to start with the discussion of a Corbino disk with outer radius $R_>$ and a hole of inner radius $R_<$ through which we apply the flux $\Phi = f \Phi_0$, see Fig.~\ref{fig:corbino}. We discuss here the most interesting case of half-integer flux $f=1/2$ and state at the appropriate places the stability of the topological states for small deviations from half-integer flux. We present the analysis of the topological states and the derivation of the effective surface Hamiltonian for the special case of the BHZ-model with Zeeman field given by Eq.~(\ref{eq:H_flux}), but note that analog considerations can be done for the more general model (\ref{eq:H_flux_general}). We start in Section~\ref{sec:topological_states_corbino} with the general setup for the differential equations to be solved for the topological states, and discuss subsequently the cases of zero Zeeman field in Section~\ref{sec:zero_B}, weak Zeeman field in Section~\ref{sec:corbino_weak_fields}, and strong Zeeman field in Section~\ref{sec:corbino_strong_fields}. For readers not interested in the technical details, the wave functions of the topological states are summarized in Section~\ref{sec:states_summary_numerics} and compared to numerical results. In Section~\ref{sec:validity} we will state the generic validity range of the low-energy theory in terms of the universal low-energy surface Hamiltonian. Based on these results we will then discuss the phase diagram in terms of the Witten index in Section~\ref{sec:phase_diagram_corbino} and again compare with numerical results.

\subsection{Topological states}
\label{sec:topological_states_corbino}

For a Corbino disk it is most convenient to represent the Hamiltonian in polar coordinates and rotate the spin in radial and angular direction locally to the $x$- and $y$-axis, respectively. This is achieved by the following unitary transformation
\begin{align}
    \label{eq:polar_H_trafo}
    \bar{H}_{1/2} =  X^\dagger W^\dagger \, U^\dagger \,\sqrt{r}\, H_{1/2} \, \frac{1}{\sqrt{r}} \, U \, W \, X \,,
\end{align}
with 
\begin{align}
    \label{eq:U_rotation}
    U &= e^{-i\frac{1}{2}(1+s_z)\varphi} \,,\\
    \label{eq:W}
    W &= \frac{1}{2}(1+\sigma_y) + s_z \frac{1}{2}(1-\sigma_y) \,,\\
    \label{eq:X}
    X &= e^{i\frac{\pi}{4}\sigma_y} = \frac{1}{\sqrt{2}}(1+i\sigma_y)\,.
\end{align}
The transformation with $\sqrt{r}$ is convenient due to the transformation of the area element $dxdy=rdrd\varphi$ and leads to the normalization condition
\begin{align}
    \label{eq:polar_normalization} 
    \int_{R_<}^{R_>} dr \,\int_0^{2\pi} d\varphi \sum_{\sigma_z,s_z=\pm} |\bar{\psi}(r,\varphi;\sigma_z,s_z)|^2 = 1 \,. 
\end{align}
for the eigenfunctions $\bar{\psi}(r,\varphi;\sigma_z,s_z)$ of $\bar{H}_{1/2}$.

\begin{figure}
	 \includegraphics[width =0.95\columnwidth]{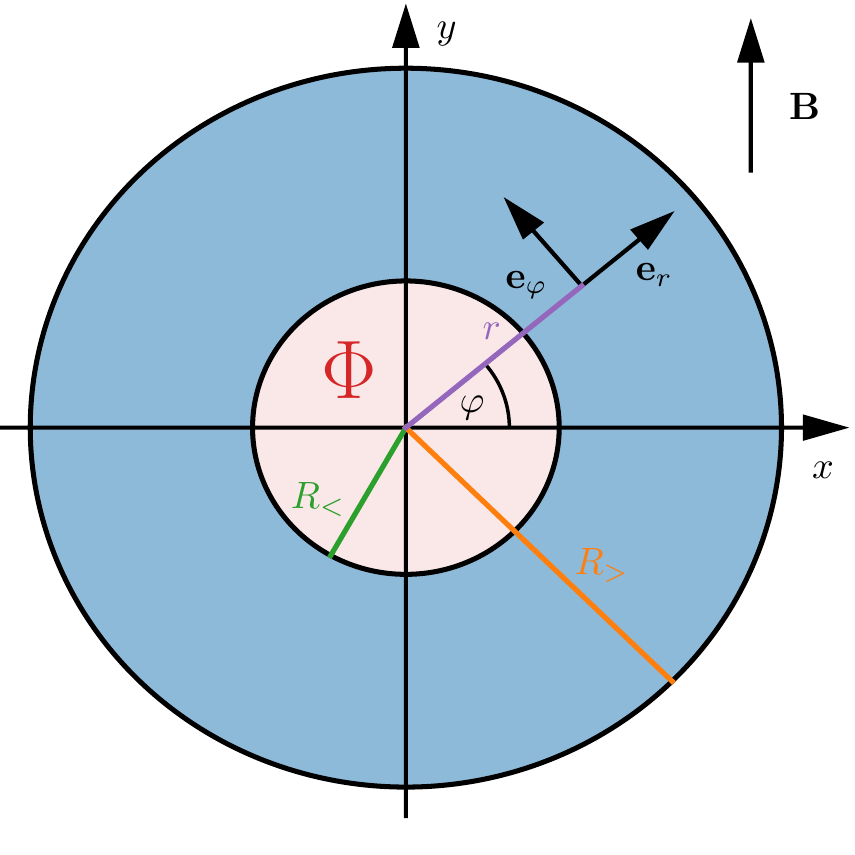}
	  \caption{Corbino disk with inner radius $R_<$ and outer radius $R_>$ (shaded area). A flux $\Phi$ is threaded through the hole.  The $y$-coordinate is chosen in the direction of the Zeeman field $\bf{B}$. Polar coordinates are denoted by $(r,\varphi)$ and ${\bf e}_{r,\varphi}$ denote the local unit vectors in radial and angular direction.    
	  } 
    \label{fig:corbino}
\end{figure}

The transformation $U$ eliminates the half-integer flux and rotates the radial and angular spin locally to $s_x$ and $s_y$, respectively, according to 
\begin{align}
    \label{eq:U_trafo}
    U^\dagger {\bf e}_r \cdot {\bf s} \,U = s_x \quad,\quad U^\dagger {\bf e}_\varphi \cdot {\bf s} \,U = s_y\,,
\end{align}
where ${\bf e}_{r,\varphi}$ are the unit vectors in radial and angular direction, see Fig.~\ref{fig:corbino}. We note that the transformation $U$ does not change the boundary conditions in angular direction since it is periodic under $\varphi\rightarrow\varphi+2\pi$ (note that $s_z=\pm 1$). 

Finally, the unitary transformations $W$ and $X$ are chosen for convenience to simplify the spinor structure. Whereas $X$ rotates the orbital spinor by $\pi/2$ around the $y$-axis
\begin{align}
    \label{eq:X_trafo}
    X^\dagger \sigma_z X = \sigma_x \quad,\quad X^\dagger \sigma_x X = -\sigma_z\,,
\end{align}
the transformation $W$ has the effect
\begin{align}
    \label{eq:W_trafo}
    s_{x,y} \xrightarrow{W} s_{x,y} \,\sigma_y \quad,\quad
    \sigma_{x,z} \xrightarrow{W} \sigma_{x,z} \,s_z \,,
\end{align}
while keeping $s_z$ and $\sigma_y$ invariant. 

A straightforward calculation gives the following result for the transformed Hamiltonian in dimensionless units
\begin{align}
    \label{eq:bar_H_polar_1}
    \bar{H}_{1/2}/E_{\rm so} &= \sigma_x \left[(-\partial^2_{\tilde{r}} - \tilde{\delta}) s_z
    + 2i \partial_{\tilde{r}}s_y) \right]\\
    \label{eq:bar_H_polar_2}
    & + \sigma_x (- \frac{2}{\tilde{r}}i\partial_\varphi s_x - \frac{1}{\tilde{r}^2}\partial_\varphi^2 s_z 
    + \frac{1}{\tilde{r}^2}i\partial_\varphi)\\
    \label{eq:bar_H_polar_3}
    & + \sigma_y \frac{1}{\tilde{l}_B^2}(s_x\,\sin{\varphi} + s_y\, \cos{\varphi} ) \,,
\end{align}
where $\tilde{r}=r/\lambda_{\rm so}$, $\tilde{\delta}=\delta/E_{\rm so}$, and $\tilde{l}_B=l_B/\lambda_{\rm so}$. 

After the transformation we get for the transformed symmetry operators
\begin{align}
    \label{eq:S_trafo}
    \bar{S} &= - \sigma_z \,,\\
    \label{eq:Pi_trafo}
    \bar{\Pi} &= - P_{\bf x} \sigma_x \,,\\
    \label{eq:SUSY_trafo}
    \bar{U}_{1/2} &= P_\varphi\sigma_z s_x \,.
\end{align}
We also note that the total angular momentum in $z$-direction $J_z=L_z+s_z/2$, with $L_z=-i\partial_\varphi$, transforms as
\begin{align}
    \label{eq:J_z_trafo}
    \bar{J}_z = L_z - \frac{1}{2}\,.
\end{align}

To discuss the appearance of topological states at zero energy we first write the Hamiltonian and the symmetry operators in the $\sigma_z$-basis as
\begin{align}
    \label{eq:bar_H_S_sigma_z}
    \bar{H}_{1/2} = \left(\begin{array}{cc}
    0 & \bar{A} \\ \bar{A}^\dagger & 0 \end{array}\right)\quad &, \quad
    \bar{S} = \left(\begin{array}{cc}
    -1 & 0 \\ 0 & 1 \end{array}\right)\,,\\
    \label{eq:bar_U_Pi_sigma_z}
    \bar{\Pi} = - P_{\bf x} \left(\begin{array}{cc}
    0 & 1 \\ 1 & 0 \end{array}\right)\quad &, \quad
    \bar{U}_{1/2} = P_\varphi s_x \left(\begin{array}{cc}
    1 & 0 \\ 0 & -1 \end{array}\right)\,,
\end{align}
with 
\begin{align}
    \nonumber
    \bar{A}/E_{\rm so} &= (-\partial^2_{\tilde{r}} - \tilde{\delta}) s_z + 2i\partial_{\tilde{r}}s_y \\
    \nonumber
    & - \frac{2}{\tilde{r}}i\partial_\varphi s_x - \frac{1}{\tilde{r}^2}\partial_\varphi^2 s_z 
    + \frac{1}{\tilde{r}^2}i\partial_\varphi\\
    \label{eq:A}
    & -i \frac{1}{\tilde{l}_B^2}(s_x\,\sin{\varphi} + s_y\, \cos{\varphi} ) \,.
\end{align}

For the zero energy states $|\bar{\psi}_{su}\rangle$ of $\bar{H}_{1/2}$, we have to solve
\begin{align}
    \label{eq:topo_A}
    \bar{A}\Phi_u = 0 \,,
\end{align}
and get from (\ref{eq:toplogical_states_SU}), (\ref{eq:change_us}), (\ref{eq:S_trafo}) and (\ref{eq:Pi_trafo})
\begin{align}
    \label{eq:topo_bar_1u}
    \bar{\psi}_{1,u}(\tilde{r},\varphi;\sigma_z,s_z) &= \Phi^{(u)}(\tilde{r},\varphi;s_z) \left(\begin{array}{c} 0 \\ 1 \end{array}\right)_{\sigma_z} \,,\\
    \nonumber
    \bar{\psi}_{-1,u}(\tilde{r},\varphi;\sigma_z,s_z) &= -(\bar{\Pi}\bar{\psi}_{1,-u})(\tilde{r},\varphi;\sigma_z,s_z) \\
    \label{eq:topo_bar_-1u}
    &\hspace{-1cm}
    = \Phi^{(-u)}(\tilde{r},\varphi+\pi;s_z) \left(\begin{array}{c} 1 \\ 0 \end{array}\right)_{\sigma_z}\,.
\end{align}
Noting that $\bar{A}$ anticommutes with $P_{\varphi}s_x$ and using $P_\varphi s_x=-\bar{U}_{1/2}$ in the subsector $\sigma_z=-1$ according to (\ref{eq:SUSY_trafo}), we can choose the two zero energy states $\Phi_u$ as eigenfunctions of $P_\varphi s_x$ with eigenvalues $-u$. This gives the following form for $\Phi_u$ 
\begin{align}
    \nonumber
    \Phi^{(u)}(\tilde{r},\varphi;s_z) &=  \frac{1}{\sqrt{2}}\left\{\chi^{(u)}_{-u}(\tilde{r},\varphi) 
    \left(\begin{array}{c} 1\\1\end{array}\right)_{s_z} + \right.\\
    \label{eq:Phi_u}
    &\hspace{1cm}
    + \left.\chi^{(u)}_{u}(\tilde{r},\varphi) \left(\begin{array}{c} 1\\-1\end{array}\right)_{s_z}\right\}\,,
\end{align}
where $\chi^{(u)}_\pm(\tilde{r},\varphi)$ are (anti-)symmetric states in angular space
\begin{align}
    \label{eq:chi_symmetry}
    \chi^{(u)}_\pm(\tilde{r},-\varphi) = \pm \chi^{(u)}_\pm(\tilde{r},\varphi)\,,
\end{align}
and normalized according to 
\begin{align}
    \label{eq:chi_normalization}
    \int_{\tilde{R}_<}^{\tilde{R}_>}d\tilde{r} \int_0^{2\pi} d\varphi 
    \left\{|\chi^{(u)}_+|^2 + |\chi^{(u)}_-|^2\right\} = 1/\lambda_{\rm so} \,,
\end{align}
with $\tilde{R}_\gtrless = R_\gtrless / \lambda_{\rm so}$. 

Inserting the form (\ref{eq:Phi_u}) in (\ref{eq:topo_A}) and using (\ref{eq:A}), we get the following two coupled differential equations to determine the functions $\chi^{(u)}_\pm(\tilde{r},\varphi)$
\begin{align}
    \nonumber
    &\left(-\partial_{\tilde{r}}^2 - \tilde{\delta} \mp 2u\partial_{\tilde{r}} \pm \frac{u}{\tilde{l}_B^2}\cos{\varphi}
    - \frac{1}{\tilde{r}^2}\partial_\varphi^2 \right)\,\chi^{(u)}_\pm \\
    \label{eq:chi_diff}
    & -\,\left(\pm 2\frac{u}{\tilde{r}}i\partial_\varphi \pm i\frac{u}{\tilde{l}_B^2}\sin{\varphi} 
    - \frac{1}{\tilde{r}^2}i\partial_\varphi \right)\,\chi^{(u)}_\mp = 0\,.
\end{align}
These two differential equations have to be solved with the boundary condition
\begin{align}
    \label{eq:chi_bc}
    \chi^{(u)}_+(\tilde{R}_\gtrless,\varphi)=\chi^{(u)}_-(\tilde{R}_\gtrless,\varphi) = 0 \,,
\end{align}
However, for states localized at the outer or inner surface, we need to consider only the boundary conditions at one of the corresponding surfaces, thereby neglecting exponentially small contributions at the other surface if 
\begin{align}
    \label{eq:delta_R_condition}
    \Delta\tilde{R} = \tilde{R}_>-\tilde{R}_< \gg 1
\end{align} 
is fulfilled, which we always assume implicitly in the following. As discussed at the end of Section~\ref{sec:SUSY}, this has the effect that the energies of the topological states are not exactly at zero energy but only at exponentially small energies (which we neglect). 

We proceed with the discussion of zero, weak and strong Zeeman field in the next subsections. A weak Zeeman field $\tilde{l}_B^2\gg 1$ allows for a clear understanding of the occurrence of topological states due to a second-order mechanism via the derivation of effective surface Hamiltonians hosting topological states in minima of effective surface potentials. The derivation of topological states at strong Zeeman field $\tilde{l}_B\sim O(1)$ is more subtle and requires a careful study of the solution of the two differential equations (\ref{eq:chi_diff}). 

\begin{figure*}[t!]
    \centering
    \includegraphics[width=\textwidth]{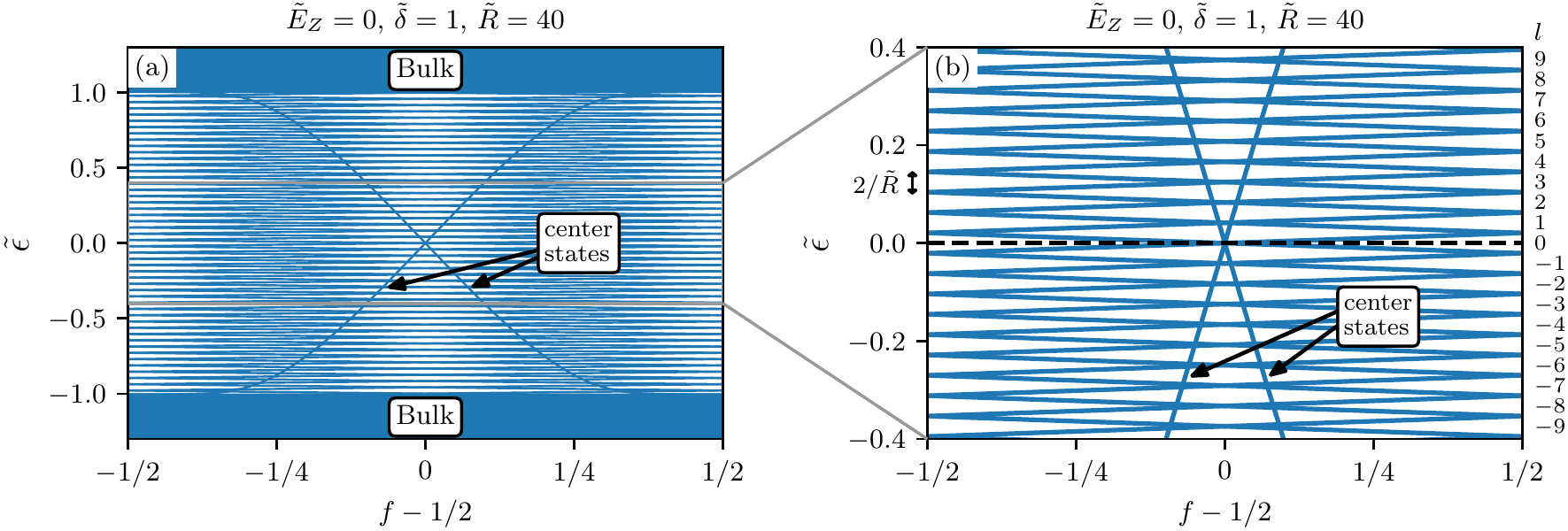}
	  \caption{Spectrum (in units of $E_{\rm so}$) at zero Zeeman field for a disk of radius $\tilde{R}=40$ as function of the deviation $f-1/2$ from half-integer flux, with $\tilde{\delta}=1$, shown on two different scales. We use tight-binding numerics as described in Appendix~\ref{app:tb_numerics} with $t=a=1$, $\alpha=2$ and $m=1/2$. The spectrum is symmetric around zero energy (due to chiral symmetry) and symmetric around $f-1/2$ (since time-reversal symmetry changes the sign of the flux, see Eq.~(\ref{eq:time_reversal})). For $f=1/2$, all states are twofold degenerate due to SUSY. Inside the bulk gap set by $\tilde{\delta}$, we find a set of discrete edge states localized at the disk boundary labeled by the angular momentum $l=0,\pm 1,\pm 2\,\dots$ with crossing linear dispersions given by $\tilde{\epsilon}^>_l=\pm 2(l+f-1/2)/\tilde{R}+O(1/\tilde{R}^2)$ (where $\pm$ refers to the $\sigma_x$-value in the transformed basis), see Eqs.~(\ref{eq:edge_state_dispersion_zero_B_large_R}) and (\ref{eq:H_zero_B}). In addition there are two center states with  a very steep slope due to their strong sensitivity to the boundary conditions, with $\epsilon\approx \mp 4\tilde{\delta}(f-1/2)$ for $|f-1/2|\ll 1$, see Eq.~(\ref{eq:E_f_small_R_l=0}).  
	  } 
    \label{fig:spectrum_disc_EZ_zero}
\end{figure*}

\subsection{Zero Zeeman field}
\label{sec:zero_B}

For the special case $B=0$, the Hamiltonian $\bar{H}^{(0)}_{1/2}=\bar{H}_{1/2}|_{B=0}$ is rotationally invariant around the $z$-axis and commutes with the angular momentum $L_z=-i\partial_\varphi$ in $z$-direction (note that $\bar{J}_z$ and $L_z$ differ only by a constant, see Eq.~(\ref{eq:J_z_trafo})). In each eigenspace of $L_z$ the angular dependence of the eigenfunctions is given by $\frac{1}{\sqrt{2\pi}}e^{il\varphi}$, where $l=0,\pm 1,\pm 2\dots$ denotes the integer eigenvalue of $L_z$. In this subspace we can replace $-i\partial_\varphi\rightarrow l$ and the radial part follows from the Hamiltonian
\begin{align}
    \label{eq:H_zero_B}
    \bar{H}^{(0)}_{1/2,l}/E_{\rm so} = \sigma_x \tilde{h}_l \,,
\end{align} 
with 
\begin{align}
    \label{eq:h_l}
    \tilde{h}_l &= (-\partial^2_{\tilde{r}} - \tilde{\delta}) s_z + 2i \partial_{\tilde{r}}s_y
    + \frac{2l}{\tilde{r}}s_x + \frac{l^2}{\tilde{r}^2} s_z - \frac{l}{\tilde{r}^2} \\
    \label{eq:h_l_SUSY}
    &=\left(\begin{array}{cc}
    \Gamma_l\Gamma_l^\dagger - \delta & 2 \Gamma_l \\
    2 \Gamma_l^\dagger & -(\Gamma_l^\dagger \Gamma_l - \delta)
    \end{array}\right) \,,    
\end{align}
where we defined
\begin{align}
    \label{eq:Gamma_l}
    \Gamma_l = \partial_{\tilde{r}}\ + \frac{l}{\tilde{r}} \quad,\quad
    \Gamma_l^\dagger = -\partial_{\tilde{r}}\ + \frac{l}{\tilde{r}} \,.
\end{align}
For a flux deviating from half-integer value we have to shift the angular momentum by $f-1/2$, i.e., we replace it by the index $\nu$ defined by
\begin{align}
    \label{eq:general_flux}
    l \rightarrow \nu = l + f - \frac{1}{2}\,.
\end{align}

The Hamiltonian $h_\nu$ written in the form (\ref{eq:h_l_SUSY}) is identical to the generalized supersymmetric Dirac Hamiltonian, as discussed e.g. in Section 9.1.1. of Ref.~[\onlinecite{junker_book_19}]. Disregarding boundary conditions it can be solved exactly for the bulk states in terms of Hankel functions, see Appendix~\ref{app:zero_B}. Taking only one of the boundary conditions at the inner or outer surface into account (which is valid under the condition (\ref{eq:delta_R_condition}), see discussion above), we will also determine in Appendix~\ref{app:zero_B} the edge states localized at the inner surface for any radius $R_<$ and the ones at the outer surface for large radius $\tilde{R}_>\gg 1$. In both cases it turns out that edge states exist only for positive band inversion parameter
\begin{align}
    \label{eq:condition_edge_states}
    \delta > 0 \,,
\end{align}
i.e., if the two bands overlap. This is standardly expected for systems involving only band inversion and spin-orbit coupling. 

If both the inner and outer radius are large, i.e., $\tilde{R}_\gtrless\gg 1$, we can neglect the last two terms of (\ref{eq:h_l}) for the determination of the edge states and approximate $2l/\tilde{r}\approx 2l/\tilde{R}_\gtrless$ at the outer/inner surface. The Hankel functions can then be replaced by plane waves and one can solve approximately the eigenvalue equation for the radial part of the edge states
\begin{align}
    \label{eq:h_l_edge_state}
    \tilde{h}_l |\bar{\psi}_n^{\gtrless}\rangle = \tilde{\epsilon}^{\gtrless}_l |\bar{\psi}_n^{\gtrless}\rangle \,.
\end{align}
This gives the following result for the dispersion
\begin{align}
    \label{eq:edge_state_dispersion_zero_B_large_R}
    \tilde{\epsilon}^{\gtrless}_l \approx \pm \frac{2\nu}{\tilde{R}_\gtrless} = \pm \frac{2(l+f-1/2)}{\tilde{R}_\gtrless} \,,
\end{align}
which differs by a sign factor for the outer/inner surface. The radial part of the edge state wave function is independent of $\nu$ and given by 
\begin{align}
    \label{eq:psi_n_corbino}
    \bar{\psi}_n^{\gtrless}(\tilde{r},s_z) \approx \frac{1}{\sqrt{2}} 
    \left(\begin{array}{c} 1 \\ \pm 1\end{array}\right)_{s_z} \bar{\Phi}_n(|\tilde{r}-\tilde{R}_\gtrless|) \,,
\end{align}
with $\tilde{r}\lessgtr\tilde{R}_\gtrless$ and 
\begin{align}
    \label{eq:Phi_n_corbino}
    \bar{\Phi}_n(\tilde{r}) = \frac{e^{-\tilde{r}}}{\sqrt{\lambda_{\rm so} N_n}} 
    \begin{cases} \sin\left[|\tilde{\delta}-1|^{1/2}\,\tilde{r}\right] & 
    \text{for}\quad \tilde{\delta} > 1 \\
    \sinh\left[|\tilde{\delta}-1|^{1/2}\,\tilde{r}\right] & 
    \text{for}\quad 0 < \tilde{\delta} < 1\end{cases}
    \,,
\end{align}
see Appendix~\ref{app:zero_B} and Section~\ref{sec:corbino_weak_fields} for details. Here, $N_n$ is a normalization factor given by 
\begin{align}
    \label{eq:N_n_corbino}
    N_n = \frac{|\tilde{\delta}-1|}{4\tilde{\delta}}\,,
\end{align}
such that the normalization condition 
\begin{align}
    \label{eq:normalization_condition_n_corbino}
    \pm\int_{\mp\infty}^{\tilde{R}_\gtrless} d\tilde{r} \sum_{s_z=\pm} |\bar{\psi}_n^{\gtrless}(\tilde{r},s_z)|^2 = 1 
\end{align}
is fulfilled. Importantly, the edge states are polarized with respect to the $x$-component of the transformed spin with eigenvalue $s_x=\pm 1$ for the outer/inner surface. This is due to the fact that $s_x$ is the chiral symmetry of the first two terms of (\ref{eq:h_l}) which determine the edge state wave function in radial direction. 

Since the total Hamiltonian is given by $\bar{H}^{(0)}_f / E_{\rm so}=\sigma_x h_\nu$, this gives rise to the two dispersions $\pm\tilde{\epsilon}_l^\gtrless$ as function of the angular momentum $l-1/2$ in $z$-direction (at fixed flux). They correspond to the two standard counter-propagating helical edge modes (labeled by the helicity $\sigma_x=\pm 1$) as known from the BHZ model \cite{bernevig_etal_science_06}. The flux dependence is shown in Fig.~\ref{fig:spectrum_disc_EZ_zero} via a numerical study of the energies of all eigenstates for a disk with radius $\tilde{R}=\tilde{R}_>=40$ and zero hole radius $\tilde{R}_<=0$ (i.e., the flux is applied through an infinitesimal small hole), with $\tilde{\delta}=1$. The spacing between adjacent levels is not precisely given by the finite size quantization $2/\tilde{R}$ since $O(1/\tilde{R}^2)$-corrections are present. However, besides this, the result of the linear dispersion of the edge modes within the bulk gap set by $\tilde{\delta}$ is perfectly reproduced. In accordance with the twofold degeneracy implied by SUSY, the two dispersions cross precisely at half-integer flux. At finite Zeeman field one obtains a repulsion of adjacent levels (without changing the degeneracy at half-integer flux due to SUSY) leading to a modified band structure as function of the flux which changes drastically when the Zeeman energy $\tilde{E}_Z$ becomes much larger than the spacing $2/\tilde{R}$, see Section~\ref{sec:corbino_weak_fields} and Fig.~\ref{fig:spectrum_disc_EZ_finite}. In this case, a surface gap of the order of $\tilde{E}_Z$ opens up, hosting bound states localized in addition in angular space. 

For the special case $\nu=0$, i.e., for $l=0$ and half-integer flux $f=1/2$, the Hamiltonian $h_0$ is translational invariant, i.e., one obtains for any hole radius $\tilde{R}_<$ the same energy $\tilde{\epsilon}^<_{l=0}=0$ and the same radial edge state wave function $\bar{\Phi}_n$ as for large hole radius. The fact that the energy of the two $\nu=0$ states must stay at zero follows also from symmetry arguments since SUSY and chiral symmetry protect their twofold degeneracy such that they can not split and must stay exactly at zero when reducing the hole radius. 

In contrast, the states at the inner surface at finite $\nu\ne 0$ have a strong flux dependence at small $\tilde{R}_<\ll 1$ since they are very sensitive to the boundary conditions. In Appendix~\ref{app:zero_B} we find that all states at finite $l\ne 0$ move out of the gap in the limit of small hole radius. An exception is the dispersion of the $l=0$ center states at the inner surface which start with zero energy at $f=1/2$ (for any hole radius, see above) but obtains a very steep slope as function of the deviation $f-1/2$ of the flux from half-integer value which remains finite in the limit $\tilde{R}_<\rightarrow 0$, see Fig.~\ref{fig:spectrum_disc_EZ_zero}. For small $|f-1/2|\ll 1$, we find the result
\begin{align}
    \nonumber
    \tilde{\epsilon}_{l=0}^<(f) &= - \frac{4\tilde{\delta}(f-1/2)}{\sqrt{|\tilde{\delta}-1}|}\\
    \label{eq:E_f_small_R_l=0}
    &\times\begin{cases}
    \arctan\sqrt{\tilde{\delta}-1} & {\rm for}\quad \tilde{\delta} > 1 \\
    \frac{1}{2}\,\ln{\frac{1+\sqrt{\tilde{\delta}-1}}{1-\sqrt{\tilde{\delta}-1}}} & {\rm for}\quad 0<\tilde{\delta} < 1
    \end{cases}\,.
\end{align}
It shows that the center states are rather unstable against the application of a flux away from half-filling. This is in contrast to the topological states for large radius in the presence of a weak magnetic field as discussed in the next subsection.

\subsection{Weak Zeeman field and the effective surface Hamiltonian}
\label{sec:corbino_weak_fields}

We continue with a discussion of a weak Zeeman field
\begin{align}
    \label{eq:weak_B}
    \tilde{l}_B^2 \gg 1 \,,
\end{align}
and consider either the occurrence of localized states at the outer or inner surface $\tilde{r}\approx \tilde{R}_\gtrless$ with a large radius
\begin{align}
    \label{eq:large_radius}
    \tilde{R} \gg 1\,.
\end{align}
Here, we use for convenience the short-hand notation $\tilde{R}\equiv\tilde{R}_\gtrless$ for the outer or inner surface, respectively. Furthermore, we assume to be deep in the gapped phase 
\begin{align}
    \label{eq:deep_gapped}
    \tilde{\delta} \gg \tilde{E}_Z=1/\tilde{l}_B^2\,.
\end{align}
This assumption is essential in the present section since the derivation is only valid if the bulk gap $\Delta_{\rm bulk}=\delta-E_Z$ is much larger than the surface gap $\Delta_{\rm surface}\approx E_Z$. Both the crossover from the gapped to the Weyl phase at $\tilde{\delta}\sim \tilde{E}_Z$ together with the regime of the Weyl phase $|\tilde{\delta}|<\tilde{E}_Z$ requires the treatment of the energy shift $\delta$ and the Zeeman term on an equal footing and will be described in the next section. 
Under these conditions we can approximate the Hamiltonian $\bar{H}_{1/2}$ first by the leading order terms (\ref{eq:bar_H_polar_1}), defining an effective Hamiltonian in normal direction to the surface
\begin{align}
    \label{eq:bar_H_n_corbino}
    \bar{H}_n/E_{\rm so} = \sigma_x \left[(-\partial^2_{\tilde{r}} - \tilde{\delta}) s_z
    + 2i \partial_{\tilde{r}}s_y) \right]\,.
\end{align}
Therefore, the radial part of the bulk Hamiltonian can be solved by plane waves $e^{i\tilde{k}\tilde{r}}$ leading to 
\begin{align}
    \label{eq:bar_H_n_k_corbino}
    \bar{H}_{n,{\rm bulk}}(\tilde{k})/E_{\rm so} = \sigma_x \left[(\tilde{k}^2 - \tilde{\delta}) s_z - 2\tilde{k}s_y\right]\,.
\end{align}
As a consequence, the bulk spectrum of the normal part is given by 
\begin{align}
    \label{eq:dispersion_normal_corbino}
    \epsilon_k/E_{\rm so} = \sqrt{(\tilde{k}^2-\tilde{\delta})^2 + 4\tilde{k}^2}\,,
\end{align}
giving rise to a bulk gap $\Delta_{\rm bulk}/E_{\rm so}=\tilde{\delta}$ for $\tilde{\delta}<2$, consistent with (\ref{eq:bulk_gap}).

Any eigenstate $\bar{\psi}_n$ of $\bar{H}_n$ localized either at the outer or inner surface $\tilde{r}\approx \tilde{R}\equiv \tilde{R}_\gtrless$ can be written as an eigenstate of $\sigma_x$ multiplied by a state $\bar{\psi}_n(\tilde{r},s_z)$, with $\tilde{r}\lessgtr\tilde{R}$, fulfilling the boundary condition  
\begin{align}
    \label{eq:phi_n_bc}
    \bar{\psi}_n(\tilde{R},s_z) = 0 \,.
\end{align}
To obtain these states we consider a linear combination of two bulk plane waves (multiplied with corresponding spinors) with different $k_{1,2}$ and finite imaginary part ${\rm Im}\,k_{1,2}\lessgtr 0$ such that the plane waves decay exponentially into the bulk. To fulfill the zero boundary condition (\ref{eq:phi_n_bc}) for both $s_z=\pm$, this is only possible if the two spinors are the same which is only the case for zero energy $\epsilon_k=0$, see Appendix~\ref{app:zero_B} for details. Using (\ref{eq:dispersion_normal_corbino}) this leads to
\begin{align}
    \label{eq:k_values_outer_corbino}
    \tilde{k}_{1,2} = \begin{cases} \pm \sqrt{\tilde{\delta}-1} + i & {\rm for} \quad \tilde{\delta}>1\gg 1/\tilde{l}_B^2 \\
    i(1\pm\sqrt{1-\tilde{\delta}}) & {\rm for} \quad 1/\tilde{l}_B^2 \ll \tilde{\delta} < 1\end{cases} 
\end{align}
for states localized at the inner surface, whereas for the ones localized at the outer surface we have to replace $\tilde{k}_{1,2}\rightarrow -\tilde{k}_{1,2}$. The eigenstate in normal direction localized at the outer/inner surface is then given by 
\begin{align}
    \label{eq:psi_n_corbino_without_normalization}
    \bar{\psi}_n^{\gtrless}(\tilde{r},s_z) \sim  
    \left(\begin{array}{c} 1 \\ \pm 1\end{array}\right)_{s_z}
    \left(e^{i\tilde{k}_1(\tilde{r}-\tilde{R})} - e^{i\tilde{k}_2(\tilde{r}-\tilde{R})}\right)
    \,,
\end{align}
which leads to Eq.~(\ref{eq:psi_n_corbino}) after normalization. From this result we get for the normal localization length $\tilde{\xi}_n = \xi_n/\lambda_{\rm so}$ the result
\begin{align}
    \label{eq:xi_n_weak_fields} 
    \frac{1}{\tilde{\xi}_n} =
    \begin{cases}
        1 & {\rm for} \,\, \tilde{\delta} > 1 \gg 1/\tilde{l}_B^2 \\
        1-\sqrt{1-\tilde{\delta}} & {\rm for} \,\, 1/\tilde{l}_B^2 \ll \tilde{\delta} < 1 
    \end{cases}\,.
\end{align}

The zero energy solutions for the normal part have an infinite degeneracy since they occur for any angle $\varphi$. The degeneracy is lifted by the other $\varphi$-dependent parts (\ref{eq:bar_H_polar_2}) and (\ref{eq:bar_H_polar_3}) of the Hamiltonian. Under the condition (\ref{eq:deep_gapped}) that the bulk gap is much larger than the surface gap, we can project the total Hamiltonian (\ref{eq:bar_H_polar_1}-\ref{eq:bar_H_polar_3}) on the zero energy solutions of the normal part. We find that in first order perturbation theory the terms involving $s_{y,z}$ do not contribute since $\bar{\psi}_n^\gtrless$ is an eigenstate of $s_x$. Furthermore, for $\tilde{R}\gg 1$, we can neglect the second term $\sim \tilde{r}^{-2}\partial_\varphi$ of (\ref{eq:bar_H_polar_2}) compared to the first term. Thus, after projection and setting $\tilde{r}\approx \tilde{R}$, we obtain the following effective surface Hamiltonian to determine the angular dependence of the edge states
\begin{align}
    \label{eq:bar_H_t_corbino}
    \pm\bar{H}_t^\gtrless/E_{\rm so} = \sigma_x \frac{2}{\tilde{R}}(-i\partial_\varphi) 
    + \sigma_y \frac{1}{\tilde{l}_B^2}\,\sin{\varphi} \,,
\end{align}
Here, the spinor operator $s_x$ is replaced by $\pm 1$ in (\ref{eq:bar_H_t_corbino}) since $\bar{\psi}_n^\gtrless$ is an eigenstate of $s_x$ with eigenvalue $\pm 1$. This sign influences only the sign of the dispersion but not the eigenstates. The total wave function localized at the outer or inner surface is then a product of the solutions along the normal and tangential direction
\begin{align}
    \label{eq:total_state_corbino}
    \bar{\psi}^\gtrless(\tilde{r},\varphi;\sigma_z,s_z) = \bar{\psi}_n^\gtrless(\tilde{r},s_z)\,
    \bar{\psi}_t(\varphi,\sigma_z)\,,
\end{align}
where $\bar{\psi}^\gtrless_n$ is given by (\ref{eq:psi_n_corbino}) and $\bar{\psi}_t$ is an eigenstate of the surface Hamiltonian (\ref{eq:bar_H_t_corbino}) normalized according to
\begin{align}
    \label{eq:bar_phi_t_normalization}
    \int_0^{2\pi}d\varphi \sum_{\sigma_z=\pm}|\bar{\psi}_t(\varphi,\sigma_z)|^2 = 1\,.
\end{align}

The surface Hamiltonian has the form of a periodic Dirac model with a potential term involving the normal component of the Zeeman field. In Section~\ref{sec:witten_model} we will see that the same result is obtained for an arbitrary smooth surface. The first term leads to a linear dispersion of two edge modes propagating in opposite directions along the surface and crossing at zero energy. The second term acts as a mass term leading to a gap in the edge state spectrum of the order of the Zeeman energy. Since the mass term changes sign at $\varphi=0,\pi$, we expect zero-energy topological bound states to appear at these positions. From the fact that the mass term changes from negative to positive values when crossing $\varphi=0$ along the surface, and vice versa for $\varphi=\pi$, we expect different chiralities $\bar{S}=-\sigma_z=\pm 1$ for the zero-energy states localized at the two positions $\varphi=0,\pi$, respectively. The angular spread $\Delta\varphi$ of the topological states can be estimated by comparing the order of magnitude of the two terms of the surface Hamiltonian (\ref{eq:bar_H_t_corbino}). This leads to  $1/(\tilde{R}\Delta\varphi)=\Delta\varphi/\tilde{l}_B^2$ or 
\begin{align}
    \label{eq:angle_spread_corbino}
    \Delta\varphi \equiv \frac{\tilde{l}_B}{\sqrt{\tilde{R}}} = \frac{l_B}{\sqrt{R\lambda_{\rm so}}} \,,
\end{align}
which gives for the tangential localization length $\xi_t$ the estimate
\begin{align}
    \label{eq:xi_t}
    \tilde{\xi}_t = \xi_t / \lambda_{\rm so} = \tilde{R}\Delta\varphi = \sqrt{\tilde{R}}\,\tilde{l}_B \,.
\end{align}
The angular spread is small compared to unity for large radius 
\begin{align}
    \label{eq:small_spread}
    \sqrt{\tilde{R}} \gg \tilde{l}_B \Leftrightarrow \Delta\varphi \ll 1 
    \Leftrightarrow \tilde{\xi}_t \ll \tilde{R}\,,
\end{align}
which is the regime of well-localized states where the tangential localization length is much smaller than the circumference of the surface. In this case the derivation of the surface Hamiltonian is systematic in the sense that it includes all sub-leading terms $\sim 1/\tilde{\xi}_t$ beyond the leading order terms $\sim O(1)$ present in the normal Hamiltonian (\ref{eq:bar_H_n_corbino}). This follows from the following estimates of the various terms present in (\ref{eq:bar_H_polar_2}) and (\ref{eq:bar_H_polar_3})
\begin{align}
    \label{eq:corbino_estimate_1}
    \frac{1}{\tilde{r}^2}\partial_\varphi^2 &\sim \frac{1}{\tilde{R}^2\Delta\varphi^2}\sim \frac{1}{\tilde{\xi}_t^2} 
    \ll \frac{1}{\tilde{\xi}_t}\,,\\
    \label{eq:corbino_estimate_2}
    \frac{1}{\tilde{r}^2}\partial_\varphi &\sim \frac{1}{\tilde{R}^2\Delta\varphi}\sim \frac{1}{\tilde{R}\tilde{\xi}_t}\,,
    \ll \frac{1}{\tilde{\xi}_t}\\
    \label{eq:corbino_estimate_3}
    \frac{1}{\tilde{l}_B^2}\sin{\varphi} &\sim \frac{\Delta\varphi}{\tilde{l}_B^2}\sim \frac{1}{\tilde{\xi}_t}\,.
\end{align}
A delicate issue is the Zeeman term in tangential direction $\sim (1/\tilde{l}_B^2)\cos{\varphi}\sim \tilde{R}/\tilde{\xi}_t^2$ which, for $\tilde{\xi}_t\ll\tilde{R}$, becomes larger than the terms considered in the surface Hamiltonian. However, since this term involves the Pauli matrix $s_y$, it can contribute only in second order perturbation theory (with bulk states of energy $\sim\tilde{\delta}\sim O(1)$ as intermediate states) and therefore contributes in order $1/\tilde{l}_B^4$ to the surface Hamiltonian. To neglect this contribution we need in addition the condition
\begin{align}
    \label{eq:additional_condition_weak_fields}
    \frac{1}{\tilde{l}_B^4} \ll \frac{1}{\tilde{\xi}_t} \Leftrightarrow \sqrt{\tilde{R}} \ll \tilde{l}_B^3 \,.
\end{align}
This means that in case of strong localization the magnetic field must be strong enough such that $\tilde{l}_B\ll\sqrt{\tilde{R}}$ but weak enough to guarantee (\ref{eq:additional_condition_weak_fields}). Otherwise, one enters the regime of a strong Zeeman field discussed in Section~\ref{sec:corbino_strong_fields}. 

We note that the additional condition (\ref{eq:additional_condition_weak_fields}) is automatically fulfilled for the case when the Zeeman field is so weak that the wave function is delocalized in angular space such that $\Delta\varphi\sim O(1)$ which happens for $\tilde{l}_B\gtrsim \sqrt{\tilde{R}}$ or $\tilde{\xi}_t\sim \tilde{R}$. In this case, we have considered consistently all subleading terms $\sim 1/\tilde{R}$ in the surface Hamiltonian, since the Zeeman term in $y$-direction gives in second-order perturbation theory a contribution $\sim 1/\tilde{l}_B^4 \lesssim 1/\tilde{R}^2$. In addition, for the opposite case of strong localization $\Delta\varphi\ll 1$, we show in Section~\ref{sec:validity}, that the tangential part of the Zeeman term can be included in the radial problem and leads only to a change of the normal localization length, without violating the validity regime of the effective surface Hamiltonian. Therefore, the additional condition (\ref{eq:additional_condition_weak_fields}) is not very restrictive, and is only relevant for the study of the extended edge states beyond the surface gap in the regime of strong localization. 

To visualize the emergence of localized states in potential minima it is instructive to square the surface Hamiltonian leading to an effective model of a particle on a ring in a periodic double sine potential  
\begin{align}
    \label{eq:double_sine_model}
    {\tilde{\cal H}}_W \equiv (\bar{H}^\gtrless_t/E_{\rm so})^2 = - \frac{4}{\tilde{R}^2}\,\partial_{\varphi}^2 + 
    \tilde{V}_W^{-\sigma_z}(\varphi)\,,  
\end{align}
where 
\begin{align}
    \nonumber
    \tilde{V}_W^{\pm}(\varphi) &= \left(V_W^{\pm}(s_t)/E_{\rm so}\right)^2 \\
    \label{eq:double_sine_potential}
    &= \frac{1}{\tilde{l}_B^4}\sin^2(\varphi) 
    \mp \frac{2}{\tilde{R}\,\tilde{l}_B^2}\cos(\varphi)  
\end{align}
are the two effective surface potentials (with dimension of energy squared) sketched in Fig.~\ref{fig:double_sine} for the two chiral sectors $\bar{S}=-\sigma_z=\pm 1$, plotted against the surface line element $s_t = R \varphi$. As one can see there are two potential minima for each chiral sector where localized states are trapped. The potential maximum is given by $1/\tilde{l}_B^4 + 1/\tilde{R}^2$ which is approximately given by the Zeeman energy squared for large radius of the surface. This shows that the potential term opens a surface gap in the effective edge state Dirac model of the order of the Zeeman energy with localized and discrete states in the gap and a continuum of edge states above the surface gap. This demonstrates the generation of localized bound states via a second-order mechanism. 

As one can see in Fig.~\ref{fig:double_sine} the lowest potential minimum is located at $\varphi=0$ for the chiral sector $s=1$ and at $\varphi=\pi$ for $s=-1$. In this minimum the lowest state is exactly at zero energy and is non-degenerate for each chiral sector. In contrast, all higher states are twofold degenerate for each chiral sector. This leads to the supersymmetric form of the spectrum for each chiral sector separately, which is consistent with the exact SUSY properties discussed in Section~\ref{sec:SUSY} for the squared Hamiltonian. In Section~\ref{sec:witten_SUSY} we will also present the exact SUSY properties of the periodic Witten model for any smooth and mirror-symmetric surface.  

Using (\ref{eq:bar_H_t_corbino}) the topological state at zero energy with chirality $\bar{S}=-\sigma_z=1$ follows from  
\begin{align}
    \label{eq:bar_phi_t_S=1_spinor}
    &\bar{\psi}_t^{(0)}(\varphi,\sigma_z) = \bar{\Phi}^{(0)}_t(\varphi)
    \frac{1}{\sqrt{2}}\left(\begin{array}{c} 0 \\ 1 \end{array}\right)_{\sigma_z}\,,\\
    \label{eq:bar_phi_t_S=1_diff}
    &\left[\frac{2}{\tilde{R}}(-i\partial_\varphi) - i \frac{1}{\tilde{l}_B^2}\sin{\varphi}\right]\bar{\Phi}_{t,1}(\varphi) = 0 \,.
\end{align}
The solution of the differential equation is given by (the superindex indicates the localization at $\varphi\approx 0$ which corresponds to $S=1$)
\begin{align}
    \label{eq:f}
    \bar{\Phi}^{(0)}_t(\varphi) = f_0(\varphi) \equiv \frac{1}{\sqrt{N_t}} \, e^{\frac{1}{2\Delta\varphi^2}\cos{\varphi}}\,,
\end{align}
where $\Delta\varphi=\tilde{l}_B/\sqrt{\tilde{R}}$ has been defined in (\ref{eq:angle_spread_corbino}) and the normalization factor $N_t$ is defined such that the normalization condition (\ref{eq:bar_phi_t_normalization}) is fulfilled. For $\Delta\varphi\ll 1$ we find that the state is localized close to $\varphi\approx 0$ and, after expanding $\cos{\varphi}\approx 1-\varphi^2/2$, we find the approximate Gaussian form
\begin{align}
    \label{eq:f_gaussian}
    f_0(\varphi) \approx \frac{1}{\sqrt{\sqrt{2\pi}\Delta\varphi}} \,e^{-\frac{1}{4}(\varphi/\Delta\varphi)^2}\,.
\end{align}
Obviously, the state for positive chirality $\bar{S}=-\sigma_z=1$ is symmetric in $\varphi$. This gives $\sigma_z=-1$, $P_\varphi=1$ and $s_x=\pm 1$ for the outer/inner surface according to (\ref{eq:psi_n_corbino}). As a consequence, the supersymmetry $\bar{U}_{1/2}=P_\varphi \sigma_z s_x = \mp 1$ according to (\ref{eq:SUSY_trafo}). This leads precisely to the two right states at the outer and inner surface shown in Fig.~\ref{fig:states_corbino}(a). The other two states with chirality $S=-1$ follow from the application of the inversion symmetry (see Eq.~(\ref{eq:change_us})) which, by using $-\bar{\Pi}=P_{\bf x}\sigma_x$ according to (\ref{eq:Pi_trafo}), leads to (here the superindex indicates the localization close to $\varphi\approx\pi$ which corresponds to $S=-1$)
\begin{align}
    \label{eq:bar_phi_t_S=-1_spinor}
    \bar{\psi}^{(\pi)}_t(\varphi,\sigma_z) &= f_0(\varphi-\pi)
    \frac{1}{\sqrt{2}}\left(\begin{array}{c} 1 \\ 0 \end{array}\right)_{\sigma_z}\,,\\
    \label{eq:f_pi}
    f_0(\varphi-\pi) &= \frac{1}{\sqrt{N_t}} \,e^{-\frac{1}{2\Delta\varphi^2}\cos{\varphi}}\,,\\
    \label{eq:f_pi_gaussian}
    &\approx \frac{1}{\sqrt{\sqrt{2\pi}\Delta\varphi}}\, e^{-\frac{1}{4}\left[(\varphi-\pi)/\Delta\varphi\right]^2}\,.
\end{align}
These two states at the outer and inner surface are localized close to $\varphi\approx\pi$ and fulfill $\sigma_z=1$, $P_{\varphi}=1$ and $s_x=\pm 1$, respectively. This leads to $S=-1$ and $U_{1/2}=\pm 1$, corresponding to the two left states of Fig.~\ref{fig:states_corbino}(a).

To calculate the excited bound states of ${\tilde{\cal H}}_W$ we consider the chirality sector $\bar{S}=-\sigma_z=1$ and start with the bound states localized close to $\varphi\approx 0$. Expanding the double sine potential $\tilde{V}^+_W(\varphi)$ shown in Fig.~\ref{fig:double_sine}(a) around $\varphi\approx 0$ we get from (\ref{eq:double_sine_potential}) for $\Delta\varphi=\tilde{l}_B/\sqrt{\tilde{R}}\ll 1$
\begin{align}
    \label{eq:double_sine_expanded_S=1_phi_0}
    \tilde{V}^+_W(\varphi) \approx \frac{\tilde{\Omega}_W^2}{4}\tilde{\varphi}^2 - \frac{\tilde{\Omega}_W^2}{2}\,,
\end{align}
where we defined the Witten frequency $\Omega_W$ for the double sine potential in dimensionless units by
\begin{align}
    \label{eq:Omega}
    \tilde{\Omega}_W = \Omega_W/E_{\rm so} = \frac{2}{\tilde{R} \Delta\varphi} = \frac{2}{\sqrt{\tilde{R}}\,\tilde{l}_B} = \frac{2}{\tilde{\xi}_t}\,, 
\end{align}
and $\tilde{\varphi}=\varphi/\Delta\varphi$, where $\Delta\varphi=\tilde{l}_B/\sqrt{\tilde{R}}$ is the angular spread defined in (\ref{eq:angle_spread_corbino}) and $\tilde{\xi}_t= \tilde{R}\Delta\varphi$ is an estimation for the tangential localization length according to (\ref{eq:xi_t}). This gives for the Hamiltonian (\ref{eq:double_sine_model}) of the double sine model 
\begin{align}
    \label{eq:double_sine_model_expanded_S=1_phi_0}
    {\tilde{\cal H}}^+_W = {\tilde{\cal H}}_W|_{\sigma_z=-1} \approx \tilde{\Omega}_W^2\left(-\partial_{\tilde{\varphi}}^2 + \frac{1}{4}\tilde{\varphi}^2 - \frac{1}{2}\right)\,,
\end{align}
which is of harmonic oscillator form 
\begin{align}
    \label{eq:h_DS_harmonic_oscillator}
    {\tilde{\cal H}}_W^+ = \tilde{\Omega}_W^2 \,a^\dagger a \,, 
\end{align}
with the annihilation/creation operators defined by
\begin{align}
    \label{eq:a_DS}
    a = \partial_{\tilde{\varphi}} + \tilde{\varphi}/2 \quad,\quad
    a^\dagger = -\partial_{\tilde{\varphi}} + \tilde{\varphi}/2 \,.
\end{align}
As a result, we find the eigenvalues
\begin{align}
    \label{eq:energies_DS}
    \tilde{E}^W_n = \tilde{\Omega}_W^2\,n \quad,\quad n=0,1,2,\dots \,,
\end{align}
and the normalized eigenstates 
\begin{align}
    \label{eq:eigenstates_DS_S=1_phi_0}
    f_n(\varphi) = \frac{1}{\sqrt{n!}}(a^\dagger)^n f_0(\varphi)\,,  
\end{align}
where $f_0(\varphi)$ defined by (\ref{eq:f_gaussian}) is the ground state of the harmonic oscillator. 

An analog result is obtained for the states localized close to the minimum $\varphi\approx\pi$, where we get 
\begin{align}
    \label{eq:double_sine_expanded_S=1_phi_pi}
    \tilde{V}^-_W(\varphi) \approx \frac{\tilde{\Omega}_W^2}{4}(\tilde{\varphi}-\pi)^2 + \frac{\tilde{\Omega}_W^2}{2}\,.
\end{align}
This gives the same eigenstates (\ref{eq:eigenstates_DS_S=1_phi_0}) but the angle is shifted by $\pi$ and the eigenvalue is shifted by $\tilde{\Omega}_W^2$, compare with Fig.~\ref{fig:double_sine}. Finally, the eigenstates of the other chiral sector $\bar{S}=-\sigma_z=-1$ follow from the ones of $\bar{S}=-\sigma_z=1$ by shifting the angle by $\pi$ and taking the same eigenvalue. 

From all eigenstates of ${\tilde{\cal H}}_W=(\bar{H}_t^\gtrless/E_{\rm so})^2$ one can construct all eigenstates of $\bar{H}^\gtrless_t$ which is possible due to chiral symmetry and since the Hamiltonian respects the periodic boundary conditions. We start with the regime $\varphi\approx 0$, where we can approximately write for (\ref{eq:bar_H_t_corbino}) in the $\sigma_z$-basis
\begin{align}
    \label{eq:H_t_harmonic_approx}
    \pm\bar{H}^\gtrless_t/E_{\rm so} \approx \tilde{\Omega}_W 
    \left(\begin{array}{cc} 0 & -i a \\ i a^\dagger & 0 \end{array}\right)\,.
\end{align}
Obviously, there is a unique zero energy state given by (the superindex indicates the regime $\varphi\approx 0$)
\begin{align}
    \label{eq:zero_state_H_t_phi_0}
    \bar{\psi}^{(0)}_t(\varphi,\sigma_z) = f_0(\varphi)
    \left(\begin{array}{c} 0 \\ 1  \end{array}\right)_{\sigma_z}\,, 
\end{align}
which has chirality $\bar{S}=-\sigma_z=1$ and agrees with (\ref{eq:bar_phi_t_S=1_spinor}) and (\ref{eq:f}). The eigenstates with non-zero energy are given by 
\begin{align}
    \label{eq:nonzero_states_H_t_phi_0}
    \bar{\psi}^{(0)}_{t,n\eta}(\varphi,\sigma_z) = \frac{1}{\sqrt{2}}
    \left(\begin{array}{c} -i \eta f_{n-1}(\varphi) \\ f_n(\varphi)  \end{array}\right)_{\sigma_z}\,, 
\end{align}
with $\eta=\pm 1$ and $n=1,2,\dots$. The corresponding energy eigenvalue of $\bar{H}_t^\gtrless/E_{\rm so}$ follows from 
\begin{align}
    \label{eq:eigenenergies_H_t}
    \tilde{\epsilon}_{n\eta}^{\gtrless} = \epsilon_{n\eta}^{\gtrless}/E_{\rm so} = \pm \eta \,\tilde{\Omega}_W\,\sqrt{n}\,.
\end{align}
Here, we note that the first sign $\pm 1$ refers to the outer/inner surface (or, equivalently, to the sign of $s_x$ of the normal part of the wave function), and the second sign $\eta$ refers to the two eigenstates (\ref{eq:nonzero_states_H_t_phi_0}) resulting from the chiral symmetry of $\bar{H}^\gtrless_t$ for each given surface. For the absolute value of the eigenenergies we get
\begin{align}
    \label{eq:eigenvalues_higher_bound_states}
    |\tilde{\epsilon}^{\gtrless}_{n\eta}| = \tilde{\Omega}_W \sqrt{n} = 
    \frac{2}{\tilde{\xi}_t} \sqrt{n} = \frac{2}{\sqrt{\tilde{R}}\,\tilde{l}_B} \sqrt{n} \,.
\end{align}
As expected, the energies scale with the inverse tangential localization length (the normal localization length $\xi_n$ given by (\ref{eq:xi_n_weak_fields}) does not appear since it is of the order of the spin-orbit length). Furthermore, since only the squared Hamiltonian is of harmonic oscillator form, they are proportional to $\sqrt{n}$. 

\begin{figure*}[t!]
    \centering
    	 \includegraphics[width =1.062\columnwidth]{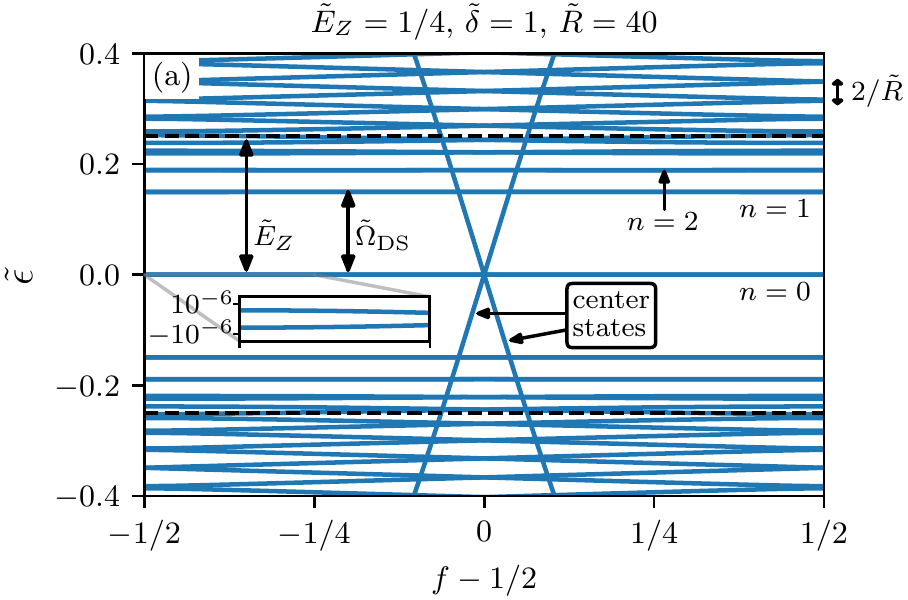} \hfill
    \includegraphics[width =0.998\columnwidth]{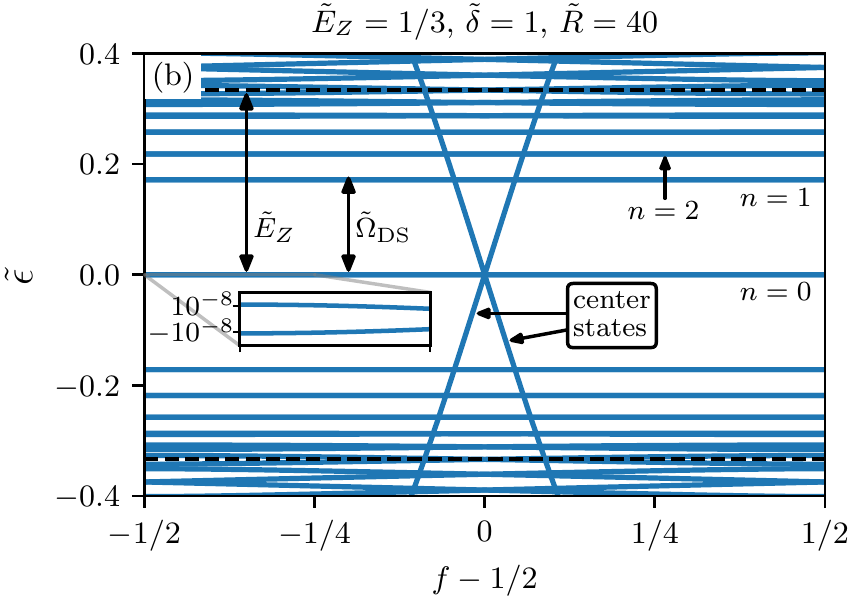}
	  \caption{Spectrum (in units of $E_{\rm so}$) of a disk with radius $\tilde{R}=40$ as function of the deviation $f-1/2$ from half-integer flux, with $\tilde{\delta}=1$ and two values of the Zeeman term $\tilde{E}_Z=1/\tilde{l}_B^2= (a) 1/4, (b) 1/3$, compare with Fig.~\ref{fig:spectrum_disc_EZ_zero} for $E_Z=0$. The surface gap is set approximately by the Zeeman energy $\tilde{E}_Z$ above which the continuum of edge states extended in angular space starts, compare with the sketch of the double sine potential in Fig.~\ref{fig:double_sine}. For $f=1/2$ all states are twofold degenerate due to SUSY with $4$ states at zero energy corresponding to two center states and two topological bound states at the disk boundary localized at $\varphi\approx 0$ and $\varphi\approx\pi$. The two center states with a very steep slope have a strong flux dependence since they are very sensitive to the boundary conditions. In contrast, the two topological states at the outer boundary are localized in angular space and rather insensitive to the boundary conditions, giving rise to a rather flat dispersion with decreasing slope for increasing Zeeman field. For increasing Zeeman field, additional discrete edge states localized in angular space at the outer boundary appear in the surface gap which are labeled by $n=1,2,\dots,N_B$. At $f=1/2$, their energy is approximately given by $\tilde{\epsilon}^>_n=\pm\tilde{\Omega}_W\sqrt{n}$, with $\tilde{\Omega}_W = 2\sqrt{\tilde{E}_Z/\tilde{R}}=\sqrt{\tilde{E}_Z/10}$, see Eqs.~(\ref{eq:eigenenergies_H_t}) and (\ref{eq:Omega}). Their number inside the surface gap can be estimated from $N_B\approx(\tilde{E}_Z/\tilde{\Omega}_W)^2=\tilde{R}\tilde{E}_Z/4=10\tilde{E}_Z$, which gives roughly $N_B\approx 2$ and $N_B\approx 3$ for $\tilde{E}_Z=1/4$ and $\tilde{E}_Z=1/3$, respectively, in rough agreement with the numerical result. At $f=1/2$ these states can be chosen as eigenfunctions of the SUSY operator and are localized either at $\varphi\approx 0$ (with SUSY eigenvalue $(-1)^{n+1}$) or at $\varphi\approx\pi$ (with SUSY eigenvalue $(-1)^n$), see Eqs.~(\ref{eq:SUSY_value_phi_0}) and (\ref{eq:SUSY_value_phi_pi}). 
	  } 
    \label{fig:spectrum_disc_EZ_finite}
\end{figure*}

For the eigenstates localized close to $\varphi\approx\pi$, we apply the inversion symmetry $\bar{\Pi}=-P_{\bf x}\sigma_x$ (see Eq.~(\ref{eq:Pi_trafo})) to (\ref{eq:zero_state_H_t_phi_0}) and (\ref{eq:nonzero_states_H_t_phi_0}). This does not change the energy but changes the states to 
\begin{align}
    \label{eq:zero_state_H_t_phi_pi}
    \bar{\psi}^{(\pi)}_t(\varphi,\sigma_z) = f_0(\varphi-\pi)
    \left(\begin{array}{c} 1 \\ 0  \end{array}\right)_{\sigma_z} 
\end{align}
for the zero energy state (which has chirality $\bar{S}=-\sigma_z=-1$ and agrees with (\ref{eq:bar_phi_t_S=-1_spinor})), and 
\begin{align}
    \label{eq:nonzero_states_H_t_phi_pi}
    \bar{\psi}^{(\pi)}_{t,n\eta}(\varphi,\sigma_z) = \frac{1}{\sqrt{2}}
    \left(\begin{array}{c} f_n(\varphi-\pi) \\ -i \eta f_{n-1}(\varphi-\pi)   \end{array}\right)_{\sigma_z} 
\end{align}
for the states with non-zero energy. 

The eigenfunctions with non-zero energy are no longer eigenstates of the chiral symmetry $\bar{S}=-\sigma_z$ but the states with different sign of $\eta$ (or different sign for the energy) are transformed into each other by the chiral symmetry
\begin{align}
    \label{eq:chiral_transform}
    \sigma_z \bar{\psi}^{(0,\pi)}_{t,n\eta} = -\bar{\psi}^{(0,\pi)}_{t,n,-\eta}\,.
\end{align}
The states are eigenstates of the SUSY operator since the property $f_n(-\varphi)=(-1)^n f_n(\varphi)$ leads to
\begin{align}
    \label{eq:SUSY_value_phi_0}
    P_\varphi \sigma_z \bar{\Psi}^{(0)}_{t,n\eta} &= - (-1)^n \bar{\Psi}^{(0)}_{t,n\eta} \,,\\
    \label{eq:SUSY_value_phi_pi}
    P_\varphi \sigma_z \bar{\Psi}^{(\pi)}_{t,n\eta} &= (-1)^n \bar{\Psi}^{(\pi)}_{t,n\eta} \,.
\end{align}
Using $\bar{U}_{1/2}=P_\varphi\sigma_z s_x$ from (\ref{eq:SUSY_trafo}), we get the SUSY eigenvalue $u=s_x(-1)^{n+1}$ and $u=s_x(-1)^n$ for the states localized at $\varphi\approx 0$ and $\varphi\approx\pi$, respectively, where $s_x=\pm 1$ corresponds to the outer/inner surface. 

Qualitatively, all our findings are perfectly reproduced by the numerical calculation of the spectrum of a disk with radius $\tilde{R}=40$ for $\tilde{\delta}=1$ and two values of the Zeeman energy $\tilde{E}_Z=1/4,1/3$, as shown in Figs.~\ref{fig:spectrum_disc_EZ_finite}(a,b) (here we use the tight-binding version described in Appendix~\ref{app:tb_numerics}). 
Compared to the case of zero Zeeman field as shown in Fig.~\ref{fig:spectrum_disc_EZ_zero}, the center states behave quite similar and show a strong dependence on the flux with a very steep slope. In contrast, for the states at the boundary of the disc, it can be clearly seen that a surface gap of the order of the Zeeman energy opens up, in which a set of bound states with energies on the scale of $\tilde{\Omega}_W$ appear. The mass term does not induce a splitting of the degenerate states at half-integer flux since SUSY protects the twofold degeneracy. It rather leads to a level repulsion between adjacent pairs pushing the states to higher energy. In contrast, a splitting occurs at integer flux,
where the twofold degeneracy is not protected at finite Zeeman energy. The exponential localization of the bound states in angular space is manifested by the very small band width as function of the flux, whereas the edge states extended in angular space with energy close or above the surface gap have a band width $\sim 1/\tilde{R}$ similar to the case for zero Zeeman field. This clearly manifests the semiclassical picture suggested by the double sine potential shown in Fig.~\ref{fig:double_sine}, hosting localized states of harmonic oscillator form in the potential minima. We note that the energy scale $\tilde{\Omega}_W$ fulfils for $\sqrt{\tilde{E}_Z}\gg 1/\sqrt{\tilde{R}}$ (which is equivalent to $\tilde{l}_B \ll \sqrt{\tilde{R}}$ or $\Delta\varphi\ll 1$) the relation
\begin{align}
    \label{eq:Omega_DS_order}
    \frac{2}{\tilde{R}} \ll \tilde{\Omega}_W = 2\left(\frac{\tilde{E}_Z}{\tilde{R}}\right)^{1/2} \ll \tilde{E}_Z\,.
\end{align}
We note that the two inequalities are equivalent since the ratios are the same
\begin{align}
    \label{eq:Omega_ratios}
    \frac{\tilde{\Omega}_W}{2/\tilde{R}} = 2 \frac{\tilde{E}_Z}{\tilde{\Omega}_W}\,.
\end{align}
The condition $\tilde{\Omega}_W\ll \tilde{E}_Z$ ensures that the number of bound states within the surface gap becomes large, whereas the relation $2/\tilde{R}\ll\tilde{\Omega}_W$ guarantees that the spacing of the localized bound states within the surface gap is much larger than the spacing $1/\tilde{R}$ of the extended states above the surface gap, see also the sketch of the spectrum in Fig.~\ref{fig:hole_system}(b). This qualitative tendency is demonstrated by comparing Fig.~\ref{fig:spectrum_disc_EZ_finite}(a) with Fig.~\ref{fig:spectrum_disc_EZ_finite}(b), where the Zeeman term increases from $\tilde{E}_Z=1/4$ to $\tilde{E}_Z=1/3$. We note that the radius $\tilde{R}=40$ used in those figures is not large enough to fulfill (\ref{eq:Omega_DS_order}) with clearly separated scales. Therefore, the scaling of the energies $\tilde{\epsilon}_n^>\approx \tilde{\Omega}_W\sqrt{n}$ of the localized bound states with $\sqrt{n}$ can not be precisely seen, only the bound state for $n=1$ has approximately the energy $\tilde{\Omega}_W$. The reason is that the spacing between the levels becomes smaller when their squared energy is close to the maximum $\sim \tilde{E}_Z^2$ of the double sine potential shown in Fig.~\ref{fig:double_sine}. Only for $\tilde{\Omega}_W$ significantly smaller than $\tilde{E}_Z$, one can demonstrate the $\sqrt{n}$-scaling, but the huge values of $\tilde{R}$ needed to fulfill this requirement are outside the scope of the numerical possibilities. 

The particle on a ring in a double sine potential is a special supersymmetric model in one dimension, occurring here for the special case of a surface in the form of a ring with a large radius. The analysis will be generalized to any smooth surface in Section~\ref{sec:witten_model} where we will see that generic periodic Witten models with supersymmetric properties can be realized. 

Furthermore, we note that the two topological bound states at the inner surface are exactly at zero energy if the radius $\tilde{R}_>$ of the outer surface tends to infinity. In this case, the SUSY is unbroken in an exact sense and two states exactly at zero energy appear in the gap. Since the degeneracy of these two states follows from SUSY, they can not split for any radius $\tilde{R}_<$ of the inner hole. Therefore, even for zero hole radius $\tilde{R}_<=0$, the two center states discussed in Section~\ref{sec:zero_B} for zero magnetic field will remain at zero energy in the presence of a finite magnetic field.

Concerning the stability of the topological bound states against deviations from half-integer flux, degenerate states at opposite positions of the same surface with different eigenvalues of $U_{1/2}$ will get coupled and split for $f\ne 1/2$.  However, under the condition (\ref{eq:small_spread}) of small angular spread, the orbital overlap of the two states is exponentially small and the splitting is negligible. This is in contrast to the two zero energy center states at small hole radius which are unstable against the application of a flux away from half-filling, see the discussion at the end of Section~\ref{sec:zero_B}.

\subsection{Strong Zeeman field}
\label{sec:corbino_strong_fields}

For a strong Zeeman field, where $\tilde{l}_B$ can approach $O(1)$, and for the discussion of the crossover to the Weyl phase with $\tilde{\delta}\sim 1/\tilde{l}_B^2$, we try to solve the differential equations (\ref{eq:chi_diff}) again with the help of clearly separated length scales. Denoting the localization lengths of the topological states in normal/tangential direction by $\tilde{\xi}_{n/t}$ (in units of the spin-orbit length $\lambda_{\rm so}$), we assume
\begin{align}
    \label{eq:xi_n_t}
    O(1)\sim\tilde{\xi}_n \ll \tilde{\xi}_t = \tilde{R}\Delta\varphi = \sqrt{\tilde{R}}\,\tilde{l}_B \ll \tilde{R} \,.
\end{align}
 Here, we have assumed that the spread $\Delta\varphi$ in angular direction is of the same order as we have found it in Eq.~(\ref{eq:angle_spread_corbino}) for the case of a weak Zeeman field, leading to the same form (\ref{eq:xi_t}) for the tangential localization length.
 This assumption is also fulfilled for the regime $\tilde{l}_B\sim O(1)$ as we will show below.
 The same holds for the normal localization length which, however, will get an additional dependence on the Zeeman field but roughly stays of the order of the spin-orbit length (except at phase transition lines where the normal localization length can diverge). We note that the conditions (\ref{eq:xi_n_t}) are equivalent to the following condition for the magnetic length
 \begin{align}
     \label{eq:condition_l_B}
     \frac{1}{\sqrt{\tilde{R}}} \ll \tilde{l}_B \ll \sqrt{\tilde{R}}\,,
 \end{align}
which can always be fulfilled for large enough radius in the thermodynamic limit for any size of the Zeeman field. The condition $\tilde{l}_B \ll \sqrt{\tilde{R}}$ of strong localization is also essential for the consistency of the following arguments to neglect various terms in the differential equations (in contrast to the previous section where this condition was not needed). However, we note that this condition is anyhow essential for the stability of the states against small deviations from half-integer flux as discussed at the end of the previous section. Furthermore, we will discuss at the end of this section that the conclusions for the existence of zero energy states do not change when $\sqrt{\tilde{R}}$ becomes of the same order or even smaller than $\tilde{l}_B$.

Assuming in addition (to be checked below) that the topological states for $S=1$ are localized at $\varphi\approx 0$ (for $S=-1$ we get a localization at $\varphi\approx\pi$) as for weak Zeeman field, we can estimate the various terms in the differential equations (\ref{eq:chi_diff}) as follows
\begin{align}
    \label{eq:estimations_1}
    \partial_{\tilde{r}}&\sim \frac{1}{\tilde{\xi}_n}\sim O(1) \quad,\quad 
    \frac{1}{\tilde{r}}\partial_{\varphi} \sim \frac{1}{\tilde{\xi}_t}\sim \frac{1}{\sqrt{\tilde{R}}\,\tilde{l}_B}\,,\\
    \label{eq:estimations_2}
    \frac{1}{\tilde{l}_B^2}\cos{\varphi} &= \frac{1}{\tilde{l}_B^2} + O\left(\frac{\Delta\varphi^2}{\tilde{l}_B^2}\right) = \frac{1}{\tilde{l}_B^2} + O\left(\frac{1}{\tilde{R}}\right)\,,\\    
    \label{eq:estimations_3}
    \frac{1}{\tilde{l}_B^2}\sin{\varphi} &\sim \frac{\Delta\varphi}{\tilde{l}_B^2} \sim \frac{1}{\sqrt{\tilde{R}}\,\tilde{l}_B}\,.    
\end{align}
Neglecting consistently all terms of $O(1/\tilde{R})$ and keeping only those of $O(1)$ and $O(1/\sqrt{\tilde{R}})$, the differential equations (\ref{eq:chi_diff}) can be approximated by
\begin{align}
    \label{eq:chi_diff_approx}
    (-i\partial_{\tilde{r}}-\tilde{k}^{(u)}_1)(-i\partial_{\tilde{r}}-\tilde{k}^{(u)}_2)
    \hat{\chi}^{(\pm u)}_\pm = \pm \frac{2iu}{\tilde{R}}\partial_\varphi\hat{\chi}^{(\mp u)}_\mp\,,
\end{align}
where
\begin{align}
    \label{eq:chi_hat}
     \chi^{(u)}_\pm(\tilde{r},\varphi) &= \hat{\chi}^{(u)}_\pm(\tilde{r},\varphi) f_0(\varphi)\,,\\
    \nonumber
     \tilde{k}^{(u)}_{1/2} &= iu \pm \sqrt{\tilde{\delta}-1-u/\tilde{l}_B^2} \,,\\
     \label{eq:k_u}
    &\hspace{-0.2cm}
    = iu \pm \sqrt{|\tilde{\delta}-1-u/\tilde{l}_B^2|}\,
    \begin{cases} 1 & {\rm for}\,\, \tilde{\delta}>1+u/\tilde{l}_B^2 \\
    iu & {\rm for}\,\, \tilde{\delta}<1+u/\tilde{l}_B^2 \end{cases}
\end{align}
and $f_0(\varphi)$ has been defined in (\ref{eq:f}) which can be approximated by the Gaussian form (\ref{eq:f_gaussian}) for $\Delta\varphi\ll 1$.

The differential equations (\ref{eq:chi_diff_approx}) can be solved exactly by a $\varphi$-independent function for $\hat{\chi}^{(u)}_+$ and a linear dependence on $\varphi$ for $\hat{\chi}^{(u)}_-$
\begin{align}
    \label{eq:hat_chi_ansatz}
    \hat{\chi}^{(u)}_+(\tilde{r},\varphi) = \hat{\chi}^{(u)}_{+,n}(\tilde{r}) \quad,\quad  
    \hat{\chi}^{(u)}_-(\tilde{r},\varphi) = (\varphi/\Delta\varphi)\,\hat{\chi}^{(u)}_{-,n}(\tilde{r}) \,. 
\end{align}
The linear dependence for $\hat{\chi}^{(u)}_-$ on $\varphi$ is needed to fulfill the antisymmetry property (\ref{eq:chi_symmetry}). We disregard here the fact that the linear function is not periodic since the angular spread is assumed to be very small. For $\hat{\chi}^{(u)}_{\pm,n}$ we find two solutions. The first one is obtained by setting $\hat{\chi}^{(u)}_{-,n}=0$ and solving
\begin{align}
    \label{eq:hat_chi_n_+_diff}
    (-i\partial_{\tilde{r}}-\tilde{k}^{(u)}_1)(-i\partial_{\tilde{r}}-\tilde{k}^{(u)}_2)
    \hat{\chi}^{(u)}_{+,n} = 0 \,.
\end{align}
Up to a normalization constant, the solution of this differential equation with zero boundary condition at $\tilde{r}=\tilde{R}$ is given by 
\begin{align}
    \label{eq:hat_chi_n_+_solution}
    \hat{\chi}_{+,n}^{(u)}(\tilde{r}) \sim e^{i\tilde{k}^{(u)}_1 (\tilde{r}-\tilde{R})}
    - e^{i\tilde{k}^{(u)}_2 (\tilde{r}-\tilde{R})}\,,
\end{align}
which is only a valid solution for the inner/outer surface if both $\tilde{k}^{(u)}_{1/2}$ have the same positive/negative sign for the imaginary part, respectively. Using (\ref{eq:k_u}) this is only the case for 
\begin{align}
    \label{eq:delta_chi_+_states}
    \tilde{\delta} > u/\tilde{l}_B^2
\end{align}
and $u=\pm 1$ corresponds to states localized at the inner/outer surface, respectively. As a consequence, we find for $\tilde{\delta} > 1/\tilde{l}_B^2$ a state with chirality $s=1$ and SUSY $u=1$ at the inner surface, and for $\tilde{\delta} > -1/\tilde{l}_B^2$ a state with chirality $s=1$ and SUSY $u=-1$ at the outer surface, both localized at $\varphi\approx 0$. This is consistent with Fig.~\ref{fig:states_corbino}, where we see that the zero energy state at the outer surface persists in the Weyl phase whereas the one at the inner surface disappears (and is replaced by another anti-symmetric one at the outer surface, see below for the second solution of the differential equations). Inserting (\ref{eq:k_u}), we can write the two solutions also as
\begin{align}
    \nonumber
    \hat{\chi}_{+,n}^{(u)}(\tilde{r}) &= 
    \frac{1}{\sqrt{\lambda_{\rm so} N^{(u)}}} \, e^{-u(\tilde{r}-\tilde{R})}\\
    \label{eq:hat_chi_n_+_explicit}
    &\hspace{-1.2cm}
    \times \begin{cases} \sin\left[q^{(u)}\,
    (\tilde{r}-\tilde{R})\right] & 
    \text{for}\,\, \tilde{\delta} > 1+u/\tilde{l}_B^2 \\
    \sinh\left[q^{(u)}\,(\tilde{r}-\tilde{R})\right] & 
    \text{for}\,\, u/\tilde{l}_B^2 < \tilde{\delta} < 1+u/\tilde{l}_B^2\end{cases}
    \,,
\end{align}
with 
\begin{align}
    \label{eq:q_u}
    q^{(u)} = |\tilde{\delta}-1 - u/\tilde{l}_B^2|^{1/2}\,,
\end{align}
 and the normalization factor 
\begin{align}
    \label{eq:N_u}
    N^{(u)} = \frac{|\tilde{\delta}-1-u/\tilde{l}_B^2|}{4|\tilde{\delta}-u/\tilde{l}_B^2|}\,,
\end{align}
to get $\pm\int_{\tilde{R}}^{\pm\infty} d\tilde{r} |\hat{\chi}_{+,n}^{(\pm)}(\tilde{r})|^2 = 1/\lambda_{\rm so}$, due to the radial part of the normalization condition (\ref{eq:chi_normalization}). These two states are consistent with the corresponding ones for weak Zeeman field in the gapped phase, compare with Eq.~(\ref{eq:psi_n_corbino}). However, for strong Zeeman field, one obtains a significant dependence of the normal localization lengths $\tilde{\xi}_n^{(su)}$ (labeled by chirality $s$ and SUSY $u$) on the Zeeman field, in contrast to the form (\ref{eq:xi_n_weak_fields}) for weak Zeeman field. For the state at the inner surface with $s=1$ and $u=1$ (only present for $\tilde{\delta}>1/\tilde{l}_B^2$ in the gapped phase) we get (the same holds for $s=-1$ and $u=-1$ since it results from applying inversion symmetry according to (\ref{eq:change_us}))
\begin{align}
    \nonumber
    \frac{1}{\tilde{\xi}_n^{<}} \equiv \frac{1}{\tilde{\xi}_n^{(11)}} &= \frac{1}{\tilde{\xi}_n^{(-1,-1)}} \\
    \label{eq:xi_n_s=1_u=1_gapped} 
    &\hspace{-0.8cm}
    = \begin{cases}
        1 & {\rm for} \,\, \tilde{\delta} > 1 + 1/\tilde{l}_B^2 \\
        1-\sqrt{1+1/\tilde{l}_B^2-\tilde{\delta}} & {\rm for} \,\, 1/\tilde{l}_B^2 < \tilde{\delta} < 1 + 1/\tilde{l}_B^2 
    \end{cases}\,.
\end{align}
At $\tilde{\delta}=1/\tilde{l}_B^2$ the normal localization length diverges and the states move over to the outer surface (see below). For the state at the outer surface with $s=1$ and $u=-1$ (present for $\tilde{\delta}>-1/\tilde{l}_B^2$ both in the gapped and in the Weyl phase) we find (the same for $s=-1$ and $u=1$)
\begin{align}
    \nonumber
    \frac{1}{\tilde{\xi}_n^{>}} \equiv \frac{1}{\tilde{\xi}_n^{(1,-1)}} &= \frac{1}{\tilde{\xi}_n^{(-1,1)}}  \\
    \label{eq:xi_n_s=1_u=-1} 
    &\hspace{-0.8cm}
    =\begin{cases}
        1 & {\rm for} \,\, \tilde{\delta} > 1 - 1/\tilde{l}_B^2 \\
        1-\sqrt{1-1/\tilde{l}_B^2-\tilde{\delta}} & {\rm for} \,\, - 1/\tilde{l}_B^2 < \tilde{\delta} < 1 - 1/\tilde{l}_B^2 
    \end{cases}\,.
\end{align}
For this state the localization length diverges for $\tilde{\delta}=-1/\tilde{l}_B^2$ at the crossover from the Weyl phase to the non-topological gapped phase, where all topological states disappear. In particular the dependence of the normal localization length of the state at the inner surface on the Zeeman field is quite useful since it allows for a tunability of the interaction between two topological states localized at different holes, possibly of interest for topological engineering, see the discussion in Section~\ref{sec:tunability}. 

The second possibility to solve the differential equations (\ref{eq:chi_diff_approx}) is to take a finite anti-symmetric part $\hat{\chi}^{(u)}_{-}(\tilde{r},\varphi)=\varphi\hat{\chi}^{(u)}_{-,n}(\tilde{r}) $ which, due to $\partial_\varphi \hat{\chi}^{(u)}_{+}=0$, has to fulfil
\begin{align}
    \label{eq:hat_chi_n_-_diff}
    (-i\partial_{\tilde{r}}-\tilde{k}^{(-u)}_1)(-i\partial_{\tilde{r}}-\tilde{k}^{(-u)}_2)
    \hat{\chi}^{(u)}_{-,n} = 0 \,.
\end{align}
In addition, the second equation of (\ref{eq:chi_diff_approx}) requires
\begin{align}
    \label{eq:second_condtion_chi_n_-}
    (-i\partial_{\tilde{r}}-\tilde{k}^{(u)}_1)(-i\partial_{\tilde{r}}-\tilde{k}^{(u)}_2)
    \hat{\chi}^{(u)}_{+,n} = iu\frac{2}{\tilde{R}}\hat{\chi}^{(u)}_{-,n} \,.
\end{align}
Whereas (\ref{eq:hat_chi_n_-_diff}) can be solved analog to (\ref{eq:hat_chi_n_+_solution}) by 
\begin{align}
    \label{eq:hat_chi_n_-_solution}
    \hat{\chi}_{-,n}^{(u)}(\tilde{r}) \sim e^{i\tilde{k}^{(-u)}_1 (\tilde{r}-\tilde{R})}
    - e^{i\tilde{k}^{(-u)}_2 (\tilde{r}-\tilde{R})}\,,
\end{align}
the solution of (\ref{eq:second_condtion_chi_n_-}) is more subtle. Since $\hat{\chi}_{-,n}^{(u)}$ contains two exponentials involving $\tilde{k}^{(-u)}_{1,2}$, the same must hold for $\hat{\chi}_{+,n}^{(u)}$. However, if only those two exponentials were present for $\hat{\chi}_{+,n}^{(u)}$, zero boundary conditions at $\tilde{r}=\tilde{R}$ require the same form (\ref{eq:hat_chi_n_-_solution}) for both $\hat{\chi}_{\pm,n}^{(u)}$ which does not solve Eq.~(\ref{eq:second_condtion_chi_n_-}). Therefore, a third exponential is needed for $\hat{\chi}_{+,n}^{(u)}$ which does not contribute to (\ref{eq:second_condtion_chi_n_-}), i.e., involves either $\tilde{k}^{(u)}_1$ or $\tilde{k}^{(u)}_2$. Since all three exponentials must decay, we need that the imaginary parts of all three momenta involved in     
$\hat{\chi}_{+,n}^{(u)}$ must have the same sign for the imaginary part. Using 
\begin{align}
    \label{eq:sign_imag_k_u}
    {\rm sign}\,{\rm Im}\tilde{k}^{(u)}_{1/2} = u\,\begin{cases}
        1 & {\rm for} \quad \tilde{\delta} > u/\tilde{l}_B^2 \\
        \pm 1 & {\rm for} \quad \tilde{\delta} < u/\tilde{l}_B^2 
    \end{cases} \,,
\end{align}
we find that this is only possible in the Weyl phase 
\begin{align}
    \label{eq:condition_antisymmetric_state}
    |\tilde{\delta}|<1/\tilde{l}_B^2    
\end{align}
by choosing three exponentials involving $\tilde{k}^{(-1)}_{1/2}$ and $\tilde{k}^{(1)}_2$, where all three imaginary parts of the momenta are negative, corresponding to a state at the outer surface with $s=1$ and $u=1$. The solution for $\hat{\chi}_{+,n}^{(1)}$ solving (\ref{eq:second_condtion_chi_n_-}) and fulfilling zero boundary conditions is then given by 
\begin{align}
    \nonumber
    \hat{\chi}_{+,n}^{(1)}(\tilde{r}) &\sim c_1 \left(e^{i\tilde{k}^{(-1)}_1 (\tilde{r}-\tilde{R})}
    - e^{i\tilde{k}^{(1)}_2 (\tilde{r}-\tilde{R})}\right) \\
    \label{eq:hat_chi_n_+_second_weyl_outer}
    & + c_2 \left(e^{i\tilde{k}^{(-1)}_2 (\tilde{r}-\tilde{R})}
    - e^{i\tilde{k}^{(1)}_2 (\tilde{r}-\tilde{R})}\right)\,,
\end{align}
with
\begin{align}
    \label{eq:c_12}
    (\tilde{k}^{(-1)}_{1/2} - \tilde{k}^{(1)}_{1})(\tilde{k}^{(-1)}_{1/2} - \tilde{k}^{(1)}_{2}) c_{1/2}  
    = \pm i \frac{2}{\tilde{R}} \,.
\end{align}
Since both $c_{1/2}\sim 1/\tilde{R}$ we can neglect $\hat{\chi}_{+}^{(1)}\sim 1/\tilde{R}$ compared to $\hat{\chi}_{-}^{(1)}\sim \Delta\varphi\sim \tilde{l}_B/\sqrt{\tilde{R}}$ for $\tilde{l}_B \gg 1/\sqrt{\tilde{R}}$. Therefore, although $\hat{\chi}_{+}^{(1)}$ is important for the discussion of the existence of a solution, it can be neglected finally. What remains is the state $\hat{\chi}_{-,n}^{(1)}$ which, according to (\ref{eq:hat_chi_n_-_solution}), is given by the state (\ref{eq:hat_chi_n_+_explicit}) with $u=-1$ 
\begin{align}
    \label{eq:hat_chi_n_-_u=1_explicit}
    \hat{\chi}_{-,n}^{(1)}(\tilde{r}) = \hat{\chi}_{+,n}^{(-1)}(\tilde{r})\,\,\theta(1/\tilde{l}_B^2-|\tilde{\delta}|)\,,
\end{align}
with the additional constraint that we are in the Weyl phase. This gives for the normal localization length of the antisymmetric state at the outer surface in the Weyl phase the result
\begin{align}
    \nonumber
    \frac{1}{\tilde{\xi}_n^{({\rm antisymm})}} &  \\
    \label{eq:xi_n_s=1_u=1_weyl} 
    &\hspace{-0.8cm}
    = \begin{cases}
        1 & {\rm for} \,\, 1/\tilde{l}_B^2 > \tilde{\delta} > 1 - 1/\tilde{l}_B^2 \\
        1-\sqrt{1-1/\tilde{l}_B^2-\tilde{\delta}} & {\rm for} \,\, -1/\tilde{l}_B^2 < \tilde{\delta} < 1 - 1/\tilde{l}_B^2 
    \end{cases}\,,
\end{align}
which is identical to the normal localization length $\tilde{\xi}_n^>$ of the states at the outer surface with $s=1$ and $u=-1$ or $s=-1$ and $u=1$, see Eq.~(\ref{eq:xi_n_s=1_u=-1}).

\begin{figure*}[t!]
    \centering
 	 \includegraphics[width =1.0\columnwidth]{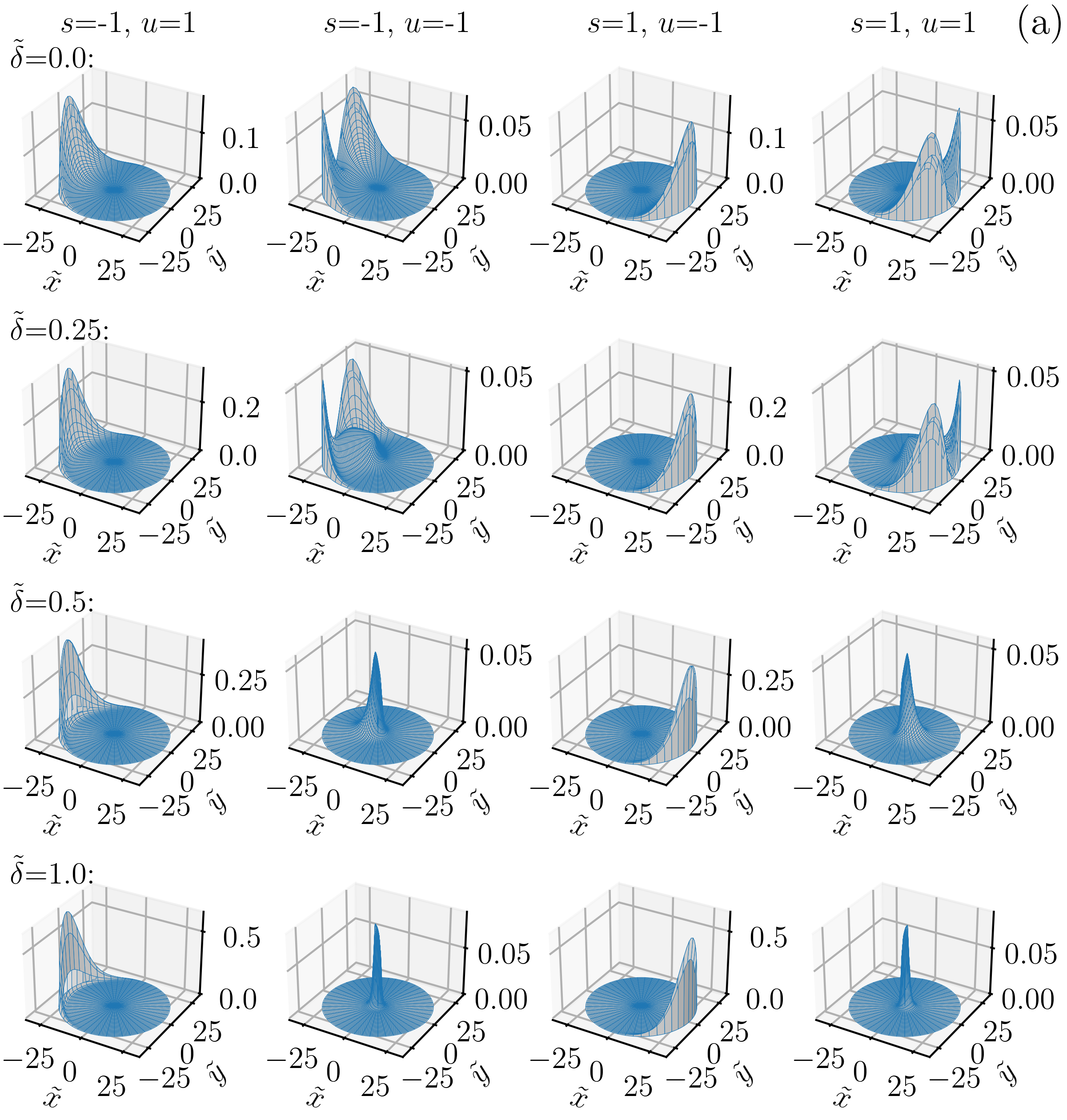} 
 	 \hspace{0.3cm}
 	 \includegraphics[width =1.0\columnwidth]{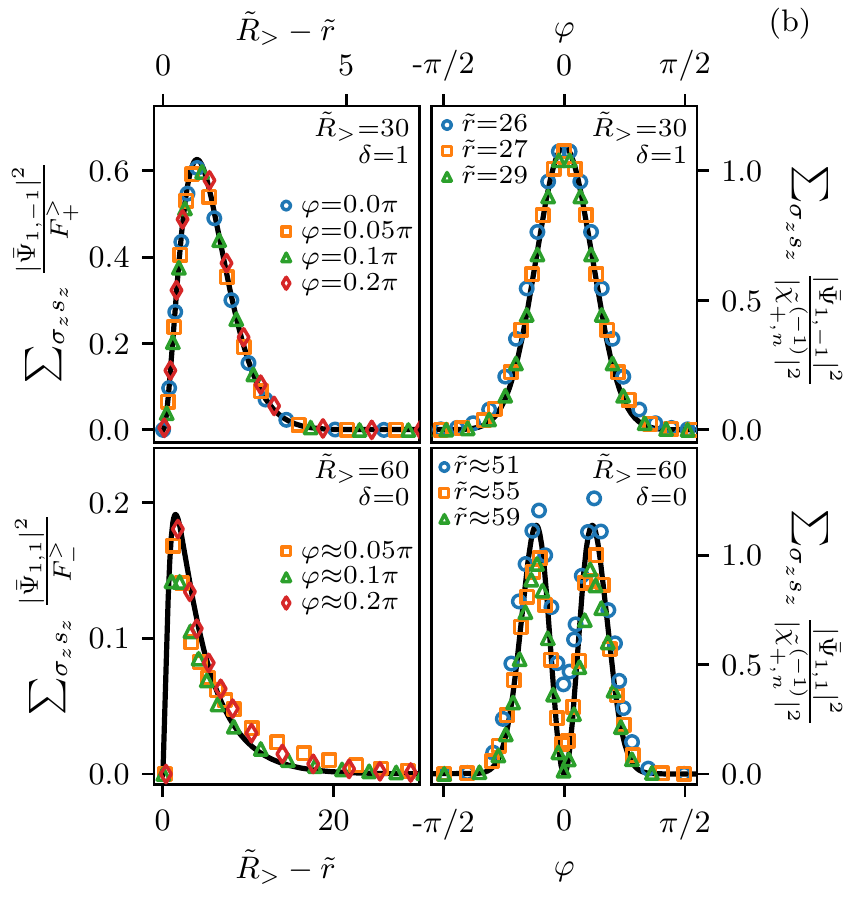}
	  \caption{(a) The topological states $\sum_{\sigma_z,s_z}|\bar{\psi}_{su}(\tilde{x},\tilde{y};\sigma_z,s_z)|^2$ as function of $\tilde{x}=x/\lambda_{\rm so}$ and $\tilde{y}=y/\lambda_{\rm so}$ in continuum numerics for a disk with radius $\tilde{R}_>=30$, $\tilde{l}_B=2$, and various values of $\tilde{\delta}$. Two upper panels of (b): A comparison of the continuum numerics for $\tilde{R}_>=30$, $\tilde{\delta}=1$ and $\tilde{l}_B=2$ to the analytical result (\ref{eq:average_square_TP_states_s=1}) for $s=1$ and $u=-1$ at the outer surface in the TP. Two lower panels of (b): A comparison of the tight-binding numerics for $\tilde{R}_>=60$, $\tilde{\delta}=0$ and $\tilde{l}_B=2$ to the analytical result (\ref{eq:average_square_WP_antisymmetric}) for $s=1$ and $u=1$ at the outer surface in the WP. In the two left panels we show various cuts for fixed $\varphi$ as function of $\tilde{R}_>-\tilde{r}$ and use the normalization to $F_\pm^>(\varphi)$ as given by Eq.~(\ref{eq:F_+}) and (\ref{eq:F_-}) for the upper/lower panel. In the two right panels, we show various cuts for fixed $\tilde{r}$ as function of $\varphi$ and use the normalization to $|\hat{\chi}_{+,n}^{(-1)}(\tilde{r})|^2$ as given by Eq.~(\ref{eq:hat_chi_n_+_explicit}). The black solid lines in the left/right panels are the theoretical results for the radial/angular part to which all cuts should collapse. 
    } 
    \label{fig:states_continuum}
\end{figure*}

The existence of two additional anti-symmetric states with $s=u=\pm 1$ at the outer surface in the Weyl phase is quite special. As sketched in Fig.~\ref{fig:states_corbino} these states have a strong orbital overlap with the states $s=-u=\pm 1$. This makes these states rather unstable against small perturbations violating chiral symmetry and SUSY (such that states with the same chirality and different SUSY eigenvalues can be coupled via the Hamiltonian). Therefore, although the Weyl phase is a regime of theoretical interest, concerning applications one should study the gapped topological phase $\tilde{\delta}>1/\tilde{l}_B^2$, where all topological states are localized at clearly separated positions for $\Delta\varphi=\tilde{l}_B/\sqrt{\tilde{R}}\ll 1$ such that weak perturbations violating symmetries have only an exponentially small effect, see also the discussion in Section~\ref{sec:stability}.   

Finally, we comment on the case when the Zeeman field is very weak such that $\sqrt{\tilde{R}}$ becomes of the same order or even much smaller than the magnetic length $\tilde{l}_B$. Since this happens first for the inner surface $\tilde{R}=\tilde{R}_<$, we discuss the case for a hole in an infinite system. For $\sqrt{\tilde{R}} \lesssim \tilde{l}_B \ll \tilde{R}$, we can still neglect the terms $\sim 1/\tilde{R}^2$ in the differential equations but can no longer expand the term $\sim (u/\tilde{l}_B^2) \cos{\varphi}$ around $\varphi\approx 0$ since $\Delta\varphi\sim O(1)$. As a result, one obtains precisely the same differential equations (\ref{eq:chi_diff_approx}) but $\tilde{k}^{(u)}_{1/2}(\varphi)$ obtains an angular dependence via
\begin{align}
     \label{eq:k_u_cos}
     \tilde{k}^{(u)}_{1/2}(\varphi) = iu \pm \sqrt{\tilde{\delta}-1-(u/\tilde{l}_B^2) \cos{\varphi}} \,.
\end{align}
This has the consequence that we find the same zero energy solution (\ref{eq:hat_chi_n_+_solution}) (up to the angular dependence via $\tilde{k}^{(u)}_{1/2}(\varphi)$) on the hole surface (i.e., $u=1$) provided that the condition  
\begin{align}
    \label{eq:delta_chi_+_states_cos}
    \tilde{\delta} > \frac{1}{\tilde{l}_B^2} \cos{\varphi}
\end{align}
is fulfilled, compare with Eq.~(\ref{eq:delta_chi_+_states}). In the TP $\tilde{\delta}>1/\tilde{l}_B^2$ this is fulfilled for all angles and we obtain a zero energy solution on the hole surface. In the NTP $\tilde{\delta}<-1/\tilde{l}_B^2$, this condition can never be fulfilled and no zero energy solution can exist. Finally, in the WP $|\tilde{\delta}|<1/\tilde{l}_B^2$, the condition can be only fulfilled for a certain angle interval but there is always a critical angle defined by $\tilde{\delta} = (1/\tilde{l}_B^2) \cos{\varphi_c}$, where the normal localization length tends to infinity. Therefore, for all angles with $|\varphi|<|\varphi_c|$, a localized solution does not exist and the state is not a true bound state but belongs to the bulk spectrum. As a consequence, zero energy topological states can not exist on the inner surface in the WP for any size of the Zeeman field.  

That a zero energy topological state can not exist in the WP on the hole surface can also be derived in an alternative and rigorous way for any hole radius. First of all we know that any outer surface infinitely away from the hole will host four zero energy states in the WP as derived above for large enough outer radius $\sqrt{\tilde{R}_>}\gg \tilde{l}_B$. Since these states do not care about the size of the inner surface (provided that $\tilde{R}_> - \tilde{R}_< \gg 1$ is not violated), it is impossible that by reducing the inner radius $\tilde{R}_<$ two of the outer states move to the inner surface. In addition, we have shown that the inner surface does not host any zero energy state in the WP for large enough inner radius $\sqrt{\tilde{R}_<}\gg \tilde{l}_B$. For these two reasons, the hole surface will not host zero energy states when reducing the inner radius, even not for the extreme case $\tilde{R}_< = 0$. Vice versa one can also argue that if two zero energy states exist at the inner surface for small enough radius then they can not go away by increasing $\tilde{R}_<$ since SUSY and chiral symmetry do not allow for a splitting of the two states. This leads to a contradiction since, for large enough radius $\sqrt{\tilde{R}_<}\gg \tilde{l}_B$, they must go away as derived above. Indeed, in the Supplemental Material \cite{SM} we show numerical results for center states at zero hole radius $\tilde{R}_<=0$ as function of the outer radius $\tilde{R}_>$ in the Weyl phase at weak Zeeman field where we confirm that the center states with $s=u=\pm 1$ move indeed to the outer surface if the radius $\tilde{R}_>$ is large enough such that $\sqrt{\tilde{R}_>}$ exceeds significantly $\tilde{l}_B$.

\subsection{Summary for topological states and numerical results}
\label{sec:states_summary_numerics}
To summarize the results of the previous section, we have found under the condition $1/\sqrt{\tilde{R}}\ll\tilde{l}_B\ll\sqrt{\tilde{R}}$ the following zero energy topological states with chirality $s$ and SUSY $u$
\begin{align}
    \label{eq:psi_bar_psi}
    \psi_{su} = \frac{1}{\sqrt{r}}\,U\,W\,X\,\bar{\psi}_{su}\,,
\end{align}
where the transformations $U$, $W$ and $X$ are given by (\ref{eq:U_rotation}), (\ref{eq:W}) and (\ref{eq:X}), respectively. A sign change of $s$ and $u$ corresponds in the transformed basis to a sign change of ${\bf x}$ and $\sigma_z$
\begin{align}
    \label{eq:sign_change_su}
    \bar{\psi}_{-s,-u}(\tilde{r},\varphi;\sigma_z,s_z) &= \bar{\psi}_{su}(\tilde{r},\varphi+\pi;-\sigma_z,s_z)\,.
\end{align}
Defining the normal part of the transformed wave function at the inner/outer surface by $\hat{\chi}_{+,n}^{(\pm 1)}(\tilde{r})$, as given by Eq.~(\ref{eq:hat_chi_n_+_explicit}), we find in the TP for $s=1$ two states localized at the inner/outer surface close to $\varphi\approx 0$
\begin{align}
    \nonumber
    \bar{\psi}_{1,\pm 1}(\tilde{r},\varphi;\sigma_z,s_z) &=
    \left(\begin{array}{c} 0 \\ 1 \end{array}\right)_{\sigma_z}\,
    \frac{1}{\sqrt{2}}\left(\begin{array}{c} 1 \\ \mp 1 \end{array}\right)_{s_z} \\
    \label{eq:TP_states_s=1}
    &\hspace{-2cm}
    \times\,\hat{\chi}_{+,n}^{(\pm 1)}(\tilde{r})\,
    \frac{1}{(2\pi)^{1/4}(\Delta\varphi_\lessgtr)^{1/2}} \,e^{-\frac{1}{4}(\varphi/\Delta\varphi_\lessgtr)^2}\,,
\end{align}
with $\Delta\varphi_\lessgtr=\tilde{l}_B/\sqrt{\tilde{R}_\lessgtr}$. The corresponding states with chirality $s=-1$ localized close to $\varphi\approx \pi$ follow from (\ref{eq:sign_change_su}). In the WP, the two states at the outer surface with $s=-u=\pm 1$ remain, but the states at the inner surface with $s=u=\pm 1$ are replaced by the two antisymmetric states in angular space at the outer surface
\begin{align}
    \nonumber
    \bar{\psi}_{11}(\tilde{r},\varphi;\sigma_z,s_z) &=
    \left(\begin{array}{c} 0 \\ 1 \end{array}\right)_{\sigma_z}\,
    \frac{1}{\sqrt{2}}\left(\begin{array}{c} 1 \\ 1 \end{array}\right)_{s_z}\\
    \label{eq:WP_antisymmetric}
    &\hspace{-2cm}
    \times\,\hat{\chi}_{+,n}^{(-1)}(\tilde{r})\,
    \frac{\varphi/\Delta\varphi_>}{(2\pi)^{1/4}(\Delta\varphi_>)^{1/2}} \,e^{-\frac{1}{4}(\varphi/\Delta\varphi_>)^2}\,.
\end{align}
together with $\bar{\psi}_{-1,-1}(\tilde{r},\varphi;\sigma_z,s_z)=\bar{\psi}_{11}(\tilde{r},\varphi+\pi;-\sigma_z,s_z)$ according to (\ref{eq:sign_change_su}).

In continuum numerics the topological states $\bar{\psi}_{su}$ in the transformed basis are shown in Fig.~\ref{fig:states_continuum}(a) for a disk with no hole and radius $\tilde{R}_> = 30$ at half-integer flux both in the TP and WP for various values of $\tilde{\delta}$ and fixed $\tilde{l}_B = 2$. Since the spinor structure agrees perfectly with the theoretical results we show only the square of the wave function averaged over the spinor labels. In the TP this gives from (\ref{eq:TP_states_s=1}) for the states with $s=1$ at the inner/outer surface
\begin{align}
    \label{eq:average_square_TP_states_s=1}
    \sum_{\sigma_z,s_z}|\bar{\psi}_{1,\pm 1}(\tilde{r},\varphi;\sigma_z,s_z)|^2 &=
    |\hat{\chi}_{+,n}^{(\pm 1)}(\tilde{r})|^2\,F^\lessgtr_+(\varphi)\,,\\
    \label{eq:F_+}
    F^\lessgtr_+(\varphi) &= \frac{e^{-\frac{1}{2}(\varphi/\Delta\varphi_\lessgtr)^2}}{(2\pi)^{1/2}\Delta\varphi_\lessgtr}\,.
\end{align}
In the WP, we get the same result for $u=-1$ at the outer surface but, for $u=1$, we have to use the result (\ref{eq:WP_antisymmetric}) for the antisymmetric state at the outer surface  
\begin{align}
    \label{eq:average_square_WP_antisymmetric}
    \sum_{\sigma_z,s_z}|\bar{\psi}_{11}(\tilde{r},\varphi;\sigma_z,s_z)|^2 &=
    |\hat{\chi}_{+,n}^{(-1)}(\tilde{r})|^2\,F^>_-(\varphi)\,,\\
    \label{eq:F_-}
    F^>_-(\varphi) &= (\varphi/\Delta\varphi_>)^2\,F^>_+(\varphi)\,.
\end{align}

A comparison of the analytical and numerical results for the topological states is shown in Fig.~\ref{fig:states_continuum}(b). The agreement is quite satisfactory although the condition $1/\sqrt{\tilde{R}}\ll\tilde{l}_B\ll\sqrt{\tilde{R}}$ is only approximately fulfilled. The analytical results predict a factorization in the thermodynamic limit into a radial and angular part. Therefore, one expects the numerical result to be approximately independent of $\varphi$/$\tilde{r}$ when normalizing to the angular/radial part (as shown in the left/right panels of Fig.~\ref{fig:states_continuum}(b)). This is confirmed in the two upper panels of Fig.~\ref{fig:states_continuum}(b) for the state $\bar{\psi}_{1,-1}$ in the TP for $\tilde{\delta}=1$ and $\tilde{l}_B=2$. One can see that all cuts of the left/right panel at fixed $\varphi/\tilde{r}$ fall almost on top of each other and agree with the theoretical results. For this comparison we used the continuum numerics and found already a good agreement for a rather small disk radius $\tilde{R}_>=30$. For the antisymmetric state $\bar{\psi}_{1,1}$ in the WP for $\tilde{\delta}=0$ and $\tilde{l}_B=2$ we needed a larger radius $\tilde{R}_>=60$ to find a good agreement, see the two lower panels of Fig.~\ref{fig:states_continuum}(b). These data have been obtained by using the tight-binding numerics. In the Supplemental Material \cite{SM} we consider also a Corbino disk with outer radius $\tilde{R}_> = 45$ and hole radius $\tilde{R}_< = 20$ by using the tight-binding numerics and find in the TP in addition a good agreement of the numerical and analytical results for the states at the inner surface.   

Concerning the spinor dependence of the topological states we see from the above formulas that the states at the outer/inner surface for chirality $s$ have the following spinor dependence in the transformed basis
\begin{align}
    \label{eq:spinor_dependence_transformed_basis}
    \sigma_z = -s \quad,\quad s_x = \pm 1 \,.
\end{align}
In the original basis, this means by reversing the unitary trafos $U$, $X$, and $W$, as defined in Eqs.~(\ref{eq:U_trafo}), (\ref{eq:X_trafo}) and (\ref{eq:W_trafo}), that we get a strong correlation between the spin and orbital part of the spinor degrees of freedom given by
\begin{align}
    \label{eq:spinor_dependence_original_basis}
    s_z = s \sigma_x \quad,\quad s_r = \pm \sigma_y \,,
\end{align}
where $s_r={\bf e}_r \cdot {\bf s}$ denotes the spin in radial direction.

\subsection{Validity range of effective surface Hamiltonian}
\label{sec:validity}

In this subsection we extend the validity range of the effective surface Hamiltonian (\ref{eq:bar_H_t_corbino}) derived in Section~\ref{sec:corbino_weak_fields} for the case of weak Zeeman field to the same regime (\ref{eq:condition_l_B}) we used in Section~\ref{sec:corbino_strong_fields} to discuss the topological states at strong Zeeman field. Besides the condition (\ref{eq:condition_l_B}), we assume in addition that the surface gap $\Delta_{\rm surface}\approx E_Z$ is below or of the order of the bulk gap, such that the complete parameter regime is defined by  
\begin{align}
    \label{eq:general_condition_H_eff}
    \frac{1}{\sqrt{\tilde{R}}} \ll \tilde{l}_B \ll \sqrt{\tilde{R}} \quad,\quad E_Z \lesssim \Delta_{\rm bulk}\,.
\end{align}
In this regime, we aim at calculating all localized bound states well below the surface gap which have an energy of the order of the Witten frequency
\begin{align}
    \label{eq:Witten_frequency_condition}
    \tilde{\Omega}_W = \frac{2}{\tilde{\xi_t}} = 2 \tilde{E}_Z \Delta\varphi \ll E_Z \lesssim \Delta_{\rm bulk}\,,
\end{align}
which is well below the bulk gap, such that first order perturbation theory will be sufficient to treat the angular part of the Hamiltonian. 

To derive the effective surface Hamiltonian, it is needed to treat the normal component of the Zeeman term 
\begin{align}
    \label{eq:H_Z_normal}
    \bar{H}_{Z,n}/E_{\rm so} = \sigma_y s_x \tilde{E}_Z \sin{\varphi} \sim  \tilde{E}_Z \Delta\varphi 
\end{align}
as a perturbation on the same footing as the angular part of the spin-orbit interaction which is $\sim 1/(\tilde{R}\Delta\varphi)\sim 1/\tilde{\xi}_t$. This requires 
\begin{align}
    \label{eq:condition_2_H_eff}
    {1\over \tilde{\xi}_t}=\tilde{E}_Z \Delta\varphi \ll 1 \,. 
\end{align}
On the other hand, the tangential component of the Zeeman term is given by 
\begin{align}
    \nonumber
    \bar{H}_{Z,t}/E_{\rm so} &= \sigma_y s_y \tilde{E}_Z \cos{\varphi} \\
    \label{eq:H_Z_tangential}
    &=  \eta_{\varphi}\sigma_y s_y \tilde{E}_Z + O(\tilde{E}_Z \Delta\varphi^2)\,,
\end{align}
where $\eta_{\varphi}=\pm 1$ if one considers the bound states localized around $\varphi=0,\pi$, respectively. It can be approximated by the first leading term for strong localization since the correction is of order $\tilde{E}_Z \Delta\varphi^2\sim 1/\tilde{R}$ and negligible against the other terms of the surface Hamiltonian. Taking the leading term together with the normal part of the Hamiltonian, we can split the Hamiltonian $\bar{H}_{1/2}\approx\bar{H}_n + \bar{H}_t$ into a normal and tangential part, thereby including consistently all terms of order $\bar{H}_n/E_{\rm so}\sim O(1)$ and $\bar{H}_t/E_{\rm so}\sim O(1/\sqrt{\tilde{R}})$, see the detailed discussion in Eqs.~(\ref{eq:estimations_1}-\ref{eq:estimations_3}). According to Eqs.~(\ref{eq:bar_H_n_corbino}) and (\ref{eq:bar_H_t_corbino}) we obtain  
\begin{align}
    \nonumber
    \bar{H}_n/E_{\rm so} &= \sigma_x \big\{(-\partial_{\tilde{r}}^2 - \tilde{\delta})s_z + 2i \partial_{\tilde{r}}s_y\big\}
    +\eta_\varphi\sigma_y s_y \tilde{E}_Z \\
    \label{eq:H_normal_general}
     &\hspace{-1cm}
     = \sigma_x \Big\{\left[-\partial_{\tilde{r}}^2 - (\tilde{\delta} + \eta_\varphi s_x\bar{S} E_Z)\right]s_z + 2i \partial_{\tilde{r}}s_y\Big\} \,,\\
    \label{eq:H_tangential_general}
    \bar{H}_t/E_{\rm so} &= s_x \big\{\sigma_x\frac{2}{\tilde{R}}(-i\partial_{\varphi}) + \sigma_y \tilde{E}_Z \sin{\varphi}\big\}\,,
\end{align}
where $\bar{S}=-\sigma_z$ is the chiral symmetry in transformed basis (see Eq.~(\ref{eq:S_trafo})), and we still include $s_x$ in $\bar{H}_t$ at this stage (like it occurs in the Hamiltonian). 

Since $\bar{H}_t$ is an order $1/\tilde{\xi}_t$ smaller than $\bar{H}_n$, we first solve for the boundary states of $\bar{H}_n$. It has the same form as the normal Hamiltonian (\ref{eq:bar_H_n_corbino}) without Zeeman field, but the band inversion parameter is shifted by
\begin{align}
    \label{eq:delta_shift}
    \tilde{\delta} \rightarrow \tilde{\delta} + \eta_\varphi s_x \bar{S} E_Z\,.
\end{align}
As a consequence, for $\delta > E_Z$, we find the same two zero-energy boundary states of $\bar{H}_n$ with chirality $s=-\sigma_z=\pm 1$ and polarization $s_x=\pm 1$ for the states at the outer/inner surface, with a wave function given analog to (\ref{eq:psi_n_corbino}) and (\ref{eq:Phi_n_corbino}) by 
\begin{align}
    \label{eq:psi_n_general}
    \bar{\psi}_{n,s}^{\gtrless}(\tilde{r},\sigma_z,s_z) = \frac{\delta_{\sigma_z,-s}}{\sqrt{2}} 
    \left(\begin{array}{c} 1 \\ \pm 1\end{array}\right)_{s_z}\, \bar{\Phi}_{n,s}^\gtrless(|\tilde{r}-\tilde{R}^\gtrless|) \,,
\end{align}
with
\begin{align}
    \label{eq:Phi_ns}
    \bar{\Phi}_{n,s}^\gtrless(\tilde{r})= \bar{\Phi}_n(\tilde{r})|_{\delta\rightarrow\delta\pm s\,\eta_{\varphi}E_Z} \,.
\end{align}
This leads precisely to the two different normal localization lengths $\xi_n^\gtrless$, as given by Eqs.~(\ref{eq:xi_n_s=1_u=-1}) and (\ref{eq:xi_n_s=1_u=1_gapped}) for the boundary states at the outer/inner surface in the topological phase, respectively. 

In the next step one considers $\bar{H}_t$ as a perturbation to calculate all localized bound states with energies of the order of the Witten frequency which is much smaller than the bulk gap according to Eq.~(\ref{eq:Witten_frequency_condition}), such that first order perturbation theory is sufficient. Due to chiral symmetry, only the nondiagonal matrix element 
\begin{align}
    \nonumber
    H^{\rm eff}_{\rm surface}/E_{\rm so}&=\langle \bar{\psi}_{n,s}^\gtrless | \hat{H}_t/E_{\rm so} | \bar{\psi}_{n,-s}^\gtrless \rangle \\ 
    \label{eq:H_t_nondiagonal_matrix_element}
    & =\pm \lambda^\gtrless \big\{\sigma_x\frac{2}{\tilde{R}}(-i\partial_{\varphi}) + \sigma_y \tilde{E}_Z \sin{\varphi}\big\}
\end{align}
enters for the effective surface Hamiltonian, where we have inserted the polarization $s_x=\pm 1$ of the bound states at the outer/inner surface and defined the matrix element
\begin{align}
    \label{eq:lambda_gtrless}
    \lambda^\gtrless = \langle \bar{\Phi}_{n,s}^\gtrless | \bar{\Phi}_{n,-s}^\gtrless \rangle\,, 
\end{align}
which is independent of $s=\pm 1$ and only rescales the Witten frequency. As a consequence, up to this trivial rescaling factor, we obtain the same effective surface Hamiltonian as for weak Zeeman field. 

In summary, we have found that the universal low-energy theory in terms of the effective surface Hamiltonian (\ref{eq:H_t_nondiagonal_matrix_element}) can be used for the calculation of all localized bound states well below the surface gap, provided that the two conditions stated in (\ref{eq:general_condition_H_eff}) are fulfilled. We emphasize that these conditions can be easily fulfilled by choosing the radius $R$ large enough and by staying not too close to the phase transition. Therefore, the universal low-energy model can be used for arbitrary spin-orbit interaction, even including the regime of weak spin-orbit $E_{\rm so}\ll E_Z$ or $\lambda_{\rm so}\gg l_B$.

\subsection{Phase diagram and comparison to numerical results}
\label{sec:phase_diagram_corbino}

In this section we state the phase diagram in terms of the Witten index based on the analytical results of Section~\ref{sec:topological_states_corbino}, together with a comparison to numerical results. Here, the Witten index $n_W$ denotes the number of zero energy states or, to be more general for a large but finite system, counts the number of states with exponentially small energies which tend to zero when increasing the system size (see the discussion at the end of Section~\ref{sec:SUSY}). Moreover, in the Weyl phase, where the bulk gap is zero, we count only topological bound states at zero energy but not a possible bulk state at zero energy. The Witten index is a standard index in supersymmetric systems distinguishing broken from unbroken SUSY, see e.g. Refs. \onlinecite{durand_vinet_pla_90,beckers_debergh_mpl_89,beckers_debergh_jmp_90}, playing here the role of the topological invariant in the presence of SUSY.
\begin{figure}
	 \includegraphics[width =1.0\columnwidth]{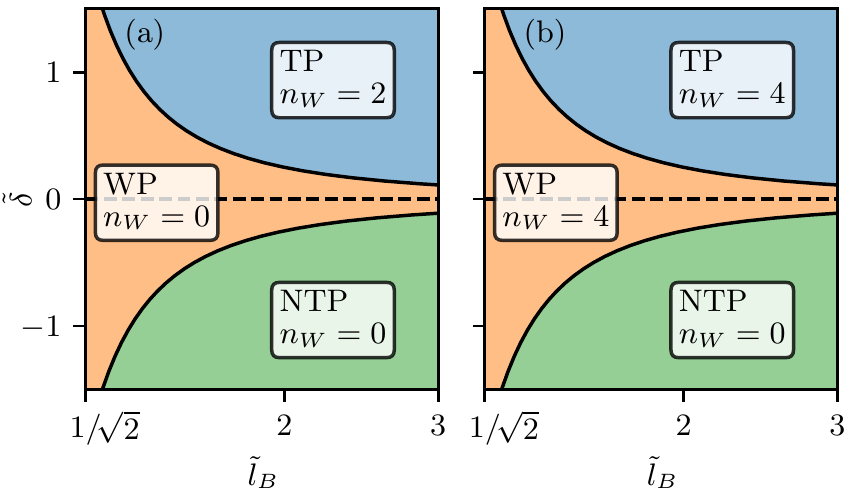} 
	  \caption{The phase diagram at half-integer flux $f=1/2$ in terms of the Witten index $n_W$ counting the number of zero energy topological states for (a) a hole with radius $\tilde{R}_<$ in an infinite system and (b) a Corbino disk with outer/inner radius $\tilde{R}_\gtrless$.
	  } 
    \label{fig:phase_diagram}
\end{figure}
\begin{figure*}[t!]
    \centering
 	 \includegraphics[width =0.95\columnwidth]{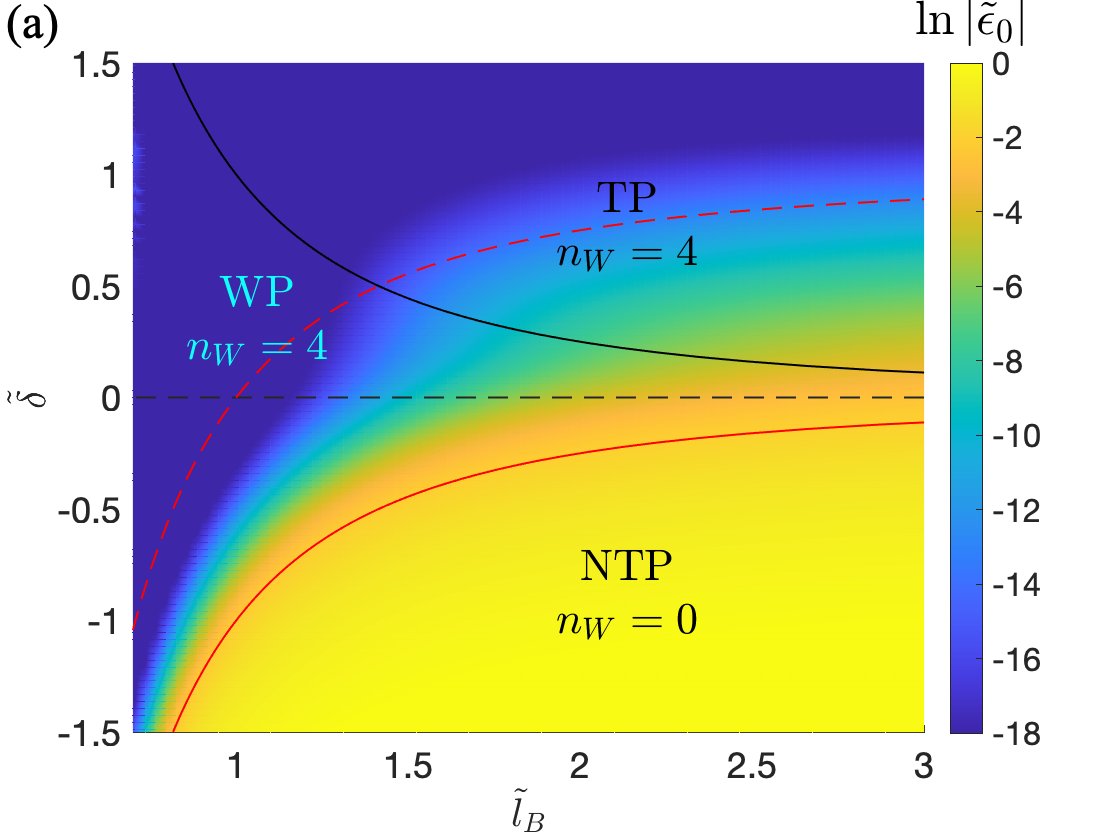}
    \hfill\includegraphics[width =0.95\columnwidth]{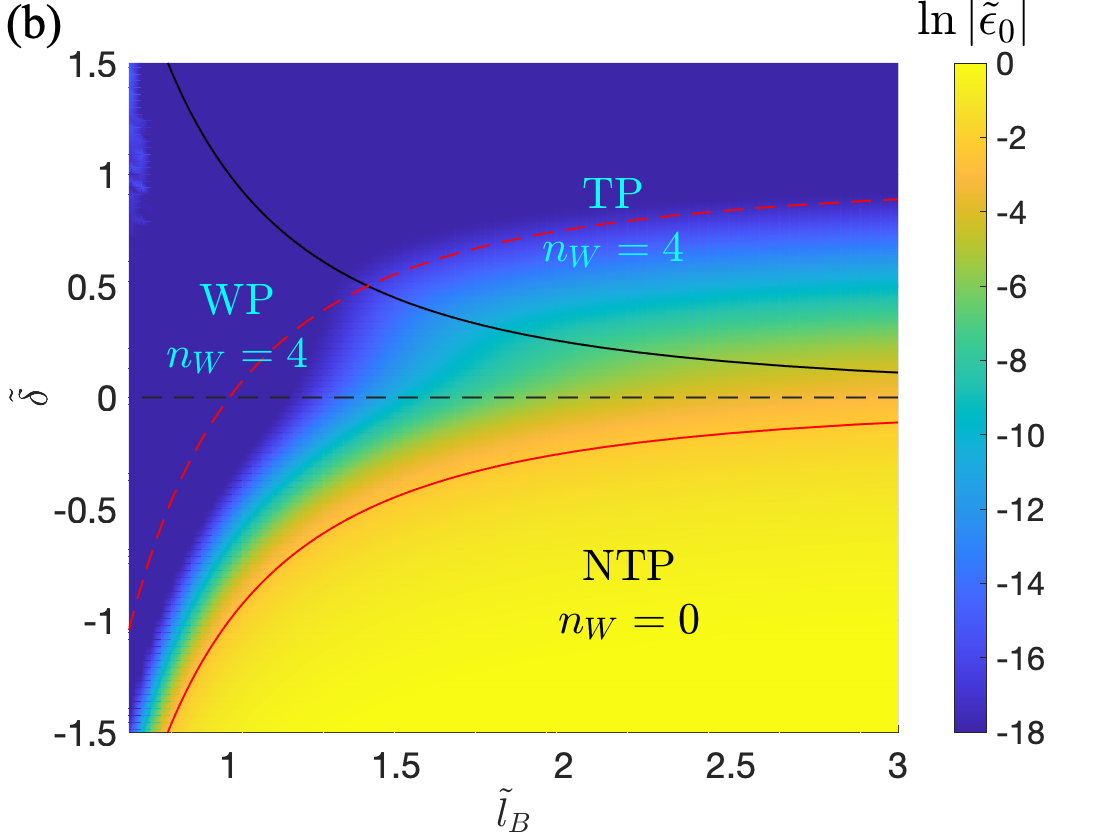}
	  \caption{Numerical results in tight-binding for the absolute value $|\tilde{\epsilon}_0|=|\epsilon_0|/E_{\rm so}$ in logarithmic color scale of the lowest absolute energy $|\epsilon_0|$ in units of the spin-orbit energy $E_{\rm so}$ as function of $\tilde{\delta}$ and magnetic length $\tilde{l}_B$. We consider half-integer flux $f=1/2$ and a Corbino disk with outer radius $\tilde{R}_>=30$ and inner radius given by (a) $\tilde{R}_< = 10$ and (b) $\tilde{R}_< = 0$. The dotted red line indicates the line $\tilde{\delta}=1-1/\tilde{l}_B^2$ above which the energy is approximately a constant since the normal localization length $\tilde{\xi}_n\sim O(1)$ of the states at the outer surface does not vary in this regime, see (\ref{eq:xi_n_s=1_u=-1}). In the region below this line the energy and the normal localization length are approximately a constant in the TP and WP along the lines $\tilde{\delta}=c-1/\tilde{l}_B^2$ with some constant $0<c<1$. For decreasing $c$ the energy becomes larger since the normal localization length increases, see (\ref{eq:xi_n_s=1_u=-1}). The deformation of these lines when crossing from the TP to the WP phase results from an increased distance between the hybridizing states since two of the states at the inner surface move to the outer surface in the WP. In the NTP there are no localized bound states and the energy is of the order of the bulk gap. When the radius of the inner surface decreases (right figure), the qualitative features remain the same but one obtains an overall decrease of the energies since the distance of the hybridizing states increases. 
    } 
    \label{fig:witten_disc_corbino}
\end{figure*}
\begin{figure*}[t!]
    \centering
 	 \includegraphics[width =0.95\columnwidth]{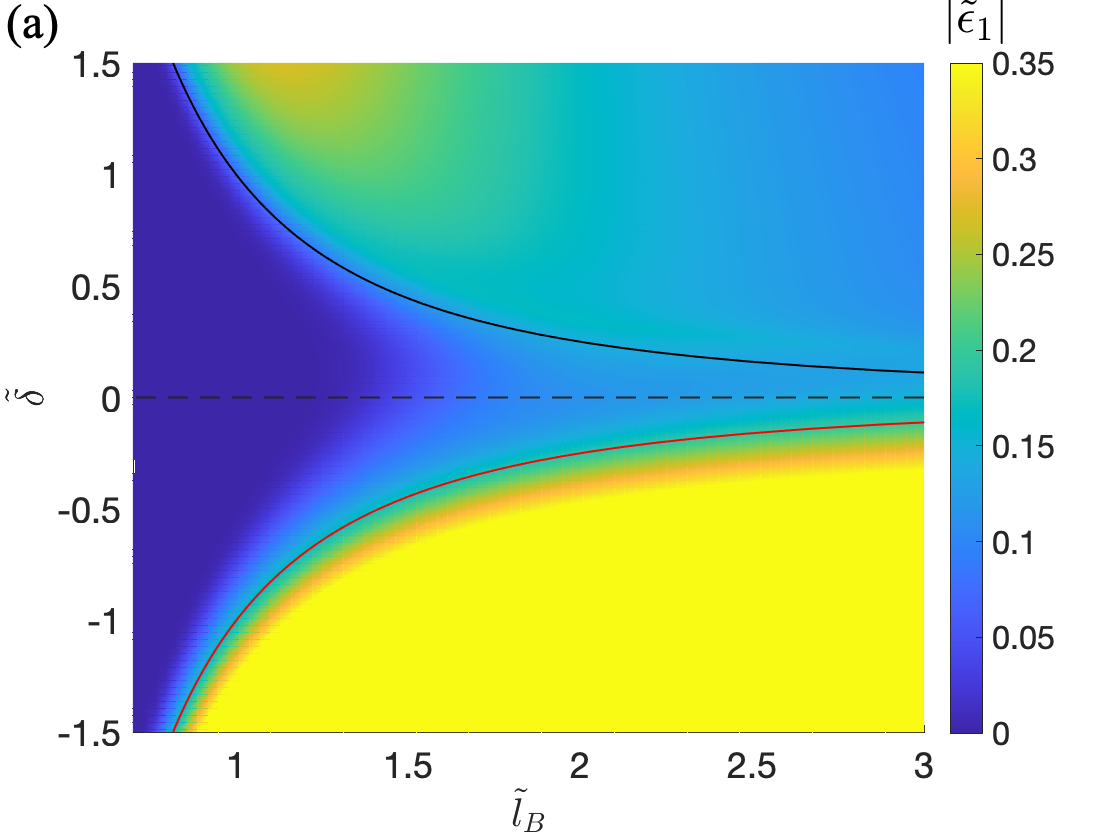}
    \hfill\includegraphics[width =0.95\columnwidth]{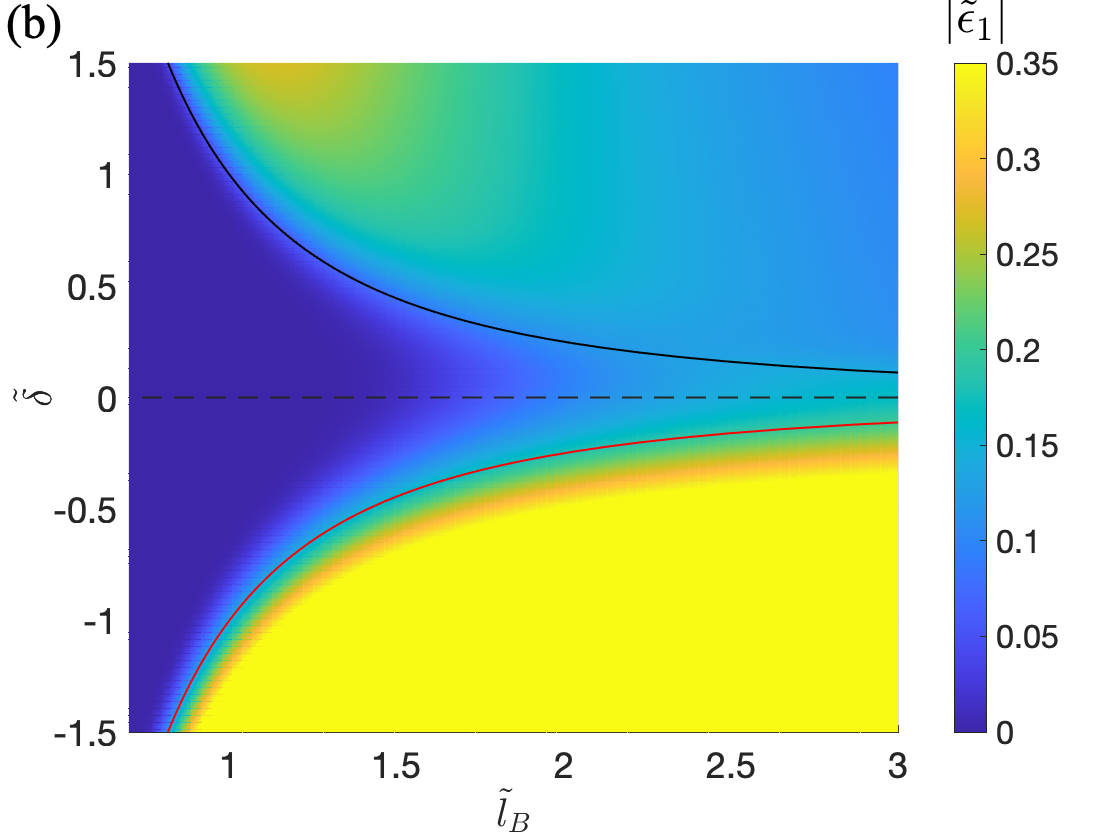}
	  \caption{The same as Fig.~\ref{fig:witten_disc_corbino} but for the second-lowest absolute energy $|\tilde{\epsilon}_1|$ in non-logarithmic scale. Roughly, this energy scales with $1/(\sqrt{\tilde{R}}\,\tilde{l}_B)$ according to (\ref{eq:eigenvalues_higher_bound_states}) in the TP, has a significantly smaller value in the ungapped WP, and is almost constant of the order of the bulk gap in the NTP.
    } 
    \label{fig:witten_disc_corbino_second_energy}
\end{figure*}

We discuss here the most interesting case close to the SUSY point of half-integer flux $f=1/2$ and the case of a Corbino disk with outer/inner radius $\tilde{R}\equiv \tilde{R}_\gtrless$ (with $\tilde{R}_> - \tilde{R}_< \gg 1$) in the regime where the magnetic length fulfils the condition (\ref{eq:condition_l_B}) 
\begin{align}
     \label{eq:condition_l_B_new}
     \frac{1}{\sqrt{\tilde{R}}} \ll \tilde{l}_B \ll \sqrt{\tilde{R}}
 \end{align}
for both $\tilde{R}=\tilde{R}_>$ and $\tilde{R}=\tilde{R}_<$. As discussed in Section~\ref{sec:corbino_strong_fields} this is a well-defined limit for large $\tilde{R}$ where the tangential localization length is much larger than the normal one (the first condition $1/\sqrt{\tilde{R}}\ll\tilde{l}_B$) and where the tangential localization length is much smaller than the circumference (i.e., small angular spread, the second condition $\tilde{l}_B\ll \tilde{R}$). As a result, the topological bound states are well-localized at the positions shown in Fig.~\ref{fig:states_corbino} with an exponentially small overlap, leading to stability against small deviations from half-integer flux or other small perturbations violating chiral symmetry or SUSY (like, e.g., weak disorder), see the detailed discussion in Section~\ref{sec:stability}.  

A sketch of the phase diagram is shown in Fig.~\ref{fig:phase_diagram} for (a) a hole in an infinite system (i.e., infinite outer radius $\tilde{R}_>=\infty$) and (b) a Corbino disc. For a hole in an infinite system we find two topological bound states ($n_W=2$) in the topological gapped phase (TP) and no topological bound states ($n_W=0$) in the Weyl phase (WP) and the non-topological gapped phase (NTP). The normal localization length $\tilde{\xi}_n$ of the two states in the TP is given by (\ref{eq:xi_n_s=1_u=1_gapped}) which diverges at the crossover line $\tilde{\delta}=1/\tilde{l}_B^2$ between TP and WP. For a Corbino disc, two additional topological bound states appear in the TP at the outer surface ($n_W=4$). Their $\tilde{\xi}_n$ is given by (\ref{eq:xi_n_s=1_u=-1}) which does {\it not} diverge at the crossover to the WP phase. These two topological bound states persist in the WP and disappear with a diverging $\tilde{\xi}_n$ at the crossover between WP and NTP. In addition, two further topological states with the same $\tilde{\xi}_n$ appear in the WP at the outer surface (such that $n_W=4$ in the WP). They are special in the sense that their $\tilde{\xi}_n$ does not diverge at the crossover between WP and TP but they disappear in the TP at the outer surface and are replaced by the two topological states at the inner surface which have a diverging $\tilde{\xi}_n$ at the crossover. Thus, at the crossover between TP and WP, two of the topological states change their position (from inner to outer surface) and change from a diverging to a finite $\tilde{\xi}_n$. Up to our knowledge this has not been found for any other topological system and seems to be a special feature generated at the crossover between a gapped topological and a gapless Weyl phase. 

In Fig.~\ref{fig:witten_disc_corbino}(a) we show the numerical results for the phase diagram for a Corbino disk with outer radius $\tilde{R}_> = 30$ and inner radius $\tilde{R}_< = 10$. To exhibit the different phase regions we plot in logarithmic scale the lowest absolute energy $|\tilde{\epsilon}_0|=|\epsilon_0|/E_{\rm so}$. Due to chiral symmetry and SUSY this energy corresponds to four states consisting of two pairs, one at positive and one at negative energy with the same absolute value. The splitting occurs since two topological states with different chiral symmetry and the same SUSY can hybridize for a finite system, leading to an exponentially small splitting of the energy. In the TP phase this splitting occurs between the right (left) states at the outer surface with $s=-u=1$ ($s=-u=-1$) and the left (right) states at the inner surface with $s=u=-1$ ($s=u=1$), see Fig.~\ref{fig:states_corbino}. The size of the splitting depends on the two normal localization lengths $\tilde{\xi}_n$ of the two states which hybridize and is expected to be exponentially small roughly $\sim e^{-(\tilde{R}_>-\tilde{R}_<)/\tilde{\xi}_n}$. Since the two hybridizing states appear at opposite angles of the two surfaces and since the angular spread is small, the $\tilde{\xi}_n$ of the states at the outer surface will dominate the orbital overlap and the splitting. This is reflected in the TP of Fig.~\ref{fig:witten_disc_corbino}(a) where the logarithm $\ln{|\epsilon_0|}$ of the lowest absolute energy follows the size of $-1/\tilde{\xi}_n$ of the states at the outer surface as given by (\ref{eq:xi_n_s=1_u=-1}): (1) in the region $\tilde{\delta}>1-1/\tilde{l}_B^2$ both the smallest energy and $\tilde{\xi}_n$ are approximately constant; (2) in the region $1/\tilde{l}_B^2 < \tilde{\delta}<1-1/\tilde{l}_B^2$, the lowest energy and $\tilde{\xi}_n$ are approximately a constant in the TP and WP on the same lines $\tilde{\delta}=1-1/\tilde{l}_B^2+c$ with $0<c<1$. In the WP all topological states appear at the outer surface with the same $\tilde{\xi}_n$. Since the hybridizing states are localized at opposite angles of the outer surface their orbital overlap is reduced compared to the TP since their distance increases from $\tilde{R}_> - \tilde{R}_<$ to $2 \tilde{R}_>$. As a consequence, the energy splitting reduces in the WP which can be seen in Fig.~\ref{fig:witten_disc_corbino}(a) by a deformation of the lines of constant energy at the crossover from TP to WP. In the NTP no topological bound states and no edge states appear in the gap and the lowest energy becomes of the order of the bulk gap, consistent with Fig.~\ref{fig:witten_disc_corbino}(a). In Fig.~\ref{fig:witten_disc_corbino}(b) we show the same but for zero hole radius $\tilde{R}_<=0$. Qualitatively the same considerations apply, leading to the same results, only the size of the lowest energy is slightly reduced in the TP since the center states are farer away from the boundary of the disc.  

In Fig.~\ref{fig:witten_disc_corbino_second_energy}(a,b) we show the same for the second-lowest absolute energy $|\tilde{\epsilon}_1|$. In contrast to the lowest one it reveals clearly both the phase transition line from TP to WP and from WP to NTP. Deep in the TP for weak Zeeman field, it is approximately given by $|\tilde{\epsilon}_1|\approx 2/(\sqrt{\tilde{R}}\,\tilde{l}_B)$, see Eq.~(\ref{eq:eigenvalues_higher_bound_states}). Therefore, it is expected to decrease with increasing magnetic length and to scale with $1/\sqrt{\tilde{R}}$, roughly consistent with Fig.~(\ref{fig:witten_disc_corbino_second_energy}). In contrast, in the WP, there is no bulk gap and therefore the second-lowest energy will drastically decrease, possibly consisting of bulk states. In the gapped NTP, there are no states in the bulk gap, and therefore the second-lowest energy behaves similar to the lowest one.

\section{Periodic Witten models}
\label{sec:witten_model}

In this section we will generalize the derivation of effective surface Hamiltonians for a Corbino disk in Section~\ref{sec:corbino_weak_fields} to the case of generic smooth surfaces. We characterize the smoothness by the local curvature radius $R$ (either for the outer or inner surface) and, in analogy to the condition (\ref{eq:large_radius}) for a Corbino disc, assume the curvature radius to be much larger than the spin-orbit length
\begin{align}
    \label{eq:small_curvature_condition}
    \tilde{R} \gg 1 \,.
\end{align}
In addition, as in Section~\ref{sec:corbino_strong_fields}, we consider weak magnetic fields deep in the TP 
\begin{align}
    \label{eq:witten_conditions}
    \tilde{l}_B^2 \gg 1 \quad,\quad \tilde{\delta}\gg 1/\tilde{l}_B^2 \,.
\end{align}
Under these conditions we will show that the whole generic class of periodic Witten models can be realized for the effective surface Hamiltonian, where the surface potential is again characterized by the normal component of the magnetic field to the surface. We outline in Section~\ref{sec:witten_derivation} the derivation of the effective surface Hamiltonian and state the central result. In Section~\ref{sec:witten_SUSY} we will exhibit the SUSY properties of the periodic Witten model together with an explicit analytical expression for the wave functions of the zero energy topological states for any shape of a smooth surface. In Section~\ref{sec:spectrum_surface_H} we will generically show how all eigenstates of the Hamiltonian can be determined from the ones of the Witten model and we will present a semiclassical analysis to calculate all localized bound states of the Witten model in the case when the tangential localization length $\xi_t$ is much smaller than the circumference of the surface. Finally, in Section~\ref{sec:peanut} we demonstrate our analytical results by a comparison to a numerical tight-binding calculation for a surface of peanut shape.

Analog to Section~\ref{sec:validity}, we note that the derivation of the effective surface Hamiltonian can be extended to the regime defined by (\ref{eq:general_condition_H_eff}) if one is only interested in the calculation of the localized bound states below the surface gap in the case of strong localization. This holds also for generic smooth surfaces, and shows the wide applicability regime of the universal low-energy theory.

\subsection{Derivation of periodic Witten models}
\label{sec:witten_derivation}

In this subsection we will outline the derivation of the effective surface Hamiltonian describing only the edge states at the boundary of the system. We summarize here the essential steps and state the final result together with the validity range of the derivation. 

\begin{figure}
	 \includegraphics[width =1.0\columnwidth]{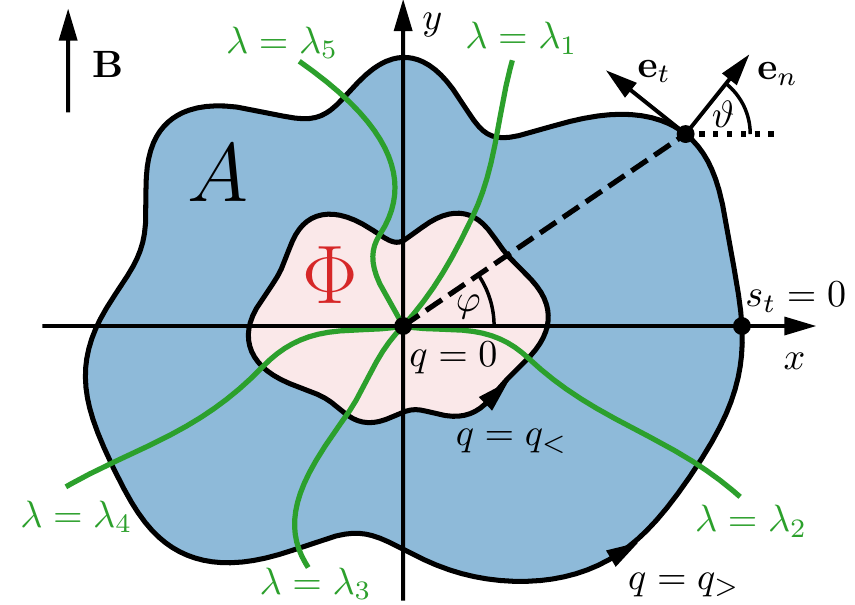}
	  \caption{Definition of generic shapes of an area $A$ via orthogonal coordinates $(q,\lambda)$. The area is defined in the region $q_< < q < q_>$. The inner and outer boundaries are defined by $q({\bf x})=q_<$ and $q({\bf x})=q_>$, respectively. The coordinate along a closed surface is denoted by $0<\lambda<\Lambda$ with periodic boundary conditions. The origin ${\bf x}=0$ corresponds to $q=0$. The two unit vectors ${\bf e}_n$ and ${\bf e}_t$ are orthogonal and tangential to the surfaces $q={\rm const}$, respectively. Whereas $\varphi$ is the polar angle of ${\bf x}=r(\cos\varphi,\sin\varphi)$, the angle $\vartheta$ denotes the angle between ${\bf e}_n$ and the $x$-axis, i.e., ${\bf e}_n {\bf e}_x = \cos{\vartheta}$ or ${\bf e}_n {\bf e}_y = \sin{\vartheta}$. The arc length $0 < s_t < L$ along a surface $q={\rm const}$ is defined by $ds_t/d\lambda=|d{\bf x}/d\lambda|$, measured in counterclockwise direction with reference point $s_t=0$ chosen on the positive $x$-axis.       
	  } 
    \label{fig:ortho}
\end{figure}

To describe arbitrary surfaces it is convenient to introduce orthogonal coordinates $(q,\lambda)$, where $q$ denotes the coordinate normal to the surface (with dimension of length) and $\lambda$ is a dimensionless angle variable, see Fig.~\ref{fig:ortho}. We do not assume at this stage that the surface is mirror-symmetric since this is not essential for the derivation of the effective surface Hamiltonian. For convenience, we will take half-integer flux $f=1/2$ since a flux deviating from $1/2$ will only change the boundary conditions for the states of the effective surface Hamiltonian.  
The two surfaces of the system describing the inner and outer surface are then given by the conditions $q=q_<$ and $q=q_>$, respectively, such that the system is present in the regime $q_< < q < q_>$ and all $\lambda$. The orthogonal coordinates are fully characterized by the relation
\begin{align}
    \label{eq:ortho_dx}
    d{\bf x} = h_n {\bf e}_n dq + h_t {\bf e}_t d\lambda \,,
\end{align}
where $h_{n,t}>0$ are the Lame coefficients, and the orthogonal unit vectors ${\bf e}_n$ and ${\bf e}_t$ are directed normal and tangential to the surface and point in the direction of increasing $q$ or $\lambda$, respectively, see Fig.~\ref{fig:ortho}. It is also convenient to introduce the line elements 
\begin{align}
    \label{eq:line_elements}
    ds_t = h_t d\lambda \quad,\quad ds_n = h_n dq     
\end{align}
along the lines where $q={\rm const}$ or $\lambda={\rm const}$, respectively. The Lame coefficient $h_n$ is dimensionless and $h_t$ is proportional to a typical length scale such that, for a given surface $q={\rm const}$, we get the order of magnitude
\begin{align}
    \label{eq:order_lame}
    h_n \sim O(1) \quad,\quad h_t \sim R \,.
\end{align}
The length scale $R$ plays a very important role and is assumed here to be at the same time the typical length scale on which the Lame coefficients vary on a given surface
\begin{align}
    \label{eq:variation_lame}
    \partial_q h_{n,t} \sim \frac{1}{R} h_{n,t} \quad,\quad \partial_\lambda h_{n,t} \sim h_{n,t}\,.
\end{align}
This means that the local curvature of the surface, defined by
\begin{align}
    \label{eq:kappa_t}
    \kappa_t = - {\bf e}_n \partial_{s_t} {\bf e}_t = \frac{1}{D}\partial_q h_t \sim O(\frac{1}{R}) \quad,\quad D=h_n h_t \,,
\end{align}
is proportional to the inverse of this length scale, i.e., we can define $R$ as the local curvature radius of the surface. Although this length scale can vary along the surface we assume that $R$ defines a lower bound for the curvature radius. We note the quantization rule of the total curvature when integrating the local curvature along the whole surface
\begin{align}
    \label{eq:curvature_quantization}
    \oint_{q={\rm const}} ds_t \,\kappa_t = 2\pi \,,
\end{align}
which follows from the relation
\begin{align}
    \label{eq:theta_kappa}
        \kappa_t = \partial_{s_t} \vartheta  \,,
\end{align}
where $\vartheta$ is the angle of the vector ${\bf e}_n$ normal to the surface and the $x$-axis, see Fig.~\ref{fig:ortho}.

In (\ref{eq:kappa_t}) we also introduced the symbol $D\sim O(R)$ which is the transformation coefficient between the area elements in cartesian and orthogonal coordinates
\begin{align}
    \label{eq:area_element}
    dx\,dy \,=\, D\,dq\,d\lambda\,.
\end{align}

To write the Hamiltonian in orthogonal coordinates, it is convenient to use an analog transformation (\ref{eq:polar_H_trafo}) as for the Corbino disc
\begin{align}
    \label{eq:ortho_H_trafo}
    \bar{H}_{1/2} =  X^\dagger W^\dagger \, U^\dagger \,\sqrt{D}\, H_{1/2} \, \frac{1}{\sqrt{D}} \, U \, W \, X \,,
\end{align}
with the only difference that $r$ is replaced by $D$, and the transformation $U$ is defined by 
\begin{align}
    \label{eq:U_rotation_ortho}
    U &= e^{-i\frac{1}{2}\varphi} \, e^{-i\frac{1}{2}s_z\vartheta} \,,
\end{align}
where $\varphi$ is the polar angle. This guarantees that the normalization is given by
\begin{align}
    \label{eq:ortho_normalization} 
    \int_{q_<}^{q_>} dq \,\oint d\lambda \sum_{\sigma_z,s_z=\pm} |\bar{\psi}(q,\lambda;\sigma_z,s_z)|^2 = 1 \,,
\end{align}
and $U$ eliminates the half-integer flux and rotates the spin to the local frame of the unit vectors ${\bf e}_n$ and ${\bf e}_t$, defining the directions normal and tangential to the surface, respectively. 

For the special case of a mirror-symmetric surface, we note the following properties for the transformation of the angle $\vartheta$ under a sign change of ${\bf x}$ or $\varphi$ 
\begin{align}
    \label{eq:theta_mirror_symmetry}
    \vartheta \xrightarrow{P_{\bf x}} \vartheta + \pi \quad,\quad
    \vartheta \xrightarrow{P_{\varphi}} -\vartheta \,.
\end{align}
This has the consequence that the chiral symmetry, the inversion symmetry, and SUSY are given by Eqs.~(\ref{eq:S_trafo}-\ref{eq:SUSY_trafo}), just as in the case of the Corbino disk after applying transformation (\ref{eq:ortho_H_trafo}). 

After a lengthy calculation we find for the transformed Hamiltonian in orthogonal coordinates and dimensionless units
\begin{widetext}
    \begin{align}
    \nonumber
        \bar{H}_{1/2}/E_{\rm so} &= \sigma_x \left\{
        \left[ -\partial_{\tilde{q}} \frac{1}{h_n^2} \partial_{\tilde{q}} 
        - \partial_\lambda \frac{1}{\tilde{h}_\lambda^2} \partial_\lambda + \frac{1}{4}(\tilde{\kappa}_n^2 + \tilde{\kappa}_t^2) + \tilde{V} - \tilde{\delta} - \frac{1}{2} \left( \{\frac{\tilde{\kappa}_t}{\tilde{h}_t},-i\partial_\lambda\}
        - \{\frac{\tilde{\kappa}_n}{h_n},-i\partial_{\tilde{q}}\} \right)  \right] s_z \right. \\
        \label{eq:bar_H_ortho}
        & \hspace{1cm} \left. + \left[ \{\frac{1}{\tilde{h}_t},-i\partial_\lambda\} s_x 
        - \{\frac{1}{h_n},-i\partial_{\tilde{q}}\} s_y\right] \right\}
        + \sigma_y \frac{1}{\tilde{l}_B^2} (s_x\,\sin{\vartheta} + s_y\,\cos{\vartheta}) \,,
    \end{align}
where $\{\cdot,\cdot\}$ denotes the anticommutator and we defined 
\begin{align}
    \label{eq:V}
    V &= - \frac{1}{2m} \frac{1}{\sqrt{D}} \left( \partial_q \frac{h_t}{h_n} \partial_q +
    \partial_\lambda \frac{h_n}{h_t} \partial_\lambda \right) \frac{1}{\sqrt{D}} \sim O(\frac{1}{m R^2})\,,\\
    \label{eq:kappa_n}
    \kappa_n &= - {\bf e}_t \partial_{s_n} {\bf e}_n = \frac{1}{D} \partial_\lambda h_n \sim O(\frac{1}{R})
\end{align}
together with the dimensionless quantities $\tilde{q}=q/\lambda_{\rm so}$, $\tilde{\kappa}_{n,t}=\kappa_{n,t}\lambda_{\rm so}$, $\tilde{h}_t = h_t/\lambda_{\rm so}$, and $\tilde{V}=V/E_{\rm so}$ (the Lame coefficient $h_q$ is already dimensionless). 
\end{widetext}

To derive the effective surface Hamiltonian for the edge states it is important that one can separate the solution for the normal and tangential part of the edges. The particle on a ring in a double sine potential is a special supersymmetric model in one dimension, occurring here for the special case of a surface in the form of a ring with a large radius. The analysis will be generalized to any smooth surface in Section~\ref{sec:witten_model} where we will see that generic periodic Witten models with supersymmetric properties can be realized. 

Furthermore, we note that the two topological bound states at the inner surface are exactly at zero energy if the radius $\tilde{R}_>$ of the outer surface tends to infinity. In this case, the SUSY is unbroken in an exact sense and two states exactly at zero energy appear in the gap. Since the degeneracy of these two states follows from SUSY, they can not split for any radius $\tilde{R}_<$ of the inner hole. Therefore, even for zero hole radius $\tilde{R}_<=0$, the two center states discussed in Section~\ref{sec:zero_B} for zero magnetic field will remain at zero energy in the presence of a finite magnetic field.

We now want to briefly discuss the stability of the topological bound states against deviations from half-integer flux. For $f\ne 1/2$ degenerate states at opposite positions of the same surface with different eigenvalues of $U_{1/2}$ will get coupled and split. However, under the condition (\ref{eq:small_spread}) of small angular spread, the orbital overlap of the two states is exponentially small and the splitting is negligible. This is in contrast to the two zero energy center states at small hole radius which are unstable against the application of a flux away from half-filling, see the discussion at the end of Section~\ref{sec:zero_B}. 
As for the Corbino disk in Section~\ref{sec:corbino_weak_fields} this is possible when the typical localization length $\xi_n$ of the edge states in normal direction is much smaller than all other characteristic length scales of the system. In units of the spin orbit length this means
\begin{align}
    \label{eq:witten_xi_n_condition}
    \tilde{\xi}_n \ll \tilde{\xi}_t\,,\,\tilde{l}_B\,,\,\tilde{R} \,,
\end{align}
where $\tilde{\xi}_t$ denotes the localization length of the edge states in tangential direction. Below we will show that both $\xi_n$ and $\xi_t$ are given by the same order as for the Corbino disk (with $R$ replaced by the curvature radius)
\begin{align}
    \label{eq:witten_xi_n_t}
    \tilde{\xi}_n \sim O(1) \quad,\quad \tilde{\xi}_t \sim \tilde{l}_B \sqrt{\tilde{R}}\,.
\end{align}
Therefore, the condition (\ref{eq:witten_xi_n_condition}) is fulfilled if $\tilde{l}_B,\tilde{R}\gg 1$. 

Based on the estimates (\ref{eq:order_lame}), (\ref{eq:variation_lame}), (\ref{eq:kappa_t}), (\ref{eq:V}), and (\ref{eq:kappa_n}), we can now determine the leading $O(1)$ and subleading $O(1/\tilde{\xi}_t)$ terms of the Hamiltonian to split the edge state wave functions in a normal and tangential one. Thereby we use in addition the property  
\begin{align}
    \label{eq:partial_wave_functions}
    \frac{1}{h_n}\partial_{\tilde{q}} \sim \frac{1}{\tilde{\xi}_n}  
    \quad,\quad
    \frac{1}{\tilde{h}_t}\partial_{\lambda} \sim \frac{1}{\tilde{\xi}_t} \,,
\end{align}
in case the differential operators act on the edge state wave function (and not on $h_q$ or $\tilde{h}_t$). Together with (\ref{eq:witten_xi_n_t}) we then get in leading order for the effective Hamiltonian in normal direction
\begin{align}
    \nonumber
    \bar{H}_n/E_{\rm so} &= \sigma_x \left[(-\partial_{\tilde{q}} \frac{1}{h_n^2} \partial_{\tilde{q}} - \tilde{\delta})s_z
    - \{\frac{1}{h_n},-i\partial_{\tilde{q}}\} s_y \right] \\
    \label{eq:bar_H_n_ortho}
    &\approx \sigma_x \left[(-\partial^2_{\tilde{s}_n} - \tilde{\delta}) s_z
    + 2i \partial_{\tilde{s}_n}s_y) \right]\,,
\end{align}
where $\tilde{s}_n(\lambda)=\tilde{q}/h_n^\gtrless(\lambda)$, with $h_n^\gtrless(\lambda)=h_n(q_\gtrless,\lambda)$, is the coordinate of the line element in normal direction involving the Lame coefficient $h_n$ projected on the considered outer or inner surface $q=q_\gtrless$. This is justified for all edge states since they are localized close to the surface. We find the same form for the normal Hamiltonian as for the Corbino disc, see Eq.~(\ref{eq:bar_H_n_corbino}), with $\tilde{r}\rightarrow\tilde{s}_n$. Therefore, for each given $\lambda$, we find a zero energy edge state in the gap and the normal part of the edge state wave function is given by \begin{align}
    \label{eq:edge_state_normal_ortho}
    \bar{\psi}_n^\gtrless(\tilde{q},\lambda;s_z) =  
    \frac{h_n^\gtrless(\lambda)^{1/2}}{\sqrt{2}} \left(\begin{array}{c} 1 \\ \pm 1\end{array}\right)_{s_z} 
    \bar{\Phi}_n\left(\frac{|\tilde{q}-\tilde{q}_\gtrless|}{h_n^\gtrless(\lambda)}\right)\,,
\end{align}
where $\bar{\Phi}_n(\tilde{r})$ is given by (\ref{eq:Phi_n_corbino}), and the prefactor accounts for the correct normalization in terms of $\tilde{q}$, see (\ref{eq:ortho_normalization}). 

Projecting the Hamiltonian on the subset defined by the normal part of the edge state wave functions, we find in analogy to the Corbino disk in first order perturbation theory for the effective surface Hamiltonian in subleading order 
\begin{align}
    \label{eq:bar_H_t_ortho}
    \pm \bar{H}_t^\gtrless/E_{\rm so} = \sigma_x \{\frac{1}{\tilde{h}_t^\gtrless},-i\partial_\lambda\} + 
    \sigma_y \frac{1}{\tilde{l}_B^2}\sin{\vartheta}\,.
\end{align}
where $\tilde{h}_t^\gtrless(\lambda) = \tilde{h}_t(q^\gtrless,\lambda)$ is the Lame coefficient projected on the corresponding surface. Thereby we note that the two terms of the Hamiltonian (\ref{eq:bar_H_ortho}) involving
\begin{align}
    \label{eq:special_term_1_ortho}
    &\{\frac{\tilde{\kappa}_n}{h_n},-i\partial_{\tilde{q}}\}\, s_z \sim \frac{1}{\tilde{R}}\,s_z \\
    \label{eq:special_term_2_ortho}
    &\frac{1}{\tilde{l}_B^2} \cos{\vartheta}\,s_y \sim \frac{1}{\tilde{l}_B^2}\,s_y \sim \frac{\tilde{R}}{\tilde{\xi}_t^2}\,s_y
\end{align}
contribute only in second-order perturbation theory since they involve $s_{y,z}$ and do not lead to a direct coupling of the states (\ref{eq:edge_state_normal_ortho}) since they are eigenfunctions of $s_x$ with eigenvalue $\pm 1$ for the outer/inner surface. Thus, the term (\ref{eq:special_term_1_ortho}) contributes in $O(1/\tilde{R}^2)$ and can be neglected, and, analog to the discussion for the Corbino disc, the neglect of the Zeeman term (\ref{eq:special_term_2_ortho}) in $y$-direction requires the additional condition (\ref{eq:additional_condition_weak_fields}).

Writing the eigenfunctions of the surface Hamiltonian as
\begin{align}
    \label{eq:bar_psi_t_ortho}
    \bar{\psi}_t^\gtrless(\lambda,\sigma_z) = \tilde{h}_t^\gtrless(\lambda)^{1/2}\,\hat{\psi}_t^\gtrless(\lambda,\sigma_z)\,,
\end{align}
we find that $\hat{\psi}_t^\gtrless$ is the eigenfunction of 
\begin{align}
    \label{eq:H_t_trafo_ortho}
    \hat{H}_t^\gtrless = 
    (\tilde{h}_t^\gtrless)^{-1/2}\,\bar{H}_t^\gtrless\,(\tilde{h}_t^\gtrless)^{1/2} \,,
\end{align}
with
\begin{align}
    \label{eq:hat_H_t_ortho}
    \pm \hat{H}_t^\gtrless/E_{\rm so} = \sigma_x 2 (-i\partial_{\tilde{s}_t}\}  + 
    \sigma_y \frac{1}{\tilde{l}_B^2}\sin{\vartheta}\,,
\end{align}
where $\partial_{\tilde{s}_t} = \tilde{h}_t^\gtrless(\lambda)^{-1}\partial_\lambda$ is the derivative with respect to the surface line element for $q=q_\gtrless$. Using (\ref{eq:line_elements}), the normalization of $\hat{\psi}_t^\gtrless$ is defined by 
\begin{align}
    \label{eq:hat_psi_normalization}
    \sum_{\sigma_z=\pm 1} \int_0^{\tilde{L}} d\tilde{s}_t |\hat{\psi}_t^\gtrless(\tilde{s}_t,\sigma_z)|^2 = 1\,, 
\end{align}
where $\tilde{L}$ is the circumference of the corresponding surface in units of the spin-orbit length. 

The effective surface Hamiltonian (\ref{eq:hat_H_t_ortho}) is the central result of this section. It has exactly the same form as the result (\ref{eq:bar_H_t_corbino}) for the Corbino disk (where $\tilde{s}_t = \tilde{R}\varphi$) and involves the normal component of the Zeeman term as the mass term. Defining the normal component along the surface in dimensionless units by 
\begin{align}
    \label{eq:Zeeman_normal}
    \tilde{E}_{Z,n}(\tilde{s}_t) = \frac{1}{\tilde{l}_B^2}\sin{\vartheta(\tilde{s}_t)} \,,
\end{align}
we obtain the generic rule that zero energy topological states will appear close to the surface points where the normal component of the Zeeman term changes sign, i.e., for $\tilde{s}_t = \tilde{s}_t^j$ with 
\begin{align}
    \label{eq:position_topological_states}
    \sin{\vartheta_j} = 0 \quad,\quad \vartheta_j = \vartheta(\tilde{s}_t^j) \,.    
\end{align}
Expanding around such a point we obtain for $\vartheta_j=0,\pi$
\begin{align}
    \label{eq:Zeeman_normal_expansion}
    \tilde{E}_{Z,n}(\tilde{s}_t) \approx \pm\frac{\tilde{\kappa}_t^j}{\tilde{l}_B^2}(\tilde{s}_t - \tilde{s}_t^j) 
    \sim \frac{\tilde{\xi}_t}{\tilde{R}\,\tilde{l}_B^2}\,,
\end{align}
with $\tilde{\kappa}_t^j=\tilde{\kappa}_t(\tilde{s}_t^j)$, where we used (\ref{eq:theta_kappa}) and the estimate $\tilde{s}_t-\tilde{s}_t^j\sim \tilde{\xi}_t$, together with the definition $\tilde{\kappa}_t = \kappa_t \lambda_{\rm so}\sim 1/\tilde{R}$. Comparing the order of magnitude of this term with the first term 
$\sim \partial_{\tilde{s}_t}\sim 1/\tilde{\xi}_t$ of the surface Hamiltonian (\ref{eq:hat_H_t_ortho}), we find precisely the result (\ref{eq:witten_xi_n_t}) for $\tilde{\xi}_t\sim \tilde{l}_B \sqrt{\tilde{R}}$. Furthermore, depending on the sign of 
\begin{align}
    \label{eq:p_j}
    p_j = {\rm sign}(\cos{\vartheta_j})\,{\rm sign}(\tilde{\kappa}_t^j)\,,
\end{align}
the mass term changes from minus to plus or vice versa when moving around the surface, giving rise to the chirality $s=p_j=\pm 1$ of the topological states, respectively.  

Squaring the Hamiltonian we obtain the generic class of periodic Witten models
\begin{align}
    \label{eq:witten_model}
    {\tilde{\cal H}}_W = (\hat{H}^\gtrless_t/E_{\rm so})^2 = - 4 \partial_{\tilde{s}_t}^2 + \tilde{V}_W^{-\sigma_z}(\tilde{s}_t)\,,
\end{align}
with the Witten potentials given by
\begin{align}
    \tilde{V}_W^{\pm}(\tilde{s}_t) = \tilde{E}_{Z,n}(\tilde{s}_t)^2 \mp 2 \tilde{E}_{Z,n}^\prime(\tilde{s}_t)\,,  
\end{align}
where $\tilde{E}_{Z,n}^\prime(\tilde{s}_t)=(d/d\tilde{s}_t)\tilde{E}_{Z,n}(\tilde{s}_t)$ denotes the derivative. We note that the study of ${\tilde{\cal H}}_W$ is sufficient to calculate the spectrum and all eigenstates of the surface Hamiltonian. This is due to chiral symmetry $\bar{S}=-\sigma_z$ and the fact that periodic boundary conditions under $\tilde{s}_t\rightarrow \tilde{s}_t + \tilde{L}$ are respected by the surface Hamiltonian. The Witten model and its relation to the surface Hamiltonian will be discussed in all detail in the next two subsections.

\subsection{Supersymmetry for the periodic Witten model}
\label{sec:witten_SUSY}

The Witten model plays a very fundamental role in the study of SUSY models \cite{junker_book_19,cooper_etal_book_01,bagchi_book_01}. The two spectra of 
\begin{align}
{\tilde{\cal H}}_W^\pm\equiv {\tilde{\cal H}}_W|_{\sigma_z=\pm} = - 4 \partial_{\tilde{s}_t}^2 + \tilde{V}_W^\pm(\tilde{s}_t) 
\end{align}
for the two chiral sectors $\bar{S}=-\sigma_z=\pm 1$ of ${\tilde{\cal H}}_W$ are exactly the same. This follows from chiral symmetry of $\hat{H}_t^\gtrless$ and the definition ${\tilde{\cal H}}_W=(\hat{H}_t^\gtrless/E_{\rm so})^2$. For each eigenstate $|\psi\rangle$ of $\hat{H}_t^\gtrless$ it follows that $\sigma_z|\psi\rangle$ is also an eigenstate with a different sign for the energy. As a consequence, the two states $|\psi\rangle \pm \sigma_z|\psi\rangle$ are eigenstates of ${\tilde{\cal H}}_W$ with the same energy belonging to two different chiral sectors. Since all eigenstates of ${\tilde{\cal H}}_W$ can be constructed in this way, the two spectra of ${\tilde{\cal H}}_W^\pm$ must be exactly the same. However, this twofold degeneracy between the spectra of the two chiral sectors is a rather trivial degeneracy for the two partner potentials $\tilde{V}_W^\pm$ and is {\it not} related to any nontrivial SUSY structure of the spectrum. 

The nontrivial SUSY properties emerge in each chiral sector separately and are associated with additional symmetries present at half-integer flux $f=1/2$ and a mirror-symmetric surface. Using (\ref{eq:theta_mirror_symmetry}), we find for a mirror-symmetric surface the following properties of the normal component $\tilde{E}_{Z,n}(\tilde{s}_t)$ of the Zeeman field along the surface
\begin{align}
    \label{eq:Zeeman_normal_reflection_symmetry}
    \tilde{E}_{Z,n}(\tilde{s}_t) &= - \tilde{E}_{Z,n}(\tilde{s}_t+\frac{\tilde{L}}{2}) \,,\\
    \label{eq:Zeeman_normal_varphi_symmetry}
    \tilde{E}_{Z,n}(\tilde{s}_t) &= - \tilde{E}_{Z,n}(\tilde{L}-\tilde{s}_t) = - \tilde{E}_{Z,n}(-\tilde{s}_t)\,,
\end{align}
where we used periodic boundary conditions $\tilde{E}_{Z,n}(\tilde{s}_t)=\tilde{E}_{Z,n}(\tilde{s}_t+\tilde{L})$ in the last equality. Using the operators $P_{\bf x}$ and $P_{\varphi}$ changing the sign of ${\bf x}$ and $\varphi$, respectively, which act within the space of a mirror-symmetric surface, we can write the symmetry properties equivalently as
\begin{align}
    \label{eq:symmetries_W}
    P_{\bf x} \,\tilde{E}_{Z,n}\, P_{\bf x} = - \tilde{E}_{Z,n} \quad,\quad P_\varphi \,\tilde{E}_{Z,n} \,P_\varphi = - \tilde{E}_{Z,n} \,.
\end{align}
Defining the supercharge operator $Q_\pm$ by
\begin{align}
    \label{eq:Q_W}
    Q_\pm = P_{\bf x} (-2i\partial_{\tilde{s}_t} \mp i \tilde{E}_{Z,n}(\tilde{s}_t)) = Q_\pm^\dagger \,,
\end{align}
and using the symmetries (\ref{eq:symmetries_W}) together with $P_{\bf x}^2=P_{\bf \varphi}^2=1$, we obtain straightforwardly the $n=1$ SUSY representation (\ref{eq:SUSY_n=1}) for each chiral sector with $K\equiv P_\varphi$
\begin{align}
    \label{eq:SUSY_n=1_W}
    {\tilde{\cal H}}_{\rm W}^\pm = \left(Q_\pm\right)^2 \quad,\quad Q_\pm P_\varphi = - P_\varphi Q_\pm \,.
\end{align}

The SUSY of both ${\tilde{\cal H}}_W^\pm$  is unbroken, i.e., a zero energy state exists in each chiral sector. This follows directly from solving $Q_\pm|\psi_{\rm W}^{(0),\pm}\rangle=0$ with the result
\begin{align}
    \label{eq:zero_energy_state_W}
    \psi_{\rm W}^{(0),\pm}(\tilde{s}_t) = \frac{1}{\sqrt{N_W^\pm}}\,e^{\mp F(\tilde{s}_t)} \,, 
\end{align}
where we defined the function
\begin{align}
    \label{eq:F_witten}
    F(\tilde{s}_t) = \frac{1}{2}\int_{-\tilde{L}/4}^{\tilde{s}_t} d\tilde{s}_t' \,\tilde{E}_{Z,n}(\tilde{s}_t')\,.
\end{align}
Here, $N_W^\pm$ is a normalization factor such that $\int_0^{\tilde{L}} d\tilde{s}_t \left[\psi_{\rm W}^\pm(\tilde{s}_t)\right]^2 = 1$. The reference point $-\tilde{L}/4$ for the integration in (\ref{eq:F_witten}) has been chosen such that the symmetries (\ref{eq:Zeeman_normal_reflection_symmetry}) and (\ref{eq:Zeeman_normal_varphi_symmetry}) for $\tilde{E}_{Z,n}(\tilde{s}_t)$ lead to the following symmetries for the function $F(\tilde{s}_t)$ 
\begin{align}
    \label{eq:F_reflection_symmetry}
    F(\tilde{s}_t) &= - F(\tilde{s}_t+\frac{\tilde{L}}{2}) \,,\\
    \label{eq:F_varphi_symmetry}
    F(\tilde{s}_t) &= F(\tilde{L}-\tilde{s}_t) = F(-\tilde{s}_t)\,,
\end{align}
or 
\begin{align}
    \label{eq:symmetries_F}
    P_{\bf x} \,F\, P_{\bf x} = - F \quad,\quad P_\varphi \,F \,P_\varphi = F \,.
\end{align}
As a consequence, the function $F(\tilde{s}_t)$ is symmetric around the $x$-axis (i.e., under $\varphi\rightarrow -\varphi$) and antisymmetric around the $y$-axis (i.e., under $\varphi\rightarrow\pi-\varphi$). For example, for the special case of the outer or inner surface of a Corbino disc, where $\tilde{s}_t=\tilde{R}_\gtrless\varphi$, $\tilde{L}=2\pi\tilde{R}_\gtrless$ and $\tilde{E}_{Z,n}(\tilde{s}_t)=(1/\tilde{l}_B^2) \sin{\varphi}$, we get $F(\tilde{s}_t)=-\tilde{R}_\gtrless/(2\tilde{l}_B^2) \cos{\varphi}=-1/(2\Delta\varphi^2)\cos{\varphi}$, consistent with (\ref{eq:f}). 

We note that the existence of the zero energy states for the two periodic Witten models does not necessarily depend on the mirror symmetry of the surface. Only the exact twofold degeneracy of all states with non-zero energy is related to SUSY present only for a mirror symmetric surface. This shows that the original model will always have states close to zero energy at half-integer flux for arbitrary smooth surfaces, provided that the conditions (\ref{eq:small_curvature_condition}) and (\ref{eq:witten_conditions}) for the derivation of the surface Hamiltonian are fulfilled. This is even true away from half-integer flux for strong localization $\xi_t \ll R_\gtrless$, where the deviation of the flux from half-integer value will only change the boundary conditions of the wave functions which is not very important for localized states with small orbital overlap. However, for a mirror-symmetric surface at half-integer flux, the two states are exactly at zero energy even when they have a strong orbital overlap for $\xi_t\sim R_\gtrless$.

\subsection{The low energy spectrum of the surface Hamiltonian}
\label{sec:spectrum_surface_H}

In this section we will discuss the low energy spectrum and the eigenstates of the surface Hamiltonian $\hat{H}_t^\gtrless$ for the outer/inner surface. We start with the zero energy solutions for any size of $\xi_t$ based on the zero energy solutions (\ref{eq:zero_energy_state_W}) of the periodic Witten models, and continue with a semiclassical calculation of the strongly localized bound states with $\xi_t \ll R_\gtrless$, in close analogy to the case of a Corbino disk discussed in Section~\ref{sec:corbino_weak_fields}. 

The zero energy solutions (\ref{eq:zero_energy_state_W}) of ${\tilde{\cal H}}_W^\eta$ give rise to two zero energy states of the surface Hamiltonian $\hat{H}_t^\gtrless$ for chirality $s=-\sigma_z$
\begin{align}
    \label{eq:zero_energy_state_surface_H_s=1}
    \hat{\psi}^\gtrless_{t,s=1}(\tilde{s}_t;\sigma_z) &= \frac{1}{\sqrt{2}}\left(\begin{array}{c} 0 \\ 1 \end{array}\right)_{\sigma_z}
    \psi_W^{(0),+}(\tilde{s}_t) \,,\\
    \label{eq:zero_energy_state_surface_H_s=-1}
    \hat{\psi}^\gtrless_{t,s=-1}(\tilde{s}_t;\sigma_z) &= \frac{1}{\sqrt{2}}\left(\begin{array}{c} 1 \\ 0 \end{array}\right)_{\sigma_z}
    \psi_W^{(0),-}(\tilde{s}_t) \,,
\end{align}
where the dependence of the right hand side of these equations on the outer/inner surface is hidden in the function $\tilde{E}_{Z,n}(\tilde{s}_t)$ defined by (\ref{eq:Zeeman_normal}), which enters into the definition of $\psi_W^{(0),\pm}$ via (\ref{eq:zero_energy_state_W}) and (\ref{eq:F_witten}). For a mirror symmetric surface we find due to (\ref{eq:zero_energy_state_W}) and (\ref{eq:symmetries_F}) that
\begin{align}
    \label{eq:symmetries_hat_psi_t}
    P_\varphi |\hat{\psi}^\gtrless_{t,s}\rangle = |\hat{\psi}^\gtrless_{t,s}\rangle \quad,\quad
    \sigma_z |\hat{\psi}^\gtrless_{t,s}\rangle = -s |\hat{\psi}^\gtrless_{t,s}\rangle \,.
\end{align}

Multiplying these solutions with the normal part (\ref{eq:edge_state_normal_ortho}) and using (\ref{eq:bar_psi_t_ortho}) we get for the total wave function of the zero energy states of $\bar{H}_{1/2}$ with chirality $s=\pm 1$ the result
\begin{align}
    \nonumber
    \bar{\psi}_s^\gtrless(\tilde{q},\lambda;\sigma_z,s_z) & \\
    \label{eq:zero_energy_state_bar_H_ortho}
    & \hspace{-1cm} 
    = \tilde{h}_t^\gtrless(\lambda)^{1/2}\,\bar{\psi}_n^\gtrless(\tilde{q},\lambda;s_z)
    \,\hat{\psi}_{t,s}^\gtrless(\tilde{s}_t(\lambda);\sigma_z) \,.   
\end{align}
This gives two zero energy states for each surface with different values for the chiral symmetry. For a mirror symmetric surface, the eigenvalue $u$ of the SUSY operator $\bar{U}_{1/2}=P_\varphi\sigma_z s_x$ is automatically fixed for given chirality. Since $h_n(\lambda)$ and $\tilde{h}_t(\lambda)$ are symmetric under $P_\varphi$ for a mirror symmetric surface, we get from (\ref{eq:edge_state_normal_ortho}) and (\ref{eq:symmetries_hat_psi_t}) that 
\begin{align}
    \label{eq:SUSY_zero_energy_state_bar_H_ortho}
    \bar{U}_{1/2} |\bar{\psi}_s^\gtrless\rangle = \mp s |\bar{\psi}_s^\gtrless\rangle\,,
\end{align}
i.e., $u=-s$ for the outer surface and $u=s$ for the inner surface, in analogy to the result for the Corbino disc, see Fig.~\ref{fig:states_corbino}. 

To find all eigenstates of the surface Hamiltonian with non-zero energy, we first note that it is sufficient to study the eigenstates of the Witten Hamiltonian. This is due to chiral symmetry and the fact that the surface Hamiltonian respects periodic boundary conditions. To see this we start from any normalized eigenstate $\psi_W^+(\tilde{s}_t)$ of ${\tilde{\cal H}}_W^+$ with positive eigenvalue $\tilde{\epsilon}^2 > 0$
\begin{align}
    \label{eq:eigenstate_H_W_+}
    {\tilde{\cal H}}_W^+ |\psi_W^+\rangle &= \tilde{\epsilon}^2 |\psi_W^+\rangle \,,\\
    \label{eq:normalization_psi_W_+}
    \langle\psi_W^+|\psi_W^+\rangle &= \int_0^{\tilde{L}} d\tilde{s}_t |\psi_W^+(\tilde{s}_t)|^2 = 1\,.
\end{align}
Writing the surface Hamiltonian (\ref{eq:hat_H_t_ortho}) in the $\sigma_z$-basis as 
\begin{align}
    \label{eq:surface_H_chiral_basis}
    \pm \hat{H}^\gtrless_t/E_{\rm so} = \left(\begin{array}{cc} 0 & \Gamma \\ \Gamma^\dagger & 0 \end{array}\right)\,, 
\end{align}   
with 
\begin{align}
    \label{eq:Gamma_surface_H}
    \Gamma = -2i\partial_{\tilde{s}_t} - i \tilde{E}_{Z,n}(\tilde{s}_t)\,,
\end{align}
we find ${\tilde{\cal H}}_W^+=\Gamma^\dagger\Gamma$ and ${\tilde{\cal H}}_W^-=\Gamma\Gamma^\dagger$ (note that the superindex of ${\tilde{\cal H}}_W^s$ refers to $s=-\sigma_z$). It is then straightforward to find two eigenstates of the surface Hamiltonian with opposite energy
\begin{align}
    \label{eq:eigenstate_surface_H}
    \pm (\hat{H}^\gtrless_t/E_{\rm so}) |\hat{\psi}_t^{\gtrless,\eta}\rangle = 
    \eta |\tilde{\epsilon}|\,|\hat{\psi}_t^{\gtrless,\eta}\rangle \,,
\end{align}
with $\eta=\pm 1$ and
\begin{align}
    \label{eq:eigenstate_surface_H_explicit}
    \hat{\psi}_t^{\gtrless,\eta}(\tilde{s}_t;\sigma_z) &= \frac{1}{\sqrt{2}}
    \left(\begin{array}{c}\eta\,\psi_W^-(\tilde{s}_t) \\ \psi_W^+(\tilde{s}_t) \end{array}\right)_{\sigma_z} \,,\\
    \label{eq:eigenstate_H_W_-}
    \psi_W^- &= \frac{1}{|\tilde{\epsilon}|}\Gamma \psi_W^+ \,.
\end{align}
Here, $\psi_W^\pm$ are by construction normalized eigenstates of ${\tilde{\cal H}}_W^\pm$, respectively, with the same eigenvalue $\tilde{\epsilon}^2$, and both respecting periodic boundary conditions $\psi_W^\pm(\tilde{s}_t)=\psi_W^\pm(\tilde{s}_t+\tilde{L})$.

The construction (\ref{eq:eigenstate_surface_H_explicit}) of the eigenstates of the surface Hamiltonian in terms of the eigenstates of the Witten Hamiltonian ${\tilde{\cal H}}_W^+$ is possible for any smooth surface, even if it is not mirror symmetric. For the special case of a mirror symmetric surface, where SUSY holds and $\bar{U}_{1/2}=P_\varphi \sigma_z s_x$ is an exact symmetry of the full Hamiltonian $\bar{H}_{1/2}$, we can choose the eigenstates of the surface Hamiltonian as eigenfunctions of $P_\varphi\sigma_z$ to fix the eigenvalue $u$ of $\bar{U}_{1/2}$ (note that the normal part of the wave function has $s_x=\pm$ for the outer/inner surface). From the symmetry (\ref{eq:symmetries_W}) we find that the Witten potential and the Witten Hamiltonian have the symmetry
\begin{align}
    \label{eq:H_witten_symmetry_P_varphi}
    \tilde{V}_W^\pm(\tilde{s}_t) = \tilde{V}_W^\pm(-\tilde{s}_t)
    \quad\Rightarrow\quad P_\varphi {\tilde{\cal H}}_W^\pm P_\varphi = {\tilde{\cal H}}_W^\pm\,.
\end{align}
Therefore, we can choose the eigenstates $\psi_W^+$ of ${\tilde{\cal H}}_W^+$ as eigenstates of $P_\varphi$ with eigenvalue $\eta_W^\varphi=\pm 1$
\begin{align}
    \label{eq:psi_W_+_P_varphi}
    P_\varphi \,|\psi_W^+\rangle = \eta_W^\varphi |\psi_W^+\rangle \,.
\end{align}
Using the symmetry 
\begin{align}
    \label{eq:Gamma_symmetry_P_varphi}
    P_\varphi \Gamma P_\varphi = - \Gamma \,,
\end{align}
we get a sign change for the eigenvalue of $P_\varphi$ for $\psi_W^-=|\tilde{\epsilon}|^{-1}\Gamma\psi_W^+$
\begin{align}
    \label{eq:psi_W_-_P_varphi}
    P_\varphi \,|\psi_W^-\rangle = -\eta_W^\varphi |\psi_W^-\rangle \,,
\end{align}
and find from (\ref{eq:eigenstate_surface_H_explicit}) that $\hat{\psi}_t^{\gtrless,\eta}$ is an eigenstate of $P_\varphi\sigma_z$ with eigenvalue $-\eta_W^\varphi$
\begin{align}
    \label{eq:psi_t_P_varphi_sigma_z}
    P_\varphi \sigma_z \,|\hat{\psi}_t^{\gtrless,\eta}\rangle = -\eta_W^\varphi |\hat{\psi}_t^{\gtrless,\eta}\rangle \,. 
\end{align}
In this way all eigenstates of $\bar{H}_{1/2}$ can be constructed from the eigenstates of the Witten Hamiltonian ${\tilde{\cal H}}_W^+$, which are at the same time eigenstates of the SUSY operator with eigenvalue given by
\begin{align}
    \label{eq:SUSY_eigenvalue_ortho}
    u = - \eta_W^\varphi s_x = \mp \eta_W^\varphi \,,
\end{align}
where we used $s_x=\pm$ for the outer/inner surface from the normal part of the eigenstates. 

In summary, for a mirror symmetric surface, we have found that the angular part of all eigenstates of the Hamiltonian $\bar{H}_{1/2}$ can be constructed as a combination of a symmetric state of ${\tilde{\cal H}}_W^+$ and an antisymmetric one of ${\tilde{\cal H}}_W^-$ with SUSY eigenvalue $u=\mp$ for the outer/inner surface, or vice versa with $u=\pm$. Furthermore, we note that two degenerate eigenstates of the Hamiltonian constructed in this way transform into each other via the inversion symmetry $\bar{\Pi}=-P_{\bf x}\sigma_x$ and have a different sign for the SUSY eigenvalue $u$. This follows since $\bar{\Pi}$ commutes with $\bar{H}_{1/2}$, whereas SUSY anticommutes with $\bar{H}_{1/2}$. On the level of the surface Hamiltonian this follows equivalenty from the symmetries
\begin{align}
    \label{eq:surface_H_inversion_symmetry}
    P_{\bf x} \sigma_x \hat{H}_t^\gtrless P_{\bf x} \sigma_x &= \hat{H}_t^\gtrless \,,\\
    \label{eq:surface_H_SUSY_symmetry}
    P_\varphi \sigma_z \hat{H}_t^\gtrless P_\varphi \sigma_z &= \hat{H}_t^\gtrless \,,\\
    (P_{\bf x} \sigma_x)(P_\varphi \sigma_z) &= - (P_\varphi \sigma_z)(P_{\bf x} \sigma_x)\,.
\end{align}

\begin{figure*}[t!]
    \centering
    \includegraphics[width =0.48\textwidth]{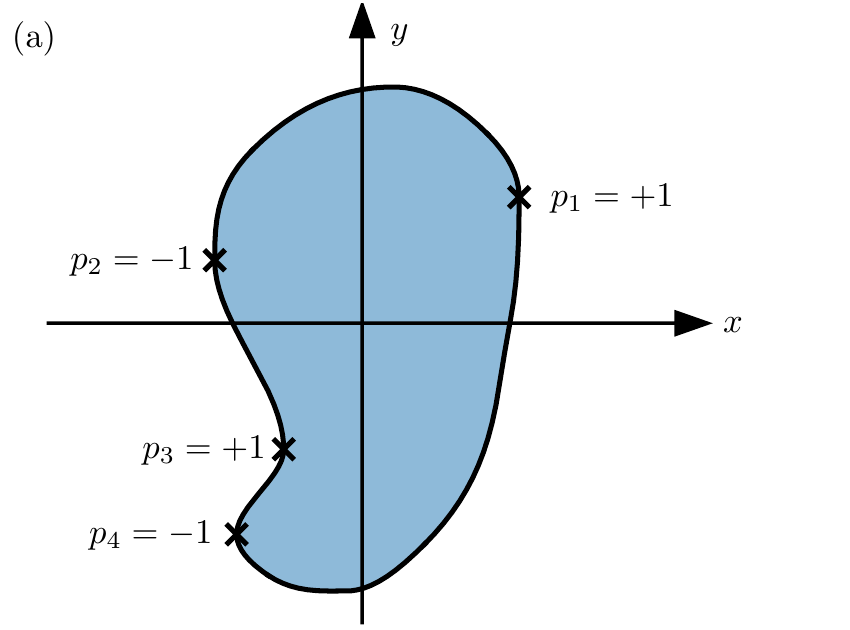}
    \includegraphics[width =0.50\textwidth]{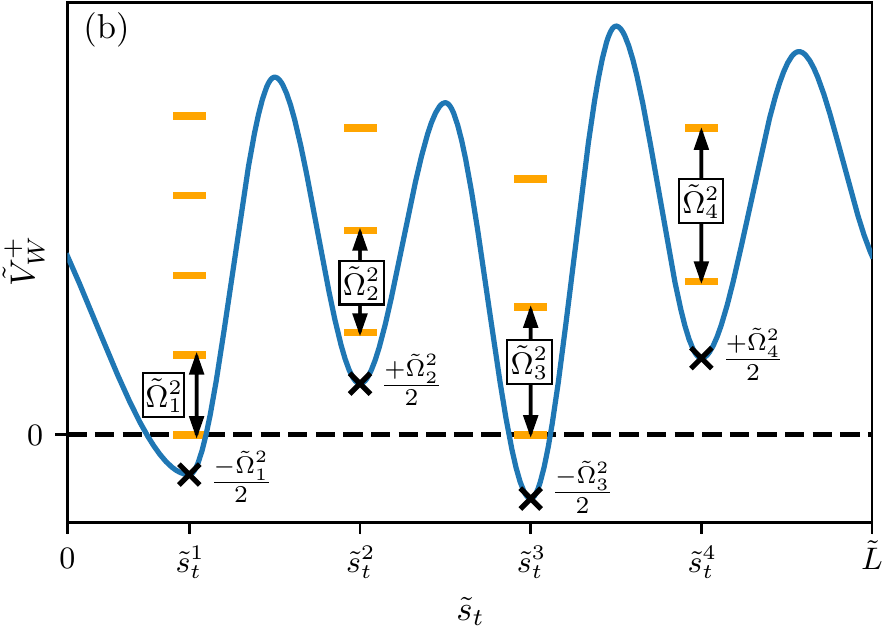}
	  \caption{(a) A generic surface which is not mirror symmetric with four points where $\vartheta=0,\pi$ and four different frequencies $\tilde{\Omega}_j = 2\sqrt{|\tilde{\kappa}_t^j|}/\tilde{l}_B$. The signs $p_j = ({\rm sign}\cos{\vartheta_j})\,({\rm sign}\tilde{\kappa}_t^j)$ are alternating when moving along the surface. This happens since (1) $\vartheta=0,\pi$ for positive/negative $x$-coordinate, and (2) the sign of the curvature $\tilde{\kappa}_t$ changes (remains the same) when the sign of the $x$-coordinate remains the same (changes) when moving from one point to the next. (b) A sketch of the corresponding Witten potential $\tilde{V}_W^+$ consisting of four harmonic oscillator potentials with potential minima at $-p_j\tilde{\Omega}_j^2/2$. In the absence of accidental degeneracies due to commensurabilities, this gives rise to a set of non-degenerate and clearly separated eigenvalues. An exception are the two zero eigenvalues which stay close to each other since the exponentially small hybridization will split them into one eigenvalue exactly at zero and another one at a very small positive value.  
	  } 
    \label{fig:surface_witten_generic}
\end{figure*}

We now turn to the explicit calculation of the localized bound states of the two Witten models ${\tilde{\cal H}}_W^\pm$ in semiclassical approximation for strong localization $\tilde{\xi}_t = \tilde{l}_B\sqrt{\tilde{R}}\ll\tilde{R}$, in analogy to the treatment described in Section~\ref{sec:corbino_weak_fields} for the Corbino disc. The existence of the unique zero energy states $\psi_W^{(0),\pm}$ for the chiral sectors $s=\pm$ of the Witten Hamiltonians ${\tilde{\cal H}}_W^\pm$ reflects the unbroken SUSY properties of the two chiral sectors of the squared Hamiltonian as described in Section~\ref{sec:SUSY}. However, for a generic smooth surface characterized by the average curvature $\tilde{\kappa}_t\sim 1/\tilde{R}\ll 1$, an arbitrary number of points with $\sin{\vartheta(\tilde{s}_{t}^j)}=0$,  $j=1,\dots,N_Z$, can occur. In the regime of strong localization, there will be $N_Z$ states exponentially close to zero energy, each of them localized at $\tilde{s}_t^j$ with tangential spread $\sim\tilde{\xi}_t$. The two states exactly at zero energy are a superposition of all these states and are given by (\ref{eq:zero_energy_state_bar_H_ortho}). In addition, there will be $N_Z-1$ other eigenstates with an exponentially small energy $\sim e^{-\tilde{R}/\tilde{\xi}_t}$ and it is in general quite difficult to calculate them exactly. Therefore, in the following we will consider the region $\tilde{s}_t\sim \tilde{s}_t^j+O(\tilde{\xi}_t)$, and study the spectrum via the harmonic oscillator eigenstates in the minima of the Witten potential by expanding the normal component $\tilde{E}_{Z,n}(\tilde{s}_t)$ of the Zeeman field up to linear order in $\tilde{s}_t-\tilde{s}_t^j$. According to (\ref{eq:Zeeman_normal_expansion}) we get
\begin{align}
    \label{eq:Zeeman_normal_expansion_Omega}
    \tilde{E}_{Z,n}(\tilde{s}_t) &\approx p_j \frac{\tilde{\Omega}_j^2}{4}(\tilde{s}_t - \tilde{s}_t^j) \,,\\
    \label{eq:V_W_expansion_Omega}
    \tilde{V}_W^\pm(\tilde{s}_t) &\approx \frac{\tilde{\Omega}_j^4}{16}(\tilde{s}_t - \tilde{s}_t^j)^2 \mp p_j \frac{\tilde{\Omega}_j^2}{2}\,, 
\end{align}
where the sign factor $p_j$ has been defined in (\ref{eq:p_j}) and $\tilde{\Omega}_j$ denotes the Witten frequency at $\tilde{s}_t=\tilde{s}_t^j$
\begin{align}
    \label{eq:Omega_j}
    \tilde{\Omega}_j &= \frac{2|\tilde{\kappa}_t^j|^{1/2}}{\tilde{l}_B} = \frac{2}{\tilde{\xi}_t^j}\,, 
\end{align}
with
\begin{align}
    \label{eq:xi_t_j}
    \tilde{\xi}_t^j = \frac{\tilde{l}_B}{|\tilde{\kappa}_t^j|^{1/2}} \sim \sqrt{\tilde{R}}\,\,\tilde{l}_B\,.
\end{align}
Defining the annihilation and creation operators by
\begin{align}
    \label{eq:annihilation_witten}
    a_j &= \frac{2}{\tilde{\Omega}_j}\partial_{\tilde{s}_t} + \frac{\tilde{\Omega}_j}{4}(\tilde{s}_t-\tilde{s}_t^j) \,,\\
    \label{eq:creation_witten}
    a_j^\dagger &= -\frac{2}{\tilde{\Omega}_j}\partial_{\tilde{s}_t} + \frac{\tilde{\Omega}_j}{4}(\tilde{s}_t-\tilde{s}_t^j) \,,
\end{align}
we find for the Witten Hamiltonian ${\tilde{\cal H}}_W^{\pm,j}$ close to $\tilde{s}_t\approx\tilde{s}_t^j$ the result
\begin{align}
    \label{eq:witten_H_expanded}
    {\tilde{\cal H}}_W^{\pm,j} \approx \tilde{\Omega}_j^2 a_j^\dagger a_j + \tilde{\Omega}_j^2
    \begin{cases} 0 & {\rm for}\quad p_j = \pm 1 \\ 1 & {\rm for}\quad p_j = \mp 1\end{cases}\,,
\end{align}
and the surface Hamiltonian (\ref{eq:hat_H_t_ortho}) close to $\tilde{s}_t\approx\tilde{s}_t^j$ can be written for $p_j=1$ in the form
\begin{align}
    \label{eq:surface_H_expanded_p=1}
    \hat{H}_t^{\gtrless,j}|_{p_j=1} \approx \pm \Omega_j
    \left(\begin{array}{cc} 0 & -i a_j \\ i a_j^\dagger & 0 \end{array}\right)\,,
\end{align}
and, for $p_j=-1$, we get
\begin{align}
    \label{eq:surface_H_expanded_p=-1}
    \hat{H}_t^{\gtrless,j}|_{p_j=-1} &\approx \pm \Omega_j
    \left(\begin{array}{cc} 0 & i a_j^\dagger \\ -i a_j & 0 \end{array}\right) \\
    \label{eq:surface_H_expanded_relation}
    &= \sigma_x (\hat{H}_t^{\gtrless,j}|_{p_j=1}) \sigma_x\,.
\end{align}
Analog to the Corbino disc, we then get for the eigenfunctions and eigenvalues of ${\tilde{\cal H}}_W^{\pm,j}$ in the semiclassical approximation the result
\begin{align}
    \label{eq:eigenstates_H_W_harmonic_approx}
    g_n^j(\tilde{s}_t) &= (\tilde{\xi}_t^j)^{-1/2} \, f_n\left((\tilde{s}_t-\tilde{s}_t^j)/\tilde{\xi}_t^j\right) \,,\\
    \label{eq:eigenvalues_H_W_harmonic_approx}
    \tilde{E}^{\pm,j}_n &= n \, \tilde{\Omega}_j^2 + \frac{1}{2}(1\mp p_j)\tilde{\Omega}_j^2\,,
\end{align}
where $f_n(\varphi)$ has been defined in (\ref{eq:eigenstates_DS_S=1_phi_0}), and $n=0,1,\dots$. The eigenstates of the surface Hamiltonian can then be written down via the explicit construction (\ref{eq:eigenstate_surface_H_explicit}) or in the same way as for the Corbino disc, see (\ref{eq:zero_state_H_t_phi_0}) and (\ref{eq:nonzero_states_H_t_phi_0}). 

The spectrum of ${\tilde{\cal H}}_W^+$ within the surface gap is then approximately given by $n\,\tilde{\Omega}_j^2$ for $p_j=1$ and by $(n+1)\,\tilde{\Omega}_j^2$ for $p_j=-1$, with $n=0,1,2,\dots$. According to the above discussion, each non-zero eigenvalue $n\,\tilde{\Omega}^2$ of ${\tilde{\cal H}}_W^+$ will lead to two energy eigenvalues $\pm\sqrt{n}\,\tilde{\Omega}_j$ for the surface Hamiltonian. This is quite analog as for the Corbino disk but the essential difference is that many different frequencies $\tilde{\Omega}_j$ can occur, see below for examples and the discussion of the qualitative form of the spectrum for specific surfaces.  

Most importantly, we find from (\ref{eq:eigenvalues_H_W_harmonic_approx}) that eigenstates of ${\tilde{\cal H}}_W^\pm$ with eigenvalue exponentially close to zero are superpositions of all states $|g_0^j\rangle$ localized at the points with $p_j=\pm 1$, i.e., for a given chiral sector $s=\pm 1$, the sign of $p_j$ is fixed to $p_j=s$. According to (\ref{eq:p_j}) this means that the signs of $\tilde{\kappa}_t^j$ and $\cos{\vartheta_j}$ for all these points must be either the same (for $s=+1$) or opposite (for $s=-1$). To determine the number of such points for a given chiral sector in the case of a mirror symmetric surface we use the properties (\ref{eq:theta_mirror_symmetry}) and (\ref{eq:theta_kappa}), and find that $\tilde{\kappa}_t$ and $\cos{\vartheta}$ transform in the following way under a sign change of $\varphi$ or ${\bf x}$
\begin{align}
    \label{eq:kappa_trafo}
    & \tilde{\kappa}_t(\tilde{s}_t) = \tilde{\kappa}_t(-\tilde{s}_t) = \tilde{\kappa}_t(\tilde{s}_t+\tilde{L}/2)\,,\\ 
    \label{eq:cos_vartheta_trafo}
    & \cos{\vartheta(\tilde{s}_t)} = \cos{\vartheta(-\tilde{s}_t)} = -\cos{\vartheta(\tilde{s}_t+\tilde{L}/2)}\,.
\end{align}
As a consequence, we get the following transformation of the sign factors $p_j=p(\tilde{s}_t^j)$ 
\begin{align}
    \label{eq:p_trafo}
    p(\tilde{s}_t^j) = p(-\tilde{s}_t^j) = - p(\tilde{s}_t^j + \tilde{L}/2) \,.
\end{align}
This means that the number of points with the same sign of $p_j$ and $\tilde{s}_t^j\ne 0,\tilde{L}/2$ is even. In addition, the two points $\tilde{s}_t=0,\tilde{L}/2$ belong to the set $\{\tilde{s}_t^j\}_j$ with different sign of $p_j$, since $\cos{\vartheta(0)}=-\cos{\vartheta(\tilde{L}/2)}=1$ and $\tilde{\kappa}_t(0)=\tilde{\kappa}_t(\tilde{L}/2)$ due to (\ref{eq:theta_mirror_symmetry}), (\ref{eq:kappa_trafo}), and (\ref{eq:cos_vartheta_trafo}). As a consequence, the number of points $\tilde{s}_t^j$ with a definite sign of $p_j$ must be odd for a mirror symmetric surface, i.e., the total number $N_Z$ of points with $\sin{\vartheta_j}=0$ is even and $N_Z/2$ is odd. This is consistent with the form of the unbroken SUSY spectrum for ${\tilde{\cal H}}_W^{s}$ since an odd number of states in the presence of SUSY can only split into one single state with zero eigenvalue and a set of twofold degenerate states with positive eigenvalues. 

Moreover, for a mirror symmetric surface, we can also predict the qualitative form of the spectrum for the bound states in the surface gap of the Witten model ${\tilde{\cal H}}_W^s$ (for fixed chirality $s=\pm 1$) significantly away from zero energy. Let us consider $M$ different frequencies $\tilde{\Omega}_k$, with $k=1,\dots,M$, with each frequency $\tilde{\Omega}_k$ occurring $m_k$ times at the points $\vartheta=0,\pi$ along the surface, such that the total number of points is decomposed as 
\begin{align}
    \label{eq:N_Z_decomposition}
    N_Z = m_1 + m_2 + \dots + m_M \,.
\end{align}
For a mirror symmetric surface, each number $m_k$ must be even due to the above discussion, with $m_k/2$ points referring to  $p_k=+1$ and $m_k/2$ ones to $p_k=-1$. Therefore, we get in the semiclassical approximation from (\ref{eq:eigenvalues_H_W_harmonic_approx}) that  $m_k/2$ harmonic oscillator potentials have eigenvalues 
$n\,\tilde{\Omega}_k^2$, with $n=0,1,\dots$, and $m_k/2$ ones have eigenvalues $n\,\tilde{\Omega}_k^2$, with $n=1,\dots$. This gives $m_k/2$  eigenvalues lying close to zero (leading in total to $N_Z/2$ states with eigenvalue close to zero in each chiral sector, see above), and $m_k$ eigenvalues lying close to $n \,\tilde{\Omega}_k^2$ for each $n=1,2,\dots$. Therefore, disregarding accidental degeneracies from commensurabilities, the spectrum of the Witten Hamiltonian ${\tilde{\cal H}}_W^s$ will show sequences of $M$ different groups (labeled by $k=1,\dots,M$) of nearly degenerate states, each of them containing $m_k$ states, except a group of $m_k/2$ states close to zero eigenvalue. An example will be discussed in all detail in the next section for a surface of peanut shape, where $N_Z=6$, $M=2$, $m_1=4$ and $m_2=2$, see Fig.~\ref{fig:peanut_different_R}. 

For generic surfaces without any symmetry one finds only an alternation of the signs of $p_j$ when moving around the surface, see an example for a surface with $N_Z=4$ and four different frequencies $\tilde{\Omega}_j$ shown in Fig.~\ref{fig:surface_witten_generic}(a), together with a sketch of the corresponding Witten potential $\tilde{V}_W^+$ in Fig.~\ref{fig:surface_witten_generic}(b). In the case where all frequencies are different only the twofold degeneracy of the zero energy state is guaranteed for the surface Hamiltonian but all non-zero energies are non-degenerate (up to accidental degeneracies due to commensurabilities between the squared frequencies).

\begin{figure}
	 \includegraphics[width =0.90\columnwidth]{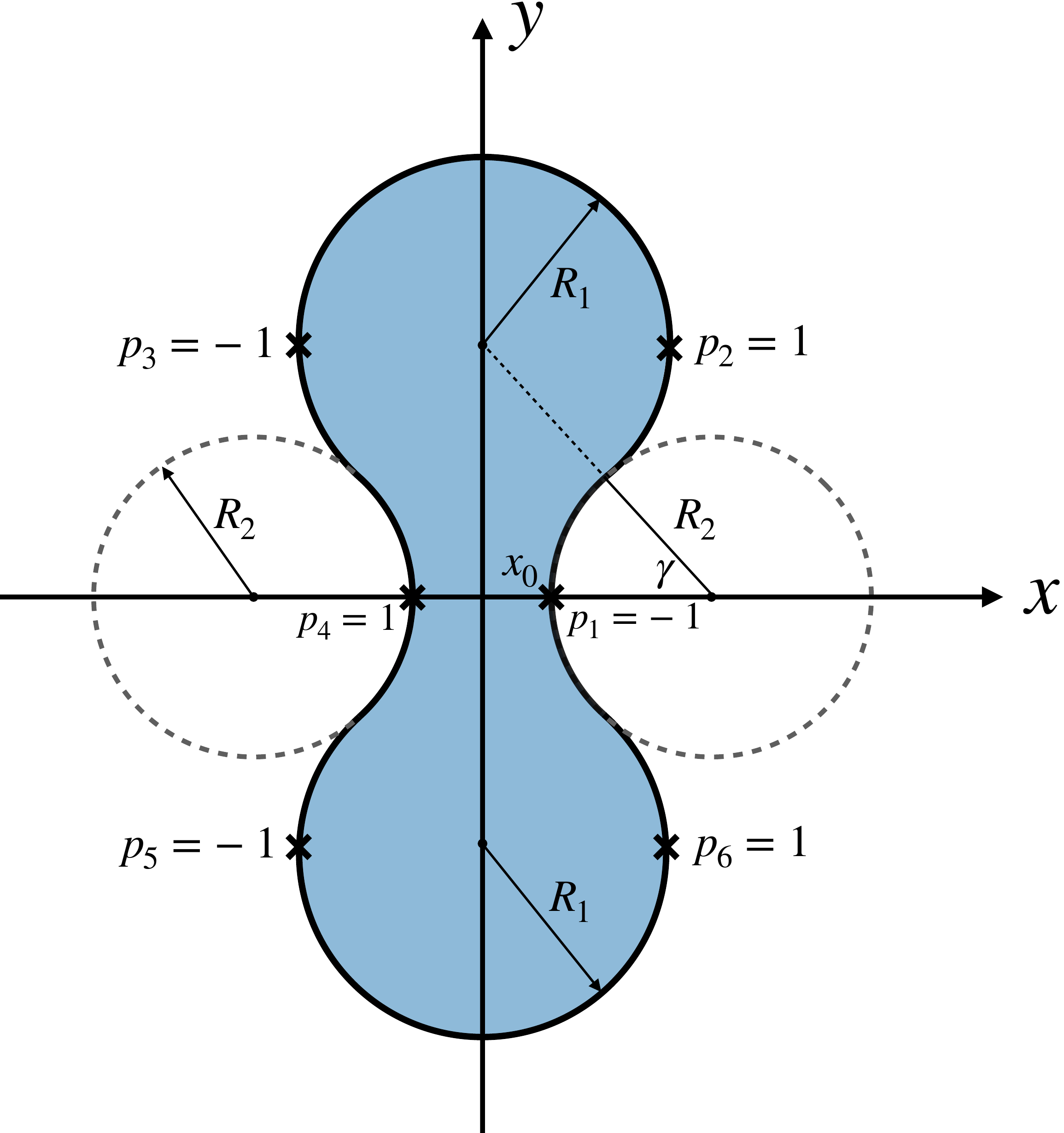}
	  \caption{Construction of a peanut shape via the parameters $R_1$, $R_2$ and $x_0$. Here, $R_1$ and $R_2$ are the radia of the circles with centers on the $y$- and $x$-axis, respectively, and $2\,x_0$ denotes the width of the peanut for $y=0$. The angle $\gamma$ determines the point where two circles meet. We get $6$ points with $\vartheta=0,\pi$, four of them having the frequency $\tilde{\Omega}_1=2/(\tilde{l}_B\sqrt{\tilde{R}}_1)$ (for $j=2,3,5,6$), and two with $\tilde{\Omega}_2=2/(\tilde{l}_B\sqrt{\tilde{R}_2})$ (for $j=1,4$). The sign factors $p_j = ({\rm sign}\cos{\vartheta_j})\,({\rm sign}\,\tilde{\kappa}_t^j)$ alternate when moving around the surface (compare with Fig.~\ref{fig:surface_witten_generic}), and each frequency has the same number of positive and negative $p_j$, i.e., $2$ for $\tilde{\Omega}_1$ and $1$ for $\tilde{\Omega}_2$. The angle $\gamma$ is the angle between the $x$-axis and the connection line between the middle points of the upper and right circle.  
	  } 
    \label{fig:peanut_construction}
\end{figure}

\begin{figure*}[t!]
    \centering
    \includegraphics[width =\textwidth]{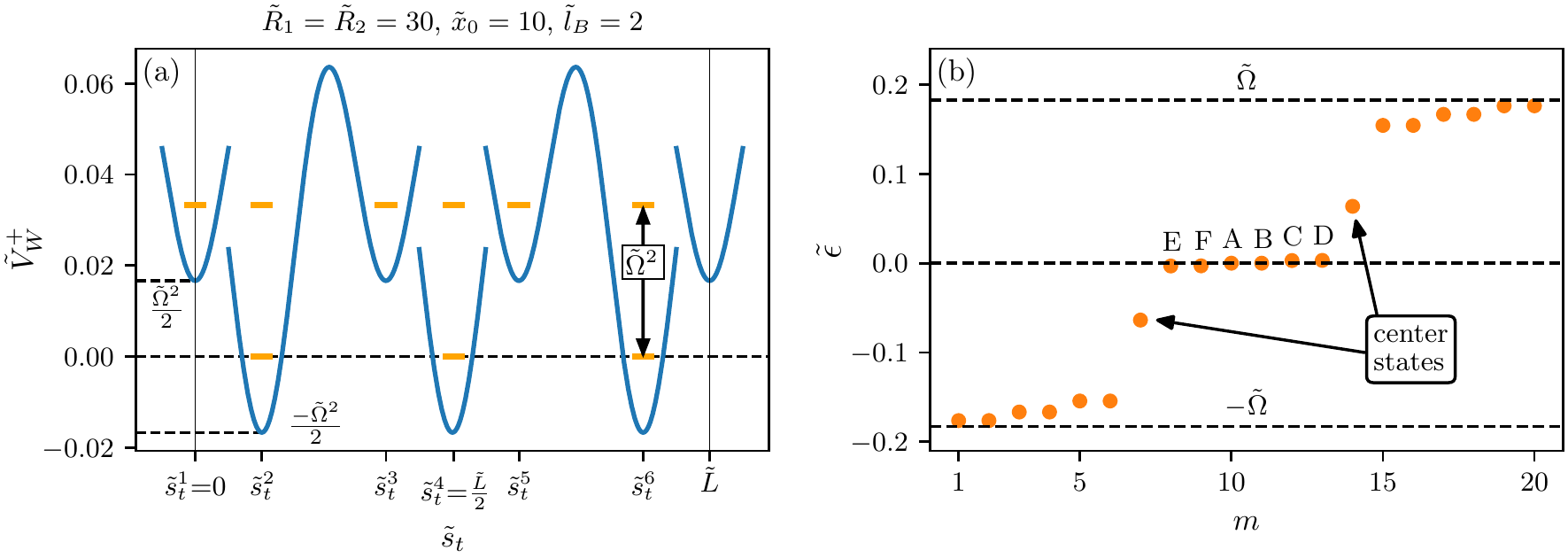}
	  \caption{(a): Witten potential $\tilde{V}_W^+$ of a peanut surface with $\tilde{R}_1=\tilde{R}_2=\tilde{R}=30$, $\tilde{x}_0=10$, and $\tilde{l}_B=2$, calculated via the formulas of Appendix~\ref{app:witten_potential_peanut}. The jumps of the potential arise from the discontinuous change of the curvature at the meeting points of two circles in the construction of the peanut surface via Fig.~\ref{fig:peanut_construction}. Based on the sign factors $p_j$ and the frequency $\tilde{\Omega}=2/(\tilde{l}_B\sqrt{\tilde{R}})=1/\sqrt{30}\approx 0.18$ introduced in Fig.~\ref{fig:peanut_construction}, we obtain $6$ harmonic oscillator potentials located at $\tilde{s}_t^j$, with potential minima at $-p_j \tilde{\Omega}^2/2$, with $p_2=p_4=p_6=1$ and $p_1=p_3=p_5=-1$. In these potentials we indicate the harmonic oscillator states in a semiclassical picture (we only indicate those below the height of the potential set by the Zeeman energy squared $\tilde{E}_Z^2=1/\tilde{l}_B^4=1/16=0.0625$). As a result, we get three states with zero eigenvalue and six states with eigenvalue $\tilde{\Omega}^2=1/30\approx 0.034$ which, due to SUSY, will split by an exponentially small hybridization into one state at zero eigenvalue and a set of twofold degenerate states. Since each state of the Witten Hamiltonian ${\tilde{\cal H}}_W^+$ with non-zero eigenvalue $\tilde{\epsilon}^2$ will lead to two corresponding states at positive and negative energy $\pm |\tilde{\epsilon}|$ for the surface Hamiltonian (see Eq.~(\ref{eq:eigenstate_surface_H_explicit})), we get the energy spectrum of the surface Hamiltonian shown in (b) (calculated numerically via the tight-binding formalism). It consists of groups of six nearly degenerate states at $\tilde{\epsilon}\approx 0, \pm\tilde{\Omega}$ (higher states at $\tilde{\epsilon}\approx\pm \sqrt{n}\,\tilde{\Omega}$, with $n=2,3,\dots$ behave differently since they are above the surface gap set by $\tilde{E}_Z$). To get the center states (indicated by an arrow) away from zero energy we have detuned the flux $f=(\pi+0.1)/(2 \pi)\approx 0.5159$ slightly away from half-integer value. This affects the bound states at the boundary of the peanut only weakly since the states with energies within the surface gap are well localized.   
	  } 
    \label{fig:peanut_same_R}
\end{figure*}

\begin{figure*}[t!]
    \centering
    \includegraphics[width =\textwidth]{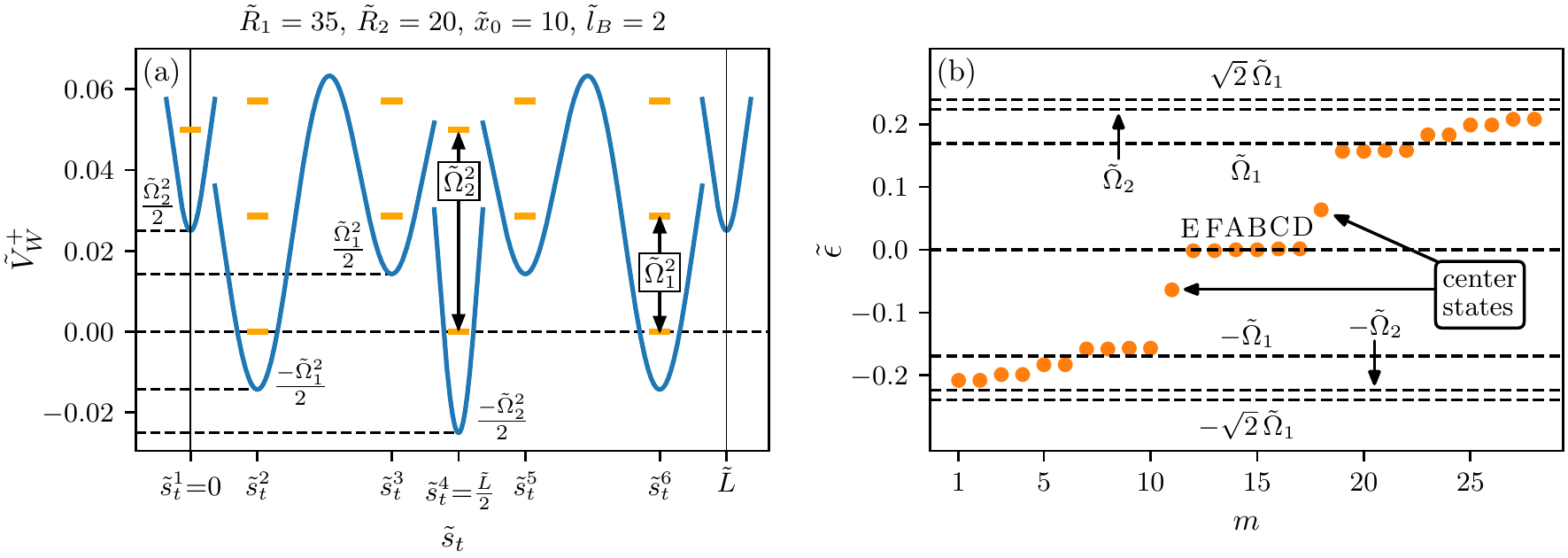}
	  \caption{
	  (a) The Witten potential $\tilde{V}_W^+$ of a peanut surface for the same parameters $\tilde{x}_0=10$ and $\tilde{l}_B=2$ as in Fig.~\ref{fig:peanut_same_R} but for two different radia $\tilde{R}_1=35$ and $\tilde{R}_2=20$. In this case there are two different frequencies $\tilde{\Omega}_1=2/(\tilde{l}_B\sqrt{\tilde{R}_1})=1/\sqrt{35}\approx 0.17$ and $\tilde{\Omega}_2=2/(\tilde{l}_B\sqrt{\tilde{R}_2})=1/\sqrt{20}\approx 0.22$. As a result, we get $4$ harmonic oscillator potentials with potential minima at $-p_j \tilde{\Omega}_1^2/2$ for $j=2,3,5,6$, and $2$ harmonic oscillator potentials with potential minima at $-p_j \tilde{\Omega}_2^2/2$ for $j=1,4$, with $p_2=p_4=p_6=1$ and $p_1=p_3=p_5=-1$. Within a semiclassical picture this gives three states with zero eigenvalue, a group of $4$ states with eigenvalue $\tilde{\Omega}_1^2=1/35\approx 0.0285$, a group of $2$ states with eigenvalue $\tilde{\Omega}_2^2=1/20=0.05$, and a group of $4$ states with eigenvalue $2\,\tilde{\Omega}_1^2=2/35\approx 0.057$ (all the other states are above the height of the Witten potential set by the squared Zeeman energy $\tilde{E}_Z^2=1/\tilde{l}_B^4=0.0625$). Due to the exponentially small splitting and SUSY, this leads to the energy spectrum of the surface Hamiltonian shown in the right figure (b) (calculated numerically via the tight-binding formalism), showing a group of $6$ states close to zero energy, and groups with four states close to $\pm\tilde{\Omega}_1$. Since the $2$ states at $\pm\tilde{\Omega}_2$ and the $4$ states at $\pm\sqrt{2}\,\tilde{\Omega}_1$ are quite close to each other, one can not distinguish them any longer after the hybridization, leading to groups of $6$ states close to these energies. All other states at $\tilde{\epsilon}\approx\pm \sqrt{n}\,\tilde{\Omega}_1$, with $n=3,4,\dots$, and $\tilde{\epsilon}\approx\pm \sqrt{n}\,\tilde{\Omega}_2$, with $n=2,3,\dots$, behave differently since they are above the surface gap set by $\tilde{E}_Z=1/\tilde{l}_B^2 = 0.25$. As in Fig.~\ref{fig:peanut_same_R} we have detuned the flux $f=(\pi+0.1)/(2 \pi)\approx 0.5159$ slightly away from half-integer value to get the center states away from zero energy.
	  }
    \label{fig:peanut_different_R}
\end{figure*}

\begin{figure*}[t!]
    \centering
    \includegraphics[width =1.0\columnwidth]{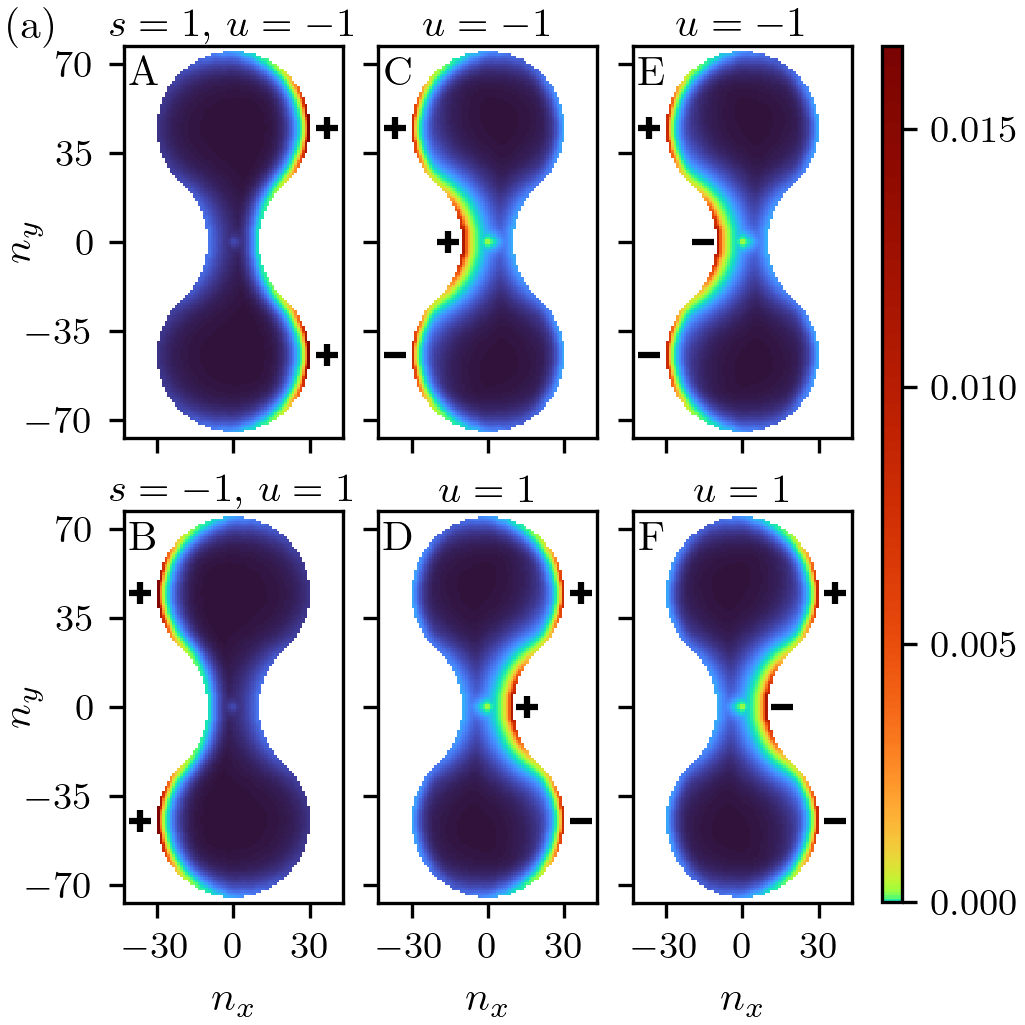}
    \hfill\includegraphics[width =1.0\columnwidth]{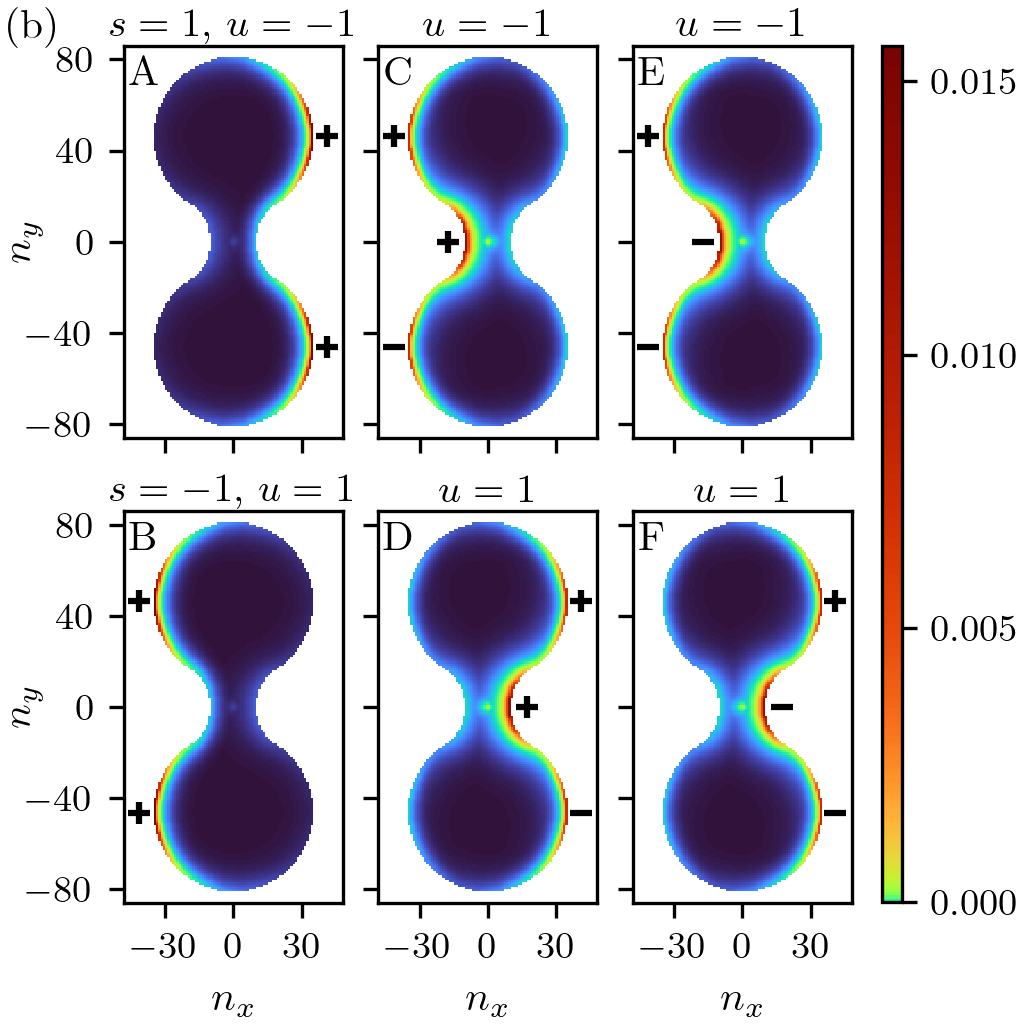}
	  \caption{Numerical calculation via the tight-binding formalism described in Appendix~\ref{app:tb_numerics} for the wave functions ``A-F'' of the $6$ states with lowest absolute value of the energy, see labeling in Figs.~\ref{fig:peanut_same_R}(b) and \ref{fig:peanut_different_R}(b). We consider a system of peanut shape as shown in Fig.~\ref{fig:peanut_construction} using the parameters $\tilde{x}_0=10$, $\tilde{l}_B=2$, $\tilde{\delta}=1$, together with (a) $\tilde{R}_1=\tilde{R}_2=\tilde{R}=30$ (left figure) and (b) $\tilde{R}_1=35$ and $\tilde{R}_2=20$ (right figure), i.e. the same parameters as in Fig.~\ref{fig:peanut_same_R} and Fig.~\ref{fig:peanut_different_R}, respectively. In the six panels we show the wave functions ``A-F'' by plotting $\sum_{\sigma_z s_z}|\psi(n_x,n_y;\sigma_z,s_z)|^2$ in color code as function of the lattice site index $(n_x,n_y)$. The signs of the wave functions before squaring them are indicated which agrees with the analytical considerations (see main text). 
	  } 
    \label{fig:peanut_wave_functions}
\end{figure*}

\begin{figure*}[t!]
    \centering
	 \includegraphics[width =\textwidth]{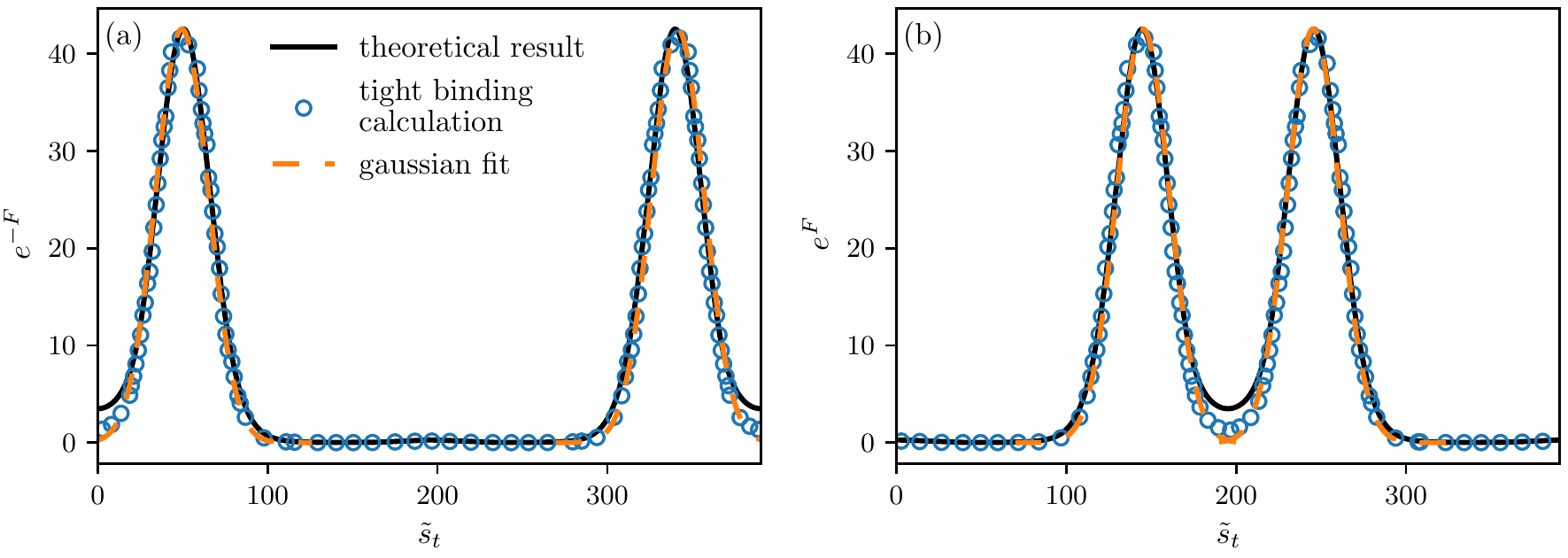}
	  \caption{The angular part of the two zero energy topological states as function of the line element $\tilde{s}_t$ along the surface for a system of peanut shape as shown in Fig.~\ref{fig:peanut_construction} with $\tilde{R}_1=\tilde{R}_2=\tilde{R}=30$, $x_0=10$, $\tilde{l}_B=2$, and $\tilde{\delta}=1$. In the left figure (a) we show the analytical result $e^{-F(\tilde{s}_t)}$ (black line), with $F(\tilde{s}_t)$ given by (\ref{eq:F_witten}) and calculated explicitly in Appendix~\ref{app:witten_potential_peanut}. The analytical result is compared to the numerical result (shown in blue) for the state labeled ``A'' in Fig.~\ref{fig:peanut_wave_functions}(a), by showing the average $(1/A)\sum_{(n_x,n_y)\in A}(\sum_{\sigma_z s_z}|\psi(x,y;\sigma_z,s_z)|^2)^{1/2}$ over a certain area $A$ defined by a fixed distance $\Delta N = 4$ away from the surface. Furthermore, to ensure for the normalization, we rescaled the numerical result to fit the peak heights. We find two peaks at the points $\tilde{s}_t^2$ and $\tilde{s}_t^6$, where the Witten Potential $\tilde{V}_W^+$ has a minimum, see Fig.~\ref{fig:peanut_same_R}. This corresponds roughly to a symmetric combination $\sim (|g_0^2\rangle + |g_0^6\rangle)$ of the two harmonic oscillator ground states localized at $\tilde{s}_t^2$ and $\tilde{s}_t^6$ (the state $|g_0^3\rangle$ of the third minimum of $\tilde{V}_W^+$ is only weakly involved). This is demonstrated by a comparison to the two Gaussian forms $g_0^{2,6}(\tilde{s}_t)\sim e^{-\frac{1}{4}((\tilde{s}_t-\tilde{s}_t^{2,6})/\tilde{\xi}_t)^2}$, with $\tilde{\xi}_t=\tilde{l}_B\sqrt{\tilde{R}}=2\sqrt{30}\approx 11$ (together with a rescaling factor to match the maximum), as shown by the dashed orange lines in the figure. In the right figure (b) we show the comparison of the analytical result for $e^{F(\tilde{s}_t)}$ with the numerical result for the state labeled ``B'' in Fig.~\ref{fig:peanut_wave_functions} and with the Gaussian fits for $g_0^{3,5}(\tilde{s}_t)$, corresponding to the two minima of the Witten potential $\tilde{V}_W^-$ at $\tilde{s}_t^{3,5}$. Here both the $x$ and $y$ coordinates of the numerical curves have been adjusted by multiplying a constant scale factor to fit the peak positions. 
	  } 
    \label{fig:psi_W}
\end{figure*}

\subsection{Peanut shape}
\label{sec:peanut}

An example for a mirror symmetric surface is shown in Fig.~\ref{fig:peanut_construction}, using a peanut shape constructed out of two circles with radia $R_1$ and $R_2$. This gives $N_Z=6$ points where $\vartheta_j=0,\pi$ and the sign factors $p_j$ are given by $p_1=p_3=p_5=-1$ and $p_2=p_4=p_6=1$, with two frequencies $\tilde{\Omega}_1=2/(\tilde{l}_B\sqrt{\tilde{R}_1})$ and $\tilde{\Omega}_2=2/(\tilde{l}_B\sqrt{\tilde{R}_2})$. The first frequency $\tilde{\Omega}_1$ occurs $m_1=4$ times at $\tilde{s}_t=\tilde{s}_t^2,\tilde{s}_t^6$ (with $p_2=p_6=1$) and $\tilde{s}_t=\tilde{s}_t^3,\tilde{s}_t^5$ (with $p_3=p_5=-1$). The second frequency $\tilde{\Omega}_2$ occurs $m_2=2$ times at $\tilde{s}_t=\tilde{s}_t^1=0$ (with $p_1=-1$) and $\tilde{s}_t=\tilde{s}_t^4=\tilde{L}/2$ (with $p_4=1$) . The qualitative form of the spectrum follows from our generic analysis in Section~\ref{sec:spectrum_surface_H}. For $R_1=R_2=R$, we get two identical frequencies $\Omega_1=\Omega_2=\Omega$ , leading to the Witten potential $\tilde{V}_W^+$ as shown in Fig.~\ref{fig:peanut_same_R}(a), based on the explicit formulas derived in Appendix~\ref{app:witten_potential_peanut}. As a consequence, the spectrum of the surface Hamiltonian (calculated numerically within the tight-binding formalism described in Appendix~\ref{app:tb_numerics}) shows groups of $6$ states lying close together in energy as shown in Fig.~\ref{fig:peanut_same_R}(b). For different $R_1 > R_2$, we see in Fig.~\ref{fig:peanut_different_R}(a) that the Witten potential $\tilde{V}_W^+$ hosts two different groups of $4$ ($2$) nearly degenerate states at $n\tilde{\Omega}_1^2$ ($n\tilde{\Omega}_2^2$), with $n=1,2,\dots$. This leads to the energy spectrum for the surface Hamiltonian shown in Fig.~\ref{fig:peanut_different_R}(b). Besides the group of $6$ states close to zero energy, one can see groups of $4$ states close to $\pm\tilde{\Omega}_1$, consistent with the semiclassical picture. Since the two energies $\pm\tilde{\Omega}_2$ and $\pm \sqrt{2}\,\tilde{\Omega}_1$ are quite close to each other for the parameters used in Fig.~\ref{fig:peanut_different_R}, one can no longer distinguish these two groups after the small hybridization, and groups of six states appear close to these energies.  

We note that only states within the surface gap set by the Zeeman energy $\tilde{E}_Z=1/\tilde{l}_B^2$ are shown in Figs.~\ref{fig:peanut_same_R} and \ref{fig:peanut_different_R}, and we detuned the flux slightly away from half-integer value to get the center states away from the energies of the bound states localized at the boundary of the peanut. Furthermore, we note that the Witten potential contains discontinuous jumps at the points $\tilde{s}_t=\pm R_2\,\gamma, \pm (\tilde{L}/2-R_2\,\gamma)$, where the curvature changes discontinuously from $-1/\tilde{R}_2$ to $1/\tilde{R}_1$. However, this is due to our special construction of the peanut shape and does not influence the low-energy wave functions significantly in the case of strong localization, since the points where the jumps of the potential appear are sufficiently away from the hotspots $\tilde{s}_t^j$, where the wave functions are localized. 

For the parameters used in Figs.~\ref{fig:peanut_same_R} and \ref{fig:peanut_different_R}, we show in Fig.~\ref{fig:peanut_wave_functions} the absolute square of the wave functions (averaged over the spinor indices) for the $6$ states lying close to zero energy (labeled by ``A-F'' in Figs.~\ref{fig:peanut_same_R}(b) and \ref{fig:peanut_different_R}(b)). For $R_1=R_2$, the two zero energy states labeled by ``A'' and ``B'' are compared with the analytical solution (\ref{eq:zero_energy_state_W}) and (\ref{eq:F_witten}) for the angular part $\psi_W^{(0),\pm}(\tilde{s}_t)\sim e^{\mp F(\tilde{s}_t)}$ in Fig.~\ref{fig:psi_W}, see Appendix~\ref{app:witten_potential_peanut} for the explicit formulas to calculate the function $F(\tilde{s}_t)$ for the peanut shape. We find two peaks at $\tilde{s}_t^{2,6}$ ($\tilde{s}_t^{3,5}$) for $\psi_W^{(0),+}$ ($\psi_W^{(0),-}$) which agrees quite nicely with the analytical prediction and are consistent with the tangential localization length $\tilde{\xi}_t=\tilde{l}_B\sqrt{\tilde{R}}$, see the Gaussian fit shown in Fig.~\ref{fig:psi_W}. Both zero energy states $\psi_W^{(0),\pm}$ are symmetric under a sign change of $\varphi$ and have chirality $s=\pm 1$ and SUSY eigenvalue $u=\mp 1$, consistent with the analytics. 

The four states ``C-F'' with finite but very small energies can be constructed from the two degenerate first excited eigenstates of the Witten Hamiltonian ${\tilde{\cal H}}_W^+$, see the detailed discussion in Section~\ref{sec:spectrum_surface_H} and the explicit formula (\ref{eq:eigenstate_surface_H_explicit}) to construct the eigenstates of the surface Hamiltonian at positive and negative energy from the ones of the Witten model ${\tilde{\cal H}}_W^+$. As shown in Fig.~\ref{fig:peanut_same_R}(a), the Witten potential $\tilde{V}_W^+$ hosts three harmonic oscillator ground states $|g_0^j\rangle$ localized at $\tilde{s}_t^j$ with $j=2,4,6$. In analogy, the Witten potential ${\tilde{\cal H}}_W^-$ hosts three harmonic oscillator ground states $|g_0^j\rangle$ for $j=1,3,5$. In Section~\ref{sec:spectrum_surface_H} we learnt for a mirror symmetric surface that each eigenstate of the surface Hamiltonian with non-zero eigenvalue can be constructed as a combination of a symmetric/antisymmetric state (with respect to $P_\varphi$) of ${\tilde{\cal H}}_W^+$ and an antisymmetric/symmetric state of ${\tilde{\cal H}}_W^-$, such that they are at the same time eigenstates of the SUSY operator with SUSY eigenvalue $u=-1/u+1$ for the outer surface. As shown in Fig.~\ref{fig:psi_W}, the symmetric combination $|g_0^2\rangle + |g_0^6\rangle$ is predominately present in the state $\psi^{(0),+}_W$ with zero eigenvalue. The other symmetric state $|g_0^4\rangle$ of ${\tilde{\cal H}}_W^+$ is combined with the antisymmetric state $|g_0^3\rangle - |g_0^5\rangle$ of ${\tilde{\cal H}}_W^-$ to form the eigenstates ``C'' and ``E'', which have SUSY eigenvalue $u=-1$. Finally, the states ``D'' and ``F'' are obtained from applying the inversion operator to ``C'' and ``E'', respectively, i.e., are combinations of the symmetric state $|g_0^1\rangle$ of ${\tilde{\cal H}}_W^-$ and the antisymmetric state  $|g_0^2\rangle - |g_0^6\rangle$ of ${\tilde{\cal H}}_W^+$, with SUSY eigenvalue $u=1$.

\vspace{1cm}

\section{Stability and topological engineering}
\label{sec:stability_tunability}

In this section we discuss the stability against various kinds of perturbations (flux away from half-integer value or penetrating into the sample, surface distortions, and random disorder), and the possibilities of how to use the topological hole states in multi-hole systems for topological engineering. 

For clarity, we summarize here again the conditions of the validity range of the analytical theory, as it was discussed in all detail in Sections~\ref{sec:corbino_weak_fields}, \ref{sec:corbino_strong_fields}, and \ref{sec:validity}. For the discussion of strongly localized bound states below the surface gap in the topological phase, we need the condition specified in (\ref{eq:general_condition_H_eff}) which, in terms of the Witten frequency $\Omega_W = 1/(m^* \lambda_{\rm so}\xi_t)$ and in dimensionfull units, can be written as 
\begin{align}
    \label{eq:general_condition_H_eff_witten_1}
    \frac{1}{m^* \lambda_{\rm so}R} &\ll \Omega_W \ll E_Z \,,\\
    \label{eq:general_condition_H_eff_witten_2}
    \Omega_W &\ll E_{\rm so} \,,\\
    \label{eq:general_condition_H_eff_3}
     E_Z \sim \Delta_{\rm surface} &\lesssim \Delta_{\rm bulk}\,.
\end{align}
The first condition (\ref{eq:general_condition_H_eff_witten_1}) is equivalent to the condition of strong localization $\xi_t = l_B \sqrt{R/\lambda_{\rm so}}\ll R$ and means that we have a clear separation of energy scales between the level spacing $1/(m^* \lambda_{\rm so} R)$ of the extended edge states, the level spacing $\Omega_W$ of the localized bound states, and the surface gap $\sim E_Z$, see Fig.~\ref{fig:hole_system}(b). We note that the two conditions $\Omega_W\gg 1/(m^*\lambda_{\rm so}R)$ and $E_Z\gg\Omega_W$ are equivalent since
\begin{align}
    \label{eq:ratio_identity}
    \frac{E_Z}{\Omega_W} = \frac{1}{2}\,\frac{\Omega_W}{1/(m^* \lambda_{\rm so}R)}\,.
\end{align}
The second condition (\ref{eq:general_condition_H_eff_witten_2}) means that we have a clear separation of length scales between the normal and tangential localization lengths $\lambda_{\rm so}\sim\xi_n\ll\xi_t$, such that we can split the Hamiltonian in a normal and tangential part, see Section~\ref{sec:validity} for the details. As already pointed out at the end of Section~\ref{sec:validity}, the two conditions can be fulfilled for sufficiently large curvature radius $R$, but do not require any condition for the ratio of spin-orbit energy $E_{\rm so}$ and Zeeman energy $E_Z$. 

To describe the topological states in the case of strong delocalization in tangential direction $\xi_t \sim R$, one needs in addition the condition of weak Zeeman energy $E_Z \ll E_{\rm so}$ as compared to spin-orbit energy, see the detailed discussion in Section~\ref{sec:corbino_weak_fields}. This is important for the study of topological engineering to generate a controlled coupling between the two topological hole states of a single hole, see below.

\subsection{Stability}
\label{sec:stability}

If the topological states are well localized in normal and angular direction, i.e., if the condition (\ref{eq:general_condition_H_eff_witten_1}) is fulfilled, only the properties of the model in a local subpart of the surface is important. This guarantees the stability of the topological states against deviations of the flux from half-integer value and against deformations of the surface. Moreover, even if the flux penetrates into the sample, the spectrum of the boundary states is not significantly changed since they are strongly localized in normal direction and feel only the total flux through the area defined by the surface. 

For the stability against random on-site disorder we discuss generic impurity potentials defined in the tight-binding version of the model by
\begin{align}
    \label{eq:impurity_potential}
    V_{\rm im} = \sum_{\alpha,\beta=0,1,2,3} \sum_{\bf n} v^{\alpha\beta}_{\bf n}|{\bf n}\rangle\langle{\bf n}| \,\sigma_\alpha \otimes s_\beta \,,
\end{align}
where ${\bf n}$ labels the lattice sites, $\sigma_0=s_0=\mathbbm{1}$, and $v^{\alpha\beta}_{\bf n}$ is randomly distributed in the interval $v^{\alpha\beta}_{\bf n}\in [-d/2,d/2]$. Here, $d$ is a measure for the impurity strength. After averaging over the disorder the self-energy will be of order $d^2/E_{\rm so}$, where $t\sim E_{\rm so}$ is the average hopping in the limit of strong spin-orbit interaction. For a generic spinor dependence of the impurity potential we then expect that stability of the topological states is guaranteed if 
\begin{align}
    \label{eq:impurity_stability}
    \frac{d^2}{E_{\rm so}} \lesssim \Omega_W \,,
\end{align}
since the Witten frequency is the energy of the first excited bound state. This is equivalent to 
\begin{align}
    \label{eq:impurity_stability_estimate}
    \left(\frac{d}{E_Z}\right)^2 \lesssim \frac{\tilde{l}_B^3}{\sqrt{\tilde{R}}}\,.
\end{align}
Thus, to achieve a stability in the regime of the surface gap $d<E_Z$, one needs to choose $\tilde{l}_B^3\gtrsim \sqrt{\tilde{R}}$. For typical parameter values $\tilde{R}\sim 30-60$ and $\tilde{l}_B\sim 2$, this can easily be achieved.

\begin{figure*}[t!]
    \centering
	 \includegraphics[width =0.65\columnwidth]{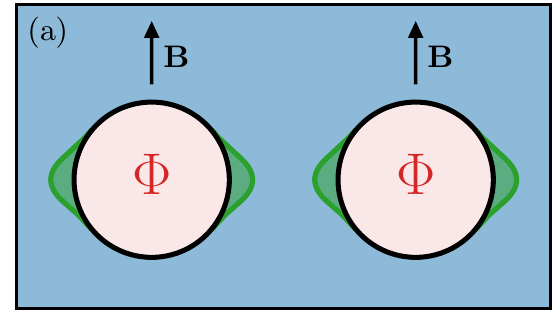}
    \hfill	 	 \includegraphics[width =0.65\columnwidth]{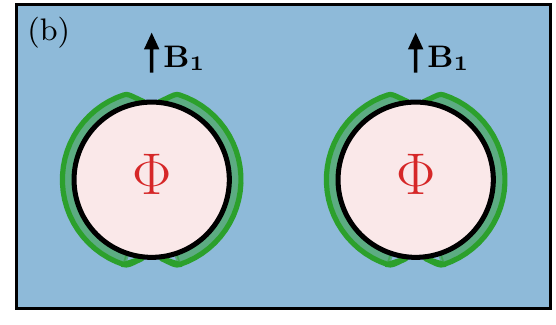}
    \hfill	 	 \includegraphics[width =0.65\columnwidth]{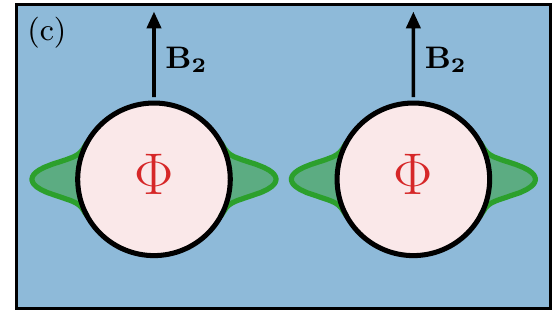}
	  \caption{Tunability of topological states localized at two holes in the system. In (b) and (c) we sketch how the shape of the topological states change qualitatively if one decreases/increases the Zeeman field $B_{1/2}\lessgtr B$ compared to (a). Whereas an increase of the Zeeman field localizes the state stronger in tangential direction (i.e., $\tilde{\xi}_t\sim \sqrt{\tilde{R}}\,\tilde{l}_B$ decreases), it delocalizes the state in normal direction since one approaches the phase transition line $\tilde{\delta}=1/\tilde{l}_B^2=\tilde{E}_Z$ at fixed $\tilde{\delta}>\tilde{E}_Z$. Therefore, a decrease of the Zeeman field increases the overlap of the two states from the {\it same} hole, such that their hybridization increases by tuning the flux away from half-integer value, see (b). In contrast, in a setup of two holes, the overlap of states from {\it neighboring} holes can be increased by increasing the Zeeman field, see (c).      
	  } 
    \label{fig:tunability}
\end{figure*}

For impurity potentials with a special spinor dependence, the stability can be even stronger. As shown in Sections~\ref{sec:corbino_weak_fields} and \ref{sec:witten_derivation} for weak Zeeman field, all edge  states are eigenfunctions of $s_x$ in the transformed basis, with eigenvalue $\pm 1$ for the outer/inner surface, see Eq.~(\ref{eq:edge_state_normal_ortho}). Therefore, if the impurity potential $\bar{V}_{\rm im}$ in the transformed basis contains only terms $\sim \sigma_\alpha s_{y,z}$, no matrix elements of the Hamiltonian are possible between any edge states within the bulk gap, leading to an increased stability 
\begin{align}
    \label{eq:impurity_stability_increased_1}
    \frac{d^2}{E_{\rm so}} \lesssim \Delta_{\rm bulk}\,,
\end{align}
i.e., the Witten frequency in (\ref{eq:impurity_stability}) is replaced by the bulk gap. Since the spinors transform under the transformation (\ref{eq:ortho_H_trafo}) as
\begin{align}
    \label{eq:spinor_trafo_1}
    s_{x,y} &\xrightarrow{UWX} e^{i\frac{1}{2}s_z\vartheta} s_{x,y} e^{-i\frac{1}{2}s_z\vartheta} \,,\\
    \label{eq:spinor_trafo_2}
    \sigma_x &\xrightarrow{UWX} -\sigma_z s_z \,,\\
    \label{eq:spinor_trafo_3}
    \sigma_z &\xrightarrow{UWX} \sigma_x s_z \,,\\
    \label{eq:spinor_trafo_4}
    s_z &\xrightarrow{UWX} s_z \quad,\quad \sigma_y \xrightarrow{UWX} \sigma_y \,,
\end{align}
we find that impurity potentials with one of the four following spinor dependencies have the increased stability regime (\ref{eq:impurity_stability_increased_1})
\begin{align}
    \label{eq:impurity_potential_increased_stability_1}
    V_{\rm im} \sim \sigma_x, \sigma_z, s_z, \sigma_y s_z \,. 
\end{align}

Furthermore, in the case of strong localization $\Delta\varphi\sim \tilde{l}_B/\sqrt{\tilde{R}}\ll 1$ around a point with $\vartheta=0$ (analog one can treat the case of localization around $\vartheta=\pi$), we can expand the rotation matrix $e^{i\frac{1}{2}s_z\vartheta}\approx 1 + i\frac{1}{2}s_z \vartheta$ and find from (\ref{eq:spinor_trafo_1})
\begin{align}
    \label{eq:spinor_trafo_5}
    s_y \xrightarrow{UWX} s_y + s_x O(\Delta\varphi) \,.
\end{align}
The first term involving $s_y$ leads to the stability regime (\ref{eq:impurity_stability_increased_1}) involving the bulk gap. The second term involves the small factor $\Delta\varphi\ll 1$ associated with $s_x$, i.e., in (\ref{eq:impurity_stability}) we have to multiply the impurity strength $d$ with this factor, leading to the increased stability region 
\begin{align}
    \label{eq:impurity_stability_increased_2}
    \frac{(\Delta\varphi\,d)^2}{E_{\rm so}} \lesssim \Omega \quad\Leftrightarrow\quad
    \left(\frac{d}{E_Z}\right)^2 \lesssim \sqrt{\tilde{R}}\,\tilde{l}_B \sim \tilde{\xi}_t \,.
\end{align}
This stability regime applies to all impurity potentials with a spinor dependence of the form
\begin{align}
    \label{eq:impurity_potential_increased_stability_2}
    V_{\rm im} \sim s_y, \sigma_x s_x, \sigma_y s_y, \sigma_z s_x \,. 
\end{align}
Since $\tilde{\xi}_t\gg 1$, the condition (\ref{eq:impurity_stability_increased_2}) leads to a stability regime for impurity strengths much beyond the surface gap.

\subsection{Topological engineering with hole states}
\label{sec:tunability}

We propose the topological hole states in the topological phase with $\tilde{E}_Z < \tilde{\delta} < 1 + \tilde{E}_Z$ to be of particular interest for topological engineering since their localization length in normal and tangential direction change in a different way when increasing the size of the Zeeman field. Whereas an increase of the normal component of the Zeeman field decreases the tangential localization length $\tilde{\xi}_t=\tilde{l}_B\sqrt{\tilde{R}}$, the normal localization length $\tilde{\xi}^<_n$ will increase according to (\ref{eq:xi_n_s=1_u=1_gapped}) since the bulk gap reduces. Therefore, by considering a multi-hole sample as sketched in Fig.~\ref{fig:tunability}(a-c), where the shape of the topological states can be controlled by  local Zeeman fields, one can increase the orbital overlap of topological states from different holes by increasing the Zeeman field, whereas a decrease of the Zeeman field leads to an increase of the orbital overlap of the topological states of the same hole. In this way, it is possible to realize controlled one- and two-hole operations. An orbital overlap of topological states from different holes will lead to an interaction since no symmetry protects the hybridization via the Hamiltonian. In contrast, for topological states from the same hole, the SUSY protects a hybridization. However, by the local Aharonov-Bohm flux through the hole, one can induce a controlled interaction between the topological states from the same hole if their wave functions have a significant orbital overlap. 

We note that the proposed scenario for two-hole operations is more difficult to realize with sharp corners, where the angle $\vartheta$ controlling the normal and tangential component of the Zeeman term changes abruptly its sign. In this case, the normal localization length is controlled by the difference $\delta - E_Z \cos{\vartheta}$ and will stay finite, even if the bulk gap closes at $\delta=E_Z$. Therefore, the normal localization length can be tuned to much larger values for smooth surfaces without closing the bulk gap. Furthermore, for sharp corners, the tangential localization length is of order $\tilde{\xi}_t\sim 1/\tilde{E}_Z = \tilde{l}_B^2$ since the normal Zeeman field is a constant along the surface, leading to an exponentially decaying wave function for the zero-energy state along the surface, according to the surface Hamiltonian (\ref{eq:hat_H_t_ortho}). As a result, one needs much weaker Zeeman fields to generate an orbital overlap between the topological states of the same hole, making one-hole operations also more difficult to realize as compared to the case of smooth surfaces where $\tilde{\xi}_t=\sqrt{\tilde{R}}\,\tilde{l}_B$ is much larger for $\sqrt{\tilde{R}}\gg\tilde{l}_B$. 

\begin{figure}
    \includegraphics[width =\columnwidth]{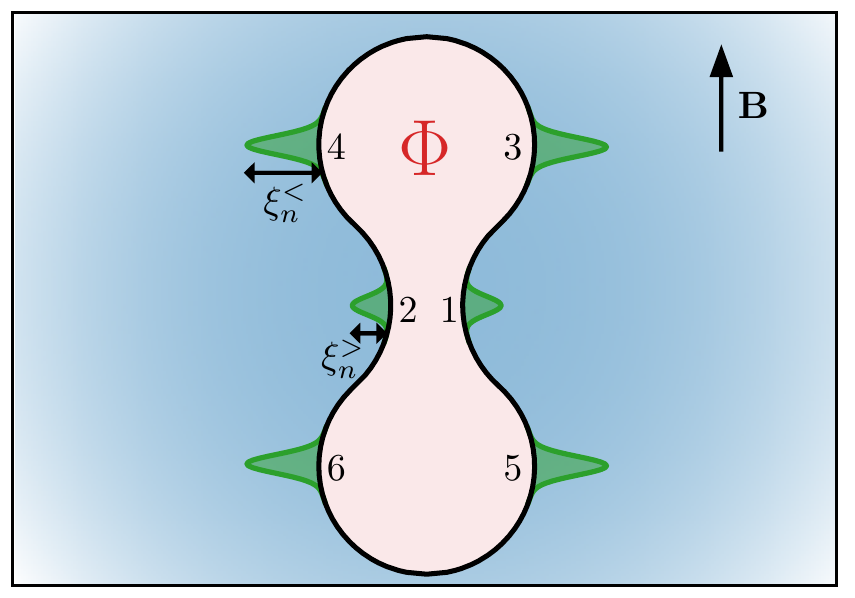}
	  \caption{Sketch of a system with a hole of peanut shape. The normal localization lengths $\xi_n^j$, with $j=1,\dots,6$, of the six indicated topological states are given by $\xi_n^{1,2}=\xi_n^>$ and $\xi_n^{3,4,5,6}=\xi_n^<$, with $\xi_n^\gtrless$ given by (\ref{eq:xi_n_s=1_u=-1}) and (\ref{eq:xi_n_s=1_u=1_gapped}), respectively. As a consequence, they behave very differently when increasing the size of the Zeeman field $B$. Whereas the states $1$ and $2$ will get more localized in normal and tangential direction, the states $3-6$ will only decrease in tangential direction but get more extended in normal direction.    
	  } 
    \label{fig:peanut_hole}
\end{figure}

Depending on the shape of the holes, a huge variety of other scenarios can be imagined for topological engineering. For example, if one takes a hole of peanut shape, as shown in Fig.~\ref{fig:peanut_hole}, both types of topological states with normal localization lengths $\xi_n^>$ and $\xi_n^<$ can be realized on the same hole. This has the effect that increasing the size of the Zeeman field, the topological states with $\xi_n^<$ (which are the states labeled by $3-6$ in Fig.~\ref{fig:peanut_hole}) will get an increased normal localization such that they can overlap with topological states from neighboring holes. In contrast, the topological states with $\xi_n^>$ will get more localized and do not participate in two-hole operations, whereas they can be used in one-hole operations, by increasing the tangential localization length by reducing the Zeeman field. Moreover, the tangential localization length can be tuned to different values by choosing different curvature radia for the various topological states. This opens up many possibilities for different protocols how interactions between various topological states can be controlled. 

\section{Summary and outlook}
\label{sec:summary}

The present work has revealed an interesting relationship between two different fields of condensed matter and high-energy physics. It has been shown that the surface spectrum of a wide class of second-order topological insulators in two dimensions has a supersymmetric structure if one applies a half-integer Aharonov-Bohm flux through the area of the surface. It was shown that the topological states are protected by supersymmetry and an effective surface Hamiltonian has been set up for smooth surfaces, revealing the whole class of supersymmetric periodic Witten models. The condition of a smooth surface is essential for a universal description. In contrast to sharp corners, where the topological states close to the corner are non-universal and not accessible to analytical approaches (except for very special cases), smooth surfaces offer the possibility for a full analytical control over all bound states localized at the surface. This has been shown via the localization of states in the minima of effective surface potentials, as they occur within the Witten models. Moreover, it has been shown that smooth surfaces offer the possibility for a more flexible tunability of the shape of the topological states in tangential and normal direction, opening up the possibility for topological engineering via one- and two-hole processes in multi-hole systems by using only magnetic fields. 

Our analysis is based on a quite generic continuum model for a second-order topological insulator, containing the basic ingredients of band inversion, Rashba spin-orbit interaction, and Zeeman field. As in previous works \cite{khalaf_prb_18,ren_etal_prl_20,laubscher_etal_prr_19,plekhanov_etal_prr_19,volpez_etal_prl_19,laubscher_etal_prb_20,laubscher_etal_prr_20,plekhanov_etal_prr_20,plekhanov_etal_prb_21}, it turns out that the Zeeman term is a particularly useful and flexible tool to induce a surface gap and to control the topological states. We have shown that the normal and tangential component of the Zeeman field play a very different role. Whereas the normal component determines the position of the topological states and controls the tangential localization length, the tangential component determines the normal localization length and contains information about occurrence of the phase transitions of the bulk. Interestingly, at the phase transition from the gapped topological to the gapless Weyl phase, it turns out that the normal localization length diverges only when the curvature of the surface is negative when looking from the side where the state is localized (e.g., for a Corbino disk, this are the states at the hole surface). It will be interesting to study how the behavior of those states is changed when considering the transition to sharp corners, where it is expected that the normal localization length will only diverge very close to the corner but not far away from the corner (where the surface is rather flat).

We note that our model can be extended to three dimensions (3D) where the topological states will change to hinge states with a dispersion as function of the perpendicular momentum which connects the conduction with the valence band. Depending on the choice of the three-dimensional area (cube, torus, sphere, etc.), a variety of anomalous Quantum Hall setups in 3D can be realized, generalizing the standard 2D Quantum Hall effect for a Corbino disk. This will be discussed in forthcoming works \cite{zhe_etal_future}.

For the future it will be interesting to study how other models of second-order topology with supersymmetric properties can be realized. The supersymmetry has the particular advantage that the two topological states are orthogonal to each other even if they have a strong orbital overlap, opening up the possibility for a controlled coupling between them by tuning the system slightly away from the supersymmetric point (which, in our case, has been achieved by changing the flux away from half-integer value). In particular, Majorana systems are of interest here, as they have been discussed in Refs.~\cite{laubscher_etal_prr_19,plekhanov_etal_prr_19,volpez_etal_prl_19,laubscher_etal_prb_20,laubscher_etal_prr_20,plekhanov_etal_prr_20,plekhanov_etal_prb_21}. It will be interesting to see which kind of Dirac model emerges for the effective surface Hamiltonian in this case and how its universal properties in the low-energy regime can be related to generalized multi-channel Witten models.

\hs{Furthermore, our findings of topologically protected zero-energy states in the Weyl phase might be another avenue for future research. As shown in Refs.~[\onlinecite{verresen_arxiv_20,verresen_etal_prl_18}] (and further references therein), general arguments have been set up to guarantee the existence of exponentially localized states in gapless systems based on topological phase transitions between systems with different half-integer toplogical invariants. Whether the same arguments can also be applied to the states generated via a second-order mechanism described in this work is an open question. As outlined in Section~\ref{sec:validity}, the effective surface Hamiltonian can only be extended to cover the regime of strong Zeeman fields for the discussion of well-localized bound states in the topological phase. It will be of interest in the future to see whether an effective surface Hamiltonian can also be derived for strong Zeeman fields to discuss all surface states, both in the topological and in the Weyl phase, to provide more insights for the phase transition and the relation to quantum critical phenomena. }

\section*{Acknowledgments}
We thank P. Brouwer, C. Bruder and H.F. Legg for fruitful discussions. This work was supported by the Deutsche Forschungsgemeinschaft via RTG 1995, the Swiss National Science Foundation (SNSF) and NCCR QSIT and by the Deutsche Forschungsgemeinschaft (DFG, German Research Foundation) under Germany's Excellence Strategy - Cluster of Excellence Matter and Light for Quantum Computing (ML4Q) EXC 2004/1 - 390534769. We acknowledge support from the Max Planck-New York City Center for Non-Equilibrium Quantum Phenomena. Simulations were performed with computing resources granted by RWTH Aachen University (projects rwth0752 and rwth0841) and at sciCORE (see Ref. \cite{scicore}) scientific computing center at University of Basel. Funding was received from the European Union's Horizon 2020 research and innovation program (ERC Starting Grant, grant agreement No 757725).

\begin{appendix}

\section{Bulk spectrum}
\label{app:bulk_spectrum}

In this Appendix we study the bulk spectrum for an infinite system where the outer surface is not present. In this case, the presence of the inner surface with the finite flux does not play any role since the vector potential is zero in the asymptotic region at large distances. Therefore, we set the flux to zero $f=0$ in the following and consider a translationally invariant system with plane waves $\sim e^{i{\bf k}{\bf x}}$ for the spatial part of the eigenfunctions. The spectrum of the Hamiltonian (\ref{eq:H_trafo_flux}), with ${\bf p}_K$ replaced by ${\bf k}$, can then be easily obtained by squaring it twice

\begin{widetext}
    
\begin{align}
    \label{eq:H_k_squared}
    H_{\bf k}^2 = \left(\frac{k^2}{2m^*}-\delta\right)^2 + \alpha^2 k^2 + E_Z^2 + 2 E_Z \left(\frac{k^2}{2m^*}-\delta\right)(\hat{\bf B}\cdot {\bf s}) \,\sigma_z + 2\alpha E_Z(\hat{\bf B}\cdot {\bf k})\sigma_x \,, 
\end{align}
where $\hat{\bf B} = {\bf B}/B$, $B=|{\bf B}|$, and $k=|{\bf k}|$. Taking the constant part of the right hand side to the left side and squaring again, we obtain for the bulk spectrum of the four bands (with $\sigma=\pm$ and $\eta=\pm$)
\begin{align}
    \label{eq:bulk_spectrum}
    \epsilon_{\bf k}^{\sigma\eta} = \sigma\sqrt{\left(\frac{k^2}{2m^*}-\delta\right)^2 + \alpha^2 k^2 + E_Z^2 + 2\eta E_Z \sqrt{\left(\frac{k^2}{2m^*}-\delta\right)^2 + \alpha^2 k^2 \cos^2\theta}}\,,
\end{align}
with $\cos{\theta}=\hat{\bf B}\cdot \hat{\bf k}$ and $\hat{\bf k}={\bf k}/k$. In dimensionless units $\tilde{k}=k/k_{\rm so}$, with $k_{\rm so}=1/\lambda_{\rm so}=|\alpha| m^*$, and using the definitions introduced at the end of Section~\ref{sec:model}, this can be rewritten as
\begin{align}
    \label{eq:bulk_spectrum_dimensionless}
    \tilde{\epsilon}_{\bf \tilde{k}}^{\sigma\eta} = \epsilon_{\bf k}^{\sigma\eta}/E_{\rm so} =
    \sigma \sqrt{\left(\sqrt{(\tilde{k}^2-\tilde{\delta})^2 + 4 \tilde{k}^2 \cos^2\theta} + \eta\tilde{E}_Z\right)^2 + 4 \tilde{k}^2 \sin^2\theta}\,.
\end{align}

\end{widetext}

To obtain the bulk gap and the gap closing point for ${\bf k}$, we consider $\sigma=+$, $\eta=-$ and $\theta=0$, and get for $\tilde{\epsilon}_{\tilde{k}} \equiv \tilde{\epsilon}_{\bf \tilde{k}}^{+-}|_{\theta=0}$ the result
\begin{align}
    \label{eq:tilde_epsilon_k}
    \tilde{\epsilon}_{\tilde{k}} =|\sqrt{(\tilde{k}^2-\tilde{\delta})^2 + 4 \tilde{k}^2} - \tilde{E}_Z|\,,
\end{align}
Using
\begin{align}
    \label{eq:G_1}
    G(\tilde{k}^2)&\equiv (\tilde{k}^2-\tilde{\delta})^2 + 4 \tilde{k}^2 \\
    \label{eq:G_2}
    &= (\tilde{k}^2 + 2 - \tilde{\delta})^2 + 4(\tilde{\delta}-1)\,,    
\end{align}
we obtain for the minimum $G_{\rm min}$ of $G(\tilde{k}^2)$ at $\tilde{k}=\tilde{k}_{\rm min}\ge 0$
\begin{align}
    \label{eq:G_min}
    G_{\rm min} = \begin{cases}
        4(\tilde{\delta}-1) & {\rm for} \quad \tilde{\delta} > 2 \\
        \tilde{\delta}^2 & {\rm for} \quad \tilde{\delta} < 2
    \end{cases}\,,\\
    \label{eq:tilde_k_min}
    \tilde{k}_{\rm min} = \begin{cases}
        \sqrt{\tilde{\delta}-2} & {\rm for} \quad \tilde{\delta} > 2 \\
        0 & {\rm for} \quad \tilde{\delta} < 2
    \end{cases}\,.
\end{align}
Together with (\ref{eq:tilde_epsilon_k}) this leads to the result (\ref{eq:bulk_gap}) for the bulk gap
\begin{align}
    \label{eq:bulk_gap_detail}
    \tilde{\Delta}_{\rm bulk} = \begin{cases}
        2\sqrt{\tilde{\delta}-1}-\tilde{E}_Z & {\rm for}\,\, \tilde{\delta}> {\rm max}\{2,1+\frac{1}{4}\tilde{E}_Z^2\} \\
        |\tilde{\delta}|-\tilde{E}_Z & {\rm for}\,\, \tilde{\delta} < 2 \,\,{\rm and}\,\, |\tilde{\delta}|> \tilde{E}_Z  \\
        0 & {\rm otherwise}
    \end{cases}
    \,.
\end{align}
In the parameter region where the bulk gap is finite, the minimum of the band dispersion occurs at $\theta=0$ and $k/k_{\rm so}=\tilde{k}_{\rm min}$, given by (\ref{eq:tilde_k_min}). The gap closes at the point $G_{\rm min}=\tilde{E}_Z^2$ which leads to
\begin{align}
    \label{eq:bulk_gap_closing_detail}
    \tilde{\delta} \,=\,
    \begin{cases} 
        1 \,+\,\frac{1}{4}\tilde{E}_Z^2 & \text{for} \,\,\tilde{\delta} > 2\\
             \pm\tilde{E}_Z &\text{for} \,\,\tilde{\delta} < 2
    \end{cases}\,,
\end{align}
and agrees with (\ref{eq:bulk_gap_closing}).

\vspace{1cm}

\section{Edge and center states for zero Zeeman field}
\label{app:zero_B}

Here we discuss the calculation of the exact edge states at zero Zeeman field $B=0$ for a Corbino disc, either localized at the inner surface with an arbitrary radius $R_<$ or at the outer surface for a large radius $\tilde{R}_>\gg 1$. We discuss only zero boundary condition at one of the surfaces, thereby neglecting exponentially small corrections at the other surface for $\tilde{R}_>-\tilde{R}_<\gg 1$, see the discussion after Eq.~(\ref{eq:delta_R_condition}) (a similar but more complicated analysis can be done by considering both boundary conditions at $\tilde{r}=\tilde{R}_\gtrless$, valid for any value of $\tilde{R}_>$). Similar to the calculation of boundary states in one-dimensional systems via linear combination of plane waves with complex momentum, we present here an analog approach for a rotationally invariant system in two dimensions via linear combinations of Hankel functions in radial direction with complex momentum. Of particular interest are the limits $\tilde{R}_>\gg 1$ and $\tilde{R}_< \ll 1$, where explicit forms can be provided for the dispersion and the wave functions of the low-energy edge states as function of the angular momentum $l=0,\pm 1,\pm 2,\dots$ in $z$-direction. In the following we use the notation $\tilde{R}\equiv\tilde{R}_<$ and discuss the edge states at the inner surface, mentioning at the appropriate places what has to be changed for the edge states at the outer surface. 

We start from the generalized supersymmetric Dirac Hamiltonian (\ref{eq:h_l_SUSY}) and replace the angular momentum $l$ by $\nu = l + f - \frac{1}{2}$ to cover also the case where the flux deviates from half-integer values. To construct the eigenstates we exploit the property that the operators $\Gamma_\nu$ and $\Gamma_\nu^\dagger$ act in the space of Hankel functions like ladder operators
\begin{align}
    \label{eq:Gamma_hankel}
    \Gamma_\nu \sqrt{\tilde{r}} H^{(1)}_{\nu+1/2}(\tilde{k}\tilde{r}) &= \tilde{k} \sqrt{\tilde{r}} H^{(1)}_{\nu-1/2}(\tilde{k}\tilde{r})\,,\\
    \label{eq:Gamma_dagger_hankel}
    \Gamma_\nu^\dagger \sqrt{\tilde{r}} H^{(1)}_{\nu-1/2}(\tilde{k}\tilde{r}) &= \tilde{k} \sqrt{\tilde{r}} H^{(1)}_{\nu+1/2}(\tilde{k}\tilde{r})\,,
\end{align}
where $H^{(1)}_\nu(z)=J_\nu(z) + i Y_\nu(z)$ is the Hankel function of first kind. Disregarding boundary conditions, it is then straightforward to see that the bulk eigenstates are given by (up to a normalization factor)
\begin{align}
    \label{eq:h_nu_eigenvalue_equation}
    \tilde{h}_\nu \,\underline{\Phi}_{\tilde{k}\nu}^{\rm bulk} &= \tilde{\epsilon} \,\underline{\Phi}_{\tilde{k}\nu}^{\rm bulk}\,,\\
    \label{eq:h_nu_bulk_states}
    \underline{\Phi}_{\tilde{k}\nu}^{\rm bulk}(\tilde{r}) &= \sqrt{\tilde{k}\tilde{r}} 
    \left(\begin{array}{c} c_{\tilde{k}} H^{(1)}_{\nu-1/2}(\tilde{k}\tilde{r}) \\
    H^{(1)}_{\nu+1/2}(\tilde{k}\tilde{r}) \end{array}\right)\,,
\end{align}
where $c_{\tilde{k}}$ is defined by
\begin{align}
    \label{eq:c_k}
    c_{\tilde{k}} = - \frac{2\tilde{k}}{\tilde{k}^2 - \tilde{\delta} - \tilde{\epsilon}}
    = \frac{\tilde{k}^2 - \tilde{\delta} + \tilde{\epsilon}}{2\tilde{k}}\,,
\end{align}
and the energy $\tilde{\epsilon}$ is related to $\tilde{k}=\tilde{k}(\tilde{\epsilon})$ by the dispersion relation 
\begin{align}
    \label{eq:dispersion_zero_B}
    \tilde{\epsilon}^2 = (\tilde{k}^2-\tilde{\delta})^2 + 4\tilde{k}^2\,.
\end{align}
For given energy $\tilde{\epsilon}$, there are four solutions for $\tilde{k}$ in the complex plane which we denote by $\tilde{k}_j(\tilde{\epsilon})$, with $j=1,2,3,4$. Since (\ref{eq:dispersion_zero_B}) depends only on $\tilde{k}^2$, we can choose $\tilde{k}_3 = -\tilde{k}_1$ and $\tilde{k}_4 = -\tilde{k}_2$, together with ${\rm Im}\tilde{k}_{1/2}\ge 0$. Furthermore, since $\tilde{\epsilon}$ is real, $\tilde{k}^*$ is also a solution of (\ref{eq:dispersion_zero_B}), such that we get either $\tilde{k}_{1/2}=-\tilde{k}_{2/1}^*$ or $\tilde{k}_{1/2}=-\tilde{k}_{1/2}^*$ (the special point $\tilde{k}_1=\tilde{k}_2$ is a bifurcation point which we disregard in the following). However, not all four solutions are allowed since the wave function should decay at large distance $\tilde{r}\gg 1$ if we consider edge states localized at the inner surface. Since the Hankel function has the asymptotic behavior 
\begin{align}
    \label{eq:hankel_nu_asymptotic}
    \sqrt{z}\,H^{(1)}_\nu(z) \xrightarrow{|z|\gg 1} \sqrt{\frac{2}{\pi}}\,\,e^{-i(\nu+1/2)(\pi/2)}\,e^{iz}\,,
\end{align}
this means that the imaginary part of $\tilde{k}$ must be strictly positive, i.e., the only allowed solutions are given by $\tilde{k}_{1/2}$. For $\tilde{\delta}<2$, we note that these solutions fulfill the useful properties 
\begin{align}
    \label{eq:product_k_12}
    \tilde{k}_1(\tilde{\epsilon}) \tilde{k}_2(\tilde{\epsilon}) &= - \sqrt{\tilde{\delta}^2 - \tilde{\epsilon}^2} \,,\\
    \label{eq:sum_k_12}
    \tilde{k}_1(\tilde{\epsilon}) + \tilde{k}_2(\tilde{\epsilon}) &= i \sqrt{2(2-\tilde{\delta}+\sqrt{\tilde{\delta}^2 - \tilde{\epsilon}^2})}\,,
\end{align}
where $|\tilde{\epsilon}|<|\tilde{\delta}|$ to guarantee that the energy of the edge state lies in the bulk gap set by $\tilde{\delta}$. For edge states localized at the outer surface only $\tilde{k}$-values with a strictly negative imaginary part are allowed, meaning that we have to replace $\tilde{k}_{1/2}\rightarrow -\tilde{k}_{1,2}$. 

For given $\nu$, the radial part of the edge state wave function $\underline{\Phi}_{\nu}^{\rm edge}(\tilde{r})$ with energy $\tilde{\epsilon}$ can be constructed by a linear combination of the two degenerate bulk eigenstates with $\tilde{k}=\tilde{k}_{1/2}(\tilde{\epsilon})$ 
\begin{align}
    \label{eq:edge_state_B=0}
    \underline{\Phi}_{\nu}^{\rm edge}(\tilde{r}) = \sum_{j=1,2} a_j\,\underline{\Phi}_{\tilde{k}_j(E),\nu}^{\rm bulk}(\tilde{r})\,,
\end{align}
such that the boundary condition at the hole surface is fulfilled
\begin{align}
    \label{eq:edge_state_bc}
    \underline{\Phi}_{\nu}^{\rm edge}(\tilde{R}) = 0 \,.
\end{align}
Inserting (\ref{eq:h_nu_bulk_states}) in (\ref{eq:edge_state_B=0}) we find that the boundary condition can only be fulfilled if
\begin{align}
    \label{eq:bc_condition_hankel}
    c_{\tilde{k}_1}\frac{H^{(1)}_{\nu-1/2}(\tilde{k}_1\tilde{R})}{H^{(1)}_{\nu+1/2}(\tilde{k}_1\tilde{R})} =
    c_{\tilde{k}_2}\frac{H^{(1)}_{\nu-1/2}(\tilde{k}_2\tilde{R})}{H^{(1)}_{\nu+1/2}(\tilde{k}_2\tilde{R})}\,,
\end{align}
and the ratio of the two coefficients is given by 
\begin{align}
    \label{eq:a_ratio}
    \frac{a_1}{a_2} =  
    -\frac{\sqrt{\tilde{k}_2\tilde{R}}\,H^{(1)}_{\nu+1/2}(\tilde{k}_2\tilde{R})} {\sqrt{\tilde{k}_1\tilde{R}}\,H^{(1)}_{\nu+1/2}(\tilde{k}_1\tilde{R})} \,.
\end{align}
With $\tilde{k}_{1/2}=\tilde{k}_{1/2}(\tilde{\epsilon})$, the condition (\ref{eq:bc_condition_hankel}) has either no solution or gives a certain value for the energy $\tilde{\epsilon}=\tilde{\epsilon}_\nu$ for a given index $\nu$. If a solution can be found, the total Hamiltonian (\ref{eq:H_zero_B}), given by $\bar{H}^{(0)}_{1/2,\nu}/E_{\rm so} = \sigma_x h_\nu$, has two solutions with energy $\pm \tilde{\epsilon}_\nu$. This provides two dispersions of edge states moving clockwise or anti-clockwise along the surface as function of the angular momentum $l=\nu-f+1/2$. In the following we will explicitly solve the condition (\ref{eq:bc_condition_hankel}) for very large and small radius $\tilde{R}$. 

{\it \underline{Large radius $\tilde{R}\gg 1$}}: For large radius we use the asymptotic form 
\begin{align}
    \label{eq:large_R_hankel_ratio}
    \frac{H^{(1)}_{\nu-1/2}(z)}{H^{(1)}_{\nu+1/2}(z)} \xrightarrow{|z|\gg 1} \frac{iz+\nu}{z}\,,
\end{align}
and find from (\ref{eq:bc_condition_hankel}) by using (\ref{eq:c_k})
\begin{align}
    \nonumber
    \tilde{\epsilon}_\nu &= \frac{\nu}{\tilde{R}} i(\tilde{k}_1 + \tilde{k}_2) - \tilde{\delta} - \tilde{k}_1 \tilde{k}_2 \,,\\
    \label{eq:E_nu_large_R_full}
    &= -\frac{\nu}{\tilde{R}}\sqrt{2(2-\tilde{\delta}+\sqrt{\tilde{\delta}^2 - \tilde{\epsilon}_\nu^2})} - \tilde{\delta}
    + \sqrt{\tilde{\delta}^2 - \tilde{\epsilon}_\nu^2}\,,
\end{align}
where we used (\ref{eq:product_k_12}) and (\ref{eq:sum_k_12}) in the second step. For $|\tilde{\epsilon}_\nu|<|\tilde{\delta}|$, this equation has only a solution for 
\begin{align}
    \label{eq:delta_condition}
    \tilde{\delta} > 0 \,,
\end{align}
and the energy is of order $\tilde{\epsilon}_\nu\sim O(1/\tilde{R})$. In leading order we obtain
\begin{align}
    \label{eq:E_nu_large_R}
    \tilde{\epsilon}_\nu = - \frac{2\nu}{\tilde{R}}\,.
\end{align}
This result is consistent with the one obtained in Section~\ref{sec:corbino_weak_fields} for the inner surface, see the first term of the effective surface Hamiltonian (\ref{eq:bar_H_t_corbino}). The same applies for edge states at the outer surface, where we have to replace $\tilde{k}_{1/2}\rightarrow -\tilde{k}_{1/2}$, giving rise to a sign change of the dispersion. 

The edge state wave function for large $\tilde{R}$ follows by using $\tilde{\epsilon}_\nu\approx 0$ in leading order (thereby neglecting corrections of $O(1/\tilde{R})$), and inserting the asymptotic form (\ref{eq:hankel_nu_asymptotic}) of the Hankel function in (\ref{eq:bc_condition_hankel}) and (\ref{eq:a_ratio}). This gives 
\begin{align}
    \label{eq:a_ratio_large_R}
    c_{\tilde{k}_1}=c_{\tilde{k}_2} \quad,\quad \frac{a_1}{a_2} = - e^{i(\tilde{k}_2-\tilde{k}_1)\tilde{R}}\,,
\end{align}
together with $\tilde{k}_{1/2}\approx\tilde{k}_{1/2}(0)$, given by
\begin{align}
    \label{eq:k_E=0}
    \tilde{k}_{1/2}(0) = i \pm \sqrt{\tilde{\delta}-1}\,,
\end{align}
where we define $\sqrt{\tilde{\delta}-1}=i\sqrt{1-\tilde{\delta}}$ for $0<\tilde{\delta}<1$. Since $\tilde{k}_{1/2}(0)^2-\delta=2i\tilde{k}_{1/2}(0)$, we find from (\ref{eq:c_k}) that 
\begin{align}
    \label{eq:c_k_large_R}
    c_{\tilde{k}_1} = c_{\tilde{k}_2} = i \,,
\end{align}
up to corrections of $O(1/\tilde{R})$. Using (\ref{eq:a_ratio_large_R}) and (\ref{eq:c_k_large_R}) in (\ref{eq:h_nu_bulk_states}) and (\ref{eq:edge_state_B=0}), we find for the radial part of the edge state wave function the $\nu$-independent result (up to a normalization factor)
\begin{align}
    \nonumber
    \underline{\Phi}_{\nu}^{\rm edge}(\tilde{r}) &= \underline{\bar{\psi}}^<_{n}(\tilde{r}) \\
    \label{eq:edge_state_inner_large_R}
    &\hspace{-1cm}
    \sim  \left(\begin{array}{c} 1 \\  -1\end{array}\right) 
    \left( e^{i\tilde{k}_1(0)(\tilde{r}-\tilde{R})} - e^{i\tilde{k}_2(0)(\tilde{r}-\tilde{R})} \right)\,,
\end{align}
which proves the result (\ref{eq:psi_n_corbino}) for the edge state at the inner surface after normalization. 

To get the edge state wave function for states localized at the outer surface we have to replace $\tilde{k}_{1/2}\rightarrow -\tilde{k}_{1/2}$ in all previous steps, leading to $c_{\tilde{k}_1}=c_{\tilde{k}_2}=-i$ instead of (\ref{eq:c_k_large_R}). This gives the result
\begin{align}
    \nonumber
    \underline{\Phi}_{\nu}^{\rm edge}(\tilde{r}) &= \underline{\bar{\psi}}^>_{n}(\tilde{r}) \\
    \label{eq:edge_state_outer_large_R}
    &\hspace{-1cm}
    \sim  \left(\begin{array}{c} 1 \\  1\end{array}\right) 
    \left( e^{-i\tilde{k}_1(0)(\tilde{r}-\tilde{R})} - e^{-i\tilde{k}_2(0)(\tilde{r}-\tilde{R})} \right)\,,
\end{align}
which proves (\ref{eq:psi_n_corbino}) for edge states at the outer surface.

{\it \underline{Small radius $\tilde{R}\ll 1$}}: For small hole radius we start with the case $|\nu|<1/2$ which is only possible for $l=0$ or $\nu=f-1/2$ (since we consider only fluxes with $0 < f <1$). In this case we get
\begin{align}
    \label{eq:small_R_hankel_ratio_l=0}
    \frac{H^{(1)}_{\nu-1/2}(z)}{H^{(1)}_{\nu+1/2}(z)} \xrightarrow{|z|\ll 1} 
    i e^{-i\nu\pi}\frac{\Gamma(1/2-\nu)}{\Gamma(1/2+\nu)} (z/2)^{2\nu}\,.
\end{align}
Using this form together with (\ref{eq:c_k}) in the condition (\ref{eq:bc_condition_hankel}), we find for the energy
\begin{align}
    \label{eq:E_l=0_small_R_1}
    \tilde{\epsilon}_\nu &= \tilde{\delta} - \frac{\tilde{k}_1^{2\nu+1}-\tilde{k}_2^{2\nu+1}}{\tilde{k}_1^{2\nu-1}-\tilde{k}_2^{2\nu-1}} \\
    \label{eq:E_l=0_small_R_2}
    &= \tilde{\delta} + \tilde{k}_1\tilde{k}_2 - (\tilde{k}_1 + \tilde{k}_2) \frac{\tilde{k}_1^{2\nu}-\tilde{k}_2^{2\nu}}{\tilde{k}_1^{2\nu-1}-\tilde{k}_2^{2\nu-1}} \,.
\end{align}
This equation is rather hard to solve when $\tilde{k}_{1/2}=\tilde{k}_{1/2}(\tilde{\epsilon}_\nu)$ depend significantly on $\tilde{\epsilon}$. Analytical results can be obtained in the limit of small $|\nu|=|f-1/2|\ll 1$, where we obtain
\begin{align}
    \label{eq:E_l=0_small_R_and_nu}
    \tilde{\epsilon}_\nu \approx \tilde{\delta} + \tilde{k}_1\tilde{k}_2 + 2\,\frac{\tilde{k}_1 + 
    \tilde{k}_2}{\tilde{k}_1 - \tilde{k}_2} \tilde{k}_1 \tilde{k}_2 \left(\ln\tilde{k}_1 - \ln\tilde{k}_2\right)\nu\,.
\end{align}
For the SUSY point $\nu=0$ we obtain
\begin{align}
    \label{eq:E_0_small_R}
    \tilde{\epsilon}_0 = \tilde{\delta} + \tilde{k}_1 \tilde{k}_2 = \tilde{\delta} - \sqrt{\tilde{\delta}^2 - \tilde{\epsilon}_0^2}\,,
\end{align}
where we used (\ref{eq:product_k_12}) in the last step. Since the solution $\tilde{\epsilon}_0=\tilde{\delta}$ is not possible (the energy is required to be in the gap), this equation has the unique solution $\tilde{\epsilon}_0 = 0$ if $\tilde{\delta}>0$. This means that $\tilde{\epsilon}_\nu\sim \nu$ has a linear slope in the deviation $\nu=f-1/2$ from half-integer flux. As a consequence, we can neglect the energy dependence of $\tilde{k}_{1/2}(E)$ in (\ref{eq:E_l=0_small_R_and_nu}) since the corrections are of quadratic order in $\tilde{\epsilon}$ and, therefore, also in $\nu$. For $\tilde{\epsilon}=0$ we insert (\ref{eq:k_E=0}) in (\ref{eq:E_l=0_small_R_and_nu}) and obtain the following dispersion $\tilde{\epsilon}(f)\equiv \tilde{\epsilon}_{f-1/2}$ for the center state with angular momentum $l=0$ as function of the flux $f$
\begin{align}
    \label{eq:E_f_small_R_and_nu}
    \tilde{\epsilon}(f) = - \frac{4\tilde{\delta}(f-1/2)}{\sqrt{|\tilde{\delta}-1}|}\,\,
    \begin{cases}
    \arctan\sqrt{\tilde{\delta}-1} & {\rm for}\quad \tilde{\delta} > 1 \\
    \frac{1}{2}\,\ln{\frac{1+\sqrt{\tilde{\delta}-1}}{1-\sqrt{\tilde{\delta}-1}}} & {\rm for}\quad 0<\tilde{\delta} < 1
    \end{cases}\,,
\end{align}
which proves Eq.~(\ref{eq:E_f_small_R_l=0}) of the main text. 

For the other angular momenta $|l|>0$ we have $|\nu|>1/2$ and use 
\begin{align}
    \label{eq:small_R_hankel_ratio_l_finite}
    \frac{H^{(1)}_{\nu-1/2}(z)}{H^{(1)}_{\nu+1/2}(z)} \xrightarrow{|z|\ll 1} 
    \begin{cases}
    z/(2\nu-1) & {\rm for} \quad \nu > 1/2 \\
    (2\nu+1)/z & {\rm for} \quad \nu < -1/2 
    \end{cases}\,.
\end{align}
After inserting this form together with (\ref{eq:c_k}) into the condition (\ref{eq:bc_condition_hankel}) we find that a solution can not be found for the energy. 

As a result, for small radius, only the center state with $l=0$ survives with energy $\tilde{\epsilon}(f)$ given by (\ref{eq:E_f_small_R_and_nu}) for small $|f-1/2|\ll 1$. The reason is that this state starts at zero energy at the SUSY point $f=1/2$ at any radius (see below) and will get a small shift when deviating from half-integer flux. In contrast, the other edge states with finite angular momentum already start at a finite energy at the SUSY point and get strongly shifted by decreasing $\tilde{R}$ such that they move out of the gap for small radius.

\vspace{1cm}

\section{Continuum numerics for a disc}
\label{app:disc_continuum_numerics}

In this Appendix we will outline the continuum numerics for a disk with radius $R$ and no hole (i.e., $R_>=R$ and $R_<=0$). We start from the Hamiltonian (\ref{eq:H_flux}) and choose a complete basis of states which diagonalizes the kinetic term ${\bf p}_K^2/(2m)$. In polar coordinates we use
\begin{align}
    \label{eq:p2_polar}
    {\bf p}_K^2 = - \frac{1}{r}\partial_r r \partial_r + \frac{1}{r^2}(-i\partial_\varphi + f)^2\,,
\end{align}
and decompose the eigenfunctions in a radial and angular part
\begin{align}
    \label{eq:ef_p2}
    \frac{{\bf p}_K^2}{2m} \, \psi_{l\beta}(r,\varphi) &= \epsilon_{l\beta} \, \psi_{l\beta}(r,\varphi)\,,\\
    \psi_{l\beta}(r,\varphi) &= u_{l\beta}(r) \, \frac{1}{\sqrt{2\pi}}e^{il\varphi}  \,,
\end{align}
where $l=0,\pm 1,\pm 2,\dots$ denotes the angular momentum and $\beta=1,2,\dots$ numerates the radial states. Using the dimensionless variable $z=\sqrt{2m\epsilon_{l\beta}}\,r$ we find that $u_{l\beta}(r)$ fulfils the differential equation for the Bessel functions
\begin{align}
    \label{eq:diff_u}
    \left[\partial_z^2 + \frac{1}{z}\partial_z + (1-\frac{(l+f)^2}{z^2})\right] u_{l\beta} = 0 \,.
\end{align}
Since $\psi_{l\beta}(r,\varphi)$ should be a well-defined wave function at $r=0$, we need the following boundary condition at $r=0$ 
\begin{align}
    \label{eq:bc_r=0}
    u_{l\beta}(r) \xrightarrow{r\rightarrow 0} 
    \begin{cases} {\rm finite\,\, value} & {\rm for} \quad l=0 \\ 0 & {\rm for}\quad l\ne 0 \end{cases} \,.
\end{align}
Furthermore, at $r=R$, the wave function should vanish
\begin{align}
    \label{eq:bc_r=R}
    u_{l\beta}(R) = 0 \,.
\end{align}
As a consequence, the normalized radial part and the energy eigenvalues are given by
\begin{align}
    \label{eq:u}
    u_{l\beta}(r) &= \frac{\sqrt{2}}{R}\,\frac{J_{|l+f|}(z_{|l+f|,\beta}\,r/R)}{|J_{|l+f|+1}(z_{|l+f|,\beta})|}\,,\\
    \label{eq:energy}
    \epsilon_{l\beta} &= \frac{z_{|l+f|,\beta}^2}{2m R^2}\,,
\end{align}
where $J_\nu(z)$ are the Bessel functions and $z_{\nu\beta}>0$ numerates the positive zero's of $J_\nu(z_{\nu,\beta})=0$ with $\beta=1,2,\dots$ in ascending order, i.e., $0 < z_{\nu,1} < z_{\nu,2} < \dots$.

In the basis of the states $\psi_{l\beta}$ we now state the matrix elements $\langle \psi_{l\beta}|{\bf p}_K|\psi_{l'\beta'}\rangle$ of the kinetic momentum operator ${\bf p}_K=\left(\begin{array}{c} p_{K,x} \\ p_{K,y}\end{array}\right)$ in order to determine the matrix elements of the spin-orbit interaction. A lengthy but straightforward calculation gives the result
\begin{align}
    \nonumber
    \langle \psi_{l\beta}|{\bf p}_K|\psi_{l'\beta'}\rangle &= \\
    \label{eq:matrix_element_pK}
    &\hspace{-2cm}
    = -\frac{1}{R} \left(\begin{array}{c} -i(\delta_{l,l'-1}+\delta_{l,l'+1}) \\ \delta_{l,l'-1}-\delta_{l,l'+1}\end{array}\right)
    P_{l\beta,l'\beta'}\,,
\end{align}
where, for $|l+f|+|l'+f|>1$, we get
\begin{align}
    \label{eq:P_1}
    P_{l\beta,l'\beta'} = (-1)^{\beta+\beta'}\,\frac{z_{|l+f|,\beta}\,z_{|l'+f|,\beta'}}{z_{|l+f|,\beta}^2 - z_{|l'+f|,\beta'}^2}\,,  
\end{align}
and for $|l+f|+|l'+f|=1$ (which, for $0<f<1$, can only happen for $l=0, l'=-1$ or $l=-1, l'=0$) we define
\begin{align}
   \nonumber
    P_{l\beta,l'\beta'} &= (-1)^{\beta+\beta'}\,\frac{z_{|l+f|,\beta}\,z_{|l'+f|,\beta'}}{z_{|l+f|,\beta}^2 - z_{|l'+f|,\beta'}^2} \\
    \label{eq:P_2}
    &\hspace{-1cm}
    \times \left\{1 - \frac{2}{\pi}\sin(\pi f) \frac{(z_{|l+f|,\beta})^{|l+f|-1}\,(z_{|l'+f|,\beta'})^{|l'+f|-1}}
    {J_{|l+f|+1}(z_{|l+f|,\beta}) \, J_{|l'+f|+1}(z_{|l'+f|,\beta'})} \right\}\,.
\end{align}
We note that, for the special case $f=1/2$, the latter matrix element can be evaluated as (for both $l=0,l'=-1$ or $l=-1,l'=0$)
\begin{align}
    \label{eq:P_2_f=1/2}
    P_{l\beta,l'\beta'} = \frac{\beta\beta'}{\beta^2-\beta'^2}\,\left((-1)^{\beta+\beta'}-1\right)\,.
\end{align}

Based on the matrix elements of ${\bf p}_K$ it is then straightforward to evaluate the matrix elements of the Hamiltonian (\ref{eq:H_flux}) which, in dimensionless units, read (with $\sigma=\pm 1$ and $s=\pm 1$ denoting the eigenvalues of $\sigma_z$ and $s_z$, respectively)
\begin{align}
     \nonumber
    \langle \psi_{l\beta},\sigma,s|H_f|\psi_{l'\beta'},\sigma',s'\rangle/E_{\rm so} &= \\
    \nonumber
    & \hspace{-3cm}
    = \delta_{ll'}\delta_{\beta\beta'}\delta_{\sigma\sigma'}\delta_{ss'}\sigma (\frac{z_{|l+f|,\beta}^2}{\tilde{R}^2}-\tilde{\delta})\\
    \nonumber
    & \hspace{-2.5cm}
    + \frac{4i}{\tilde{R}}\,\delta_{l,l'-s}\delta_{\sigma,-\sigma'}\delta_{s,-s'} P_{l\beta,l'\beta'}\\
    \label{eq:matrix_elements_H}
    & \hspace{-2.5cm}
    - \frac{i}{\tilde{l}_B^2}\,\delta_{ll'}\delta_{\beta\beta'}\delta_{\sigma\sigma'}\delta_{s,-s'}s\,.
\end{align}

For the numerical implementation we take a cutoff $N_L$ for the allowed values of the angular momentum in the following way
\begin{align}
    \label{eq:l_cutoff}
    - N_L +1 \le l \le N_L \,.
\end{align}
We have chosen it in the prescribed way to ensure that the SUSY operator $U_{1/2}$ at $f=1/2$ stays in the space of chosen states. This follows from the matrix elements (valid only for $f=1/2$)
\begin{align}
    \label{eq:SUSY_matrix_element}
    \langle \psi_{l\beta}|e^{-i\varphi}P_\varphi|\psi_{l'\beta'}\rangle = \delta_{l+l'+1,0}\delta_{\beta\beta'}\,.
\end{align}
In the chosen way, the Hamiltonian matrix fulfils exactly all symmetries for any cutoff (i.e., chiral and inversion symmetry, and, for the special case $f=1/2$, the supersymmetry). 

In addition, we have to choose a cutoff $N_Z$ for the number of zero's of the Bessel function, i.e., $\beta=1,2,\dots,N_Z$. Whereas convergence is quickly reached in the cutoff $N_L$ for the angular momentum, the cutoff $N_Z$ has to be taken rather large for increasing $\tilde{R}$ such that $N_Z\sim\tilde{R}$ to get values of the energy $z_{|l+f|,\beta}^2/{\tilde{R}^2}$ beyond $\tilde{\delta}$. This restricts the numerical implementation to values $\tilde{R}\sim 30-40$ for a good efficiency. 

We note that the matrix elements $P_{l\beta,l'\beta'}$ can be of order $O(\tilde{R})$ for $z_{|l+f|,\beta},z_{|l'+f|,\beta'}\sim O(\tilde{R})$ and $|z_{|l+f|,\beta}-z_{|l'+f|,\beta'}|\sim O(1)$. Therefore, the spin-orbit interaction is not $\sim 1/\tilde{R}$ as one might conclude from (\ref{eq:matrix_elements_H}) but has a weight $\sim O(1)$. This makes analytical treatments as function of the cutoff $N_Z$ rather difficult and requires a careful study of convergence in numerics. The data shown in this paper were calculated with cutoffs $N_L=10$ and $N_Z=150$ and it was verified that they were already converged with respect to both cutoffs.

\section{Tight-binding numerics}
\label{app:tb_numerics}

To set up the numerics on a discrete two-dimensional quadratic lattice with lattice spacing $a$, we first consider zero flux $f=0$ and replace in the low-energy regime the momentum dependent terms in the Hamiltonian $H_0$ by
\begin{align}
    \label{eq:p2_lattice}
    p^2 &\rightarrow \int_{1.\rm{B.Z}}d{\bf k} |{\bf k}\rangle\langle{\bf k}|\left\{-\frac{2}{a^2}\left(\cos{k_x} + \cos{k_y}\right) + \frac{4}{a^2}\right\}\,,\\
    \label{eq:so_lattice}
    \bf{s}\cdot\bf{p} &\rightarrow \int_{1.\rm{B.Z}}d{\bf k} |{\bf k}\rangle\langle{\bf k}|\frac{1}{a}\left(s_x\sin{k_x} + s_y\sin{k_y} \right)\,,
\end{align}
where ${\bf k}=(k_x,k_y)$ denotes the dimensionless quasimomentum vector with $-\pi < k_{x,y} < \pi$ defining the first Brioullin zone ($1.$ BZ.). Here, we have used the plane wave states
\begin{align}
    |k_{x,y}\rangle = \frac{1}{\sqrt{2\pi}} \sum_{n_{x,y}} e^{i k_{x,y} n_{x,y}} |n_{x,y}\rangle\,,
\end{align}
where $|{\bf n}\rangle$ denotes the state with an electron on lattice site ${\bf n}=(n_x,n_y)$, with $n_{x,y}=0,\pm 1,\pm 2, \dots$. 

To obtain the tight-binding Hamiltonian in real space representation we use
\begin{align}
    \sum_{\bf n} |{\bf n}+{\bf e}_{x,y}\rangle\langle {\bf n}| + \rm{h.c.} &= 2 \int_{1.\rm{B.Z}}d{\bf k} \cos(k_{x,y})|{\bf k}\rangle\langle{\bf k}|\,,\\
    i\sum_{\bf n} |{\bf n}+{\bf e}_{x,y}\rangle\langle {\bf n}| + \rm{h.c.} &= 2\int_{1.\rm{B.Z}}d{\bf k} \sin(k_{x,y})|{\bf k}\rangle\langle{\bf k}|\,,\\\end{align}
where ${\bf e}_x=(1,0)$ and ${\bf e}_y=(0,1)$. Inserting these relationships in (\ref{eq:p2_lattice}) and (\ref{eq:so_lattice}), we find for the Hamiltonian (\ref{eq:H_flux}) at zero flux the nearest-neighbor tight-binding model
\begin{align}
    H_0 &= -t \sigma_z \sum_{\bf n}\sum_{i=x,y}\left(|{\bf n}+{\bf e}_i\rangle\langle {\bf n}|+\rm{h.c.}\right) \\
    &+\frac{i\alpha}{2a}\sigma_x\sum_{\bf n}\sum_{i=x,y}\left(s_i|{\bf n}+{\bf e}_i\rangle\langle {\bf n}| -\rm{h.c.}\right)\\
    &+(4t-\delta)\sigma_z +\frac{1}{2}g\mu_B{\bf B}\cdot{\bf s}\,,
\end{align}
with $t=1/(2m^* a^2)$. For the case of strong spin-orbit interaction as considered in the present work, the spin-orbit length $\lambda_{\rm so}$ is the smallest length scale in the continuum model. Therefore, we choose the lattice spacing $a$ in the discretized model of the order of the spin-orbit length $a\sim\lambda_{so}\sim 1/(\alpha m^*)$, which gives 
\begin{align}
    \label{eq:hopping}
    t=1/(2m^* a^2)\sim E_{\rm so}\,.
\end{align}

In the presence of a finite flux $f$ we use
\begin{align}
    H_f = e^{-if\varphi} H_0 e^{if\varphi}\,,
\end{align}
which leads to the following change of the tight-binding Hamiltonian
\begin{align}
    |{\bf n}\rangle\langle {\bf m}|\rightarrow
    e^{-if\Delta\varphi_{{\bf n},{\bf m}}}|{\bf n}\rangle\langle {\bf m}|\,,
\end{align}
where $\Delta\varphi_{{\bf n},{\bf m}}=\varphi_{\bf n}-\varphi_{\bf m}\,\,{\rm mod}(2\pi)$ and $0\le\varphi_{\bf n}<2\pi$ denotes the polar angle (measured relative to the positive $x$-axis in anti-clockwise direction) of the lattice site at position ${\bf x}=a(n_x,n_y)$ (hereby, we assume that no lattice site is located on the $x$-axis). In order to get the total phase factor $e^{-i2\pi f}$ when hopping in a closed loop anti-clockwise around the origin we have to choose the ${\rm mod} (2\pi)$-contribution equal to $2\pi$ ($-2\pi$), when the hopping occurs in $y$-direction at $x>0$ in positive (negative) $y$-direction. With this definition it is guaranteed that $|\Delta\varphi_{{\bf n},{\bf m}}|<\pi$ is always the smallest possible one. 

We note that the tight-binding model fulfils exactly the same symmetries (inversion, chiral, SUSY) as the continuum model. For inversion symmetry it is essential that the shape of the lattice is symmetric under ${\bf x}\rightarrow -{\bf x}$ and for SUSY we need half-integer flux $f=1/2$ and a shape which is symmetric under both $x\rightarrow -x$ or $y\rightarrow -y$.

\vspace{1cm}

\section{Witten potential and zero energy wave functions for peanut shape}
\label{app:witten_potential_peanut}

In this Appendix we derive the formulas needed to calculate the Witten potential $\tilde{V}_W^+$ and the zero energy wave functions $\psi_W^{(0)\pm}$ for a surface of peanut shape. From the construction shown in Fig.~\ref{fig:peanut_construction} we see that the peanut shape is completely determined by the three parameters $R_1$, $R_2$ and $x_0$. The point with $\tilde{s}_t=\tilde{R}_2\gamma$ determines the position on the surface where the curvature changes discontinuously from $\tilde{\kappa}_t=-1/\tilde{R}_2$ to $\tilde{\kappa}_t=1/\tilde{R}_1$. This gives rise to discontinuities in the Witten potential which can be washed out and do not influence the low-lying states too much if the radia $\tilde{R}_{1,2}\ll 1$ are large enough. The angle $\gamma$ is determined from
\begin{align}
    \label{eq:gamma_peanut}
    \cos{\gamma} = \frac{x_0 + R_2}{R_1 + R_2}\,.
\end{align}
The Witten potential follows from the formulas
\begin{align}
    \label{eq:witten_potential_peanut}
    \tilde{V}_W^+(\tilde{s}_t) &= \tilde{E}_{Z,n}(\tilde{s}_t)^2 - 2 \tilde{E}_{Z,n}^\prime(\tilde{s}_t) \,,\\
    \label{eq:witten_Zeeman_normal_peanut}
    \tilde{E}_{Z,n}(\tilde{s}_t) &= \frac{1}{\tilde{l}_B^2}\sin{\vartheta(\tilde{s}_t)} \,,\\
    \label{eq:witten_Zeeman_normal_der_peanut}
    \tilde{E}_{Z,n}^\prime(\tilde{s}_t) &= - 2\frac{\tilde{\kappa}_t(\tilde{s}_t)}{\tilde{l}_B^2}\cos{\vartheta(\tilde{s}_t)}\,,
\end{align}
where the angle $\vartheta(\tilde{s}_t)$ of the normal vector ${\bf e}_n$ with the $x$-axis and the curvature $\tilde{\kappa}_t(\tilde{s}_t)=\frac{d}{d\tilde{s}_t}\vartheta(\tilde{s}_t)$ fulfill the symmetries
\begin{align}
    \label{eq:vartheta_symmetries_peanut}
    \vartheta(\tilde{s}_t) &= \pi - \vartheta(\tilde{L}/2 - \tilde{s}_t) = - \vartheta(\tilde{L}-\tilde{s}_t)\,,\\
    \label{eq:kappa_symmetries_peanut}
    \tilde{\kappa}_t(\tilde{s}_t) &= \tilde{\kappa}_t(\tilde{L}/2 - \tilde{s}_t) = \tilde{\kappa}_t(\tilde{L}-\tilde{s}_t)\,,
\end{align}
which lead to the relations 
\begin{align}
    \label{eq:Zeeman_normal_symmetries_peanut}
    \tilde{E}_{Z,n}(\tilde{s}_t) &= \tilde{E}_{Z,n}(\tilde{L}/2-\tilde{s}_t) = - \tilde{E}_{Z,n}(\tilde{L}-\tilde{s}_t)\,,\\
    \label{eq:Zeeman_normal_der_symmetries_peanut}
    \tilde{E}_{Z,n}^\prime(\tilde{s}_t) &= - \tilde{E}_{Z,n}^\prime(\tilde{L}/2-\tilde{s}_t) = \tilde{E}_{Z,n}^\prime(\tilde{L}-\tilde{s}_t)\,.
\end{align}
Therefore, it is sufficient to know $\vartheta(\tilde{s}_t)$ and $\tilde{\kappa}_t(\tilde{s}_t)$ in the regime $0<\tilde{s}_t<\tilde{L}/4$. For the total length $\tilde{L}$ we obtain
\begin{align}
    \label{eq:L_peanut}
    \tilde{L} = 4(\tilde{R}_1 + \tilde{R}_2)\gamma + 2\pi \tilde{R}_1 \,.
\end{align}
For $0<\tilde{s}_t<\tilde{L}/4$, a straightforward analysis gives for $\vartheta(\tilde{s}_t)$ the result 
\begin{align}
    \nonumber
    \vartheta(\tilde{s}_t) & \\
    \label{eq:vartheta_peanut}
    &\hspace{-0.5cm}
    = \begin{cases} - \tilde{s}_t/\tilde{R}_2 & {\rm for}\quad 0 < \tilde{s}_t < \tilde{R}_2 \,\gamma \\
    \left[\tilde{s}_t - (\tilde{R}_1 + \tilde{R}_2)\gamma\right]/\tilde{R}_1 & {\rm for} 
    \quad \tilde{R}_2\,\gamma < \tilde{s}_t < \tilde{L}/4    \end{cases} \,,
\end{align}
and from the derivative we find for the curvature
\begin{align}
    \label{eq:kappa_t_peanut}
    \tilde{\kappa}_t(\tilde{s}_t) =
    \begin{cases} - 1/\tilde{R}_2 & {\rm for}\quad 0 < \tilde{s}_t < \tilde{R}_2 \,\gamma \\
    1/\tilde{R}_1 & {\rm for} 
    \quad \tilde{R}_2\,\gamma < \tilde{s}_t < \tilde{L}/4    \end{cases} \,.
\end{align}

From (\ref{eq:vartheta_peanut}), (\ref{eq:witten_Zeeman_normal_peanut}) and (\ref{eq:F_witten}), we can also calculate the function $F(\tilde{s}_t)$ entering the zero energy wave functions $\psi_W^{(0)\pm}(\tilde{s}_t)$ of the Witten Hamiltonian via (\ref{eq:zero_energy_state_W}). Using the symmetries (\ref{eq:F_reflection_symmetry}) and (\ref{eq:F_varphi_symmetry}), we get
\begin{align}
    \label{eq:F_symmetries}
    F(\tilde{s}_t) = - F(\tilde{L}/2 - \tilde{s}_t) = F(\tilde{L} - \tilde{s}_t) \,.
\end{align}
Therefore, it is sufficient to calculate $F(\tilde{s}_t)$ for $0 < \tilde{s}_t < \tilde{L}/4$. Using $R(\tilde{s}_t)=-R(-\tilde{s}_t)$, we can replace the lower integration limit in (\ref{eq:F_witten}) by $\tilde{L}/4$ and find
\begin{align}
    \label{eq:F_witten_L/4}
    F(\tilde{s}_t) = -\frac{1}{2}\int_{\tilde{s}_t}^{\tilde{L}/4} d\tilde{s}_t' \,\tilde{E}_{Z,n}(\tilde{s}_t')\,.
\end{align}
Therefore, for $0 < \tilde{s}_t < \tilde{L}/4$, we need the function $\tilde{E}_{Z,n}(\tilde{s}_t)$ only in the same regime. Inserting (\ref{eq:witten_Zeeman_normal_peanut}) and (\ref{eq:vartheta_peanut}) in (\ref{eq:F_witten_L/4}) we find for $0<\tilde{s}_t<R_2\,\gamma$
\begin{align}
    \label{eq:F_peanut_first_regime}
    F(\tilde{s}_t) = \frac{1}{2\tilde{l}_B^2} \left[ \tilde{R}_2 \cos\left(\frac{\tilde{s}_t}{\tilde{R}_2}\right)
    - (\tilde{R}_1 + \tilde{R}_2) \cos{\gamma} \right] \,,
\end{align}
and for $R_2\,\gamma < \tilde{s}_t < \tilde{L}/4$
\begin{align}
    \label{eq:F_peanut_second_regime}
    F(\tilde{s}_t) = -\frac{1}{2\tilde{l}_B^2} \tilde{R}_1 
    \cos\left(\frac{\tilde{s}_t-(\tilde{R}_1 + \tilde{R}_2)\gamma}{\tilde{R}_1}\right) \,.
\end{align}

\end{appendix}

\bibliography{citations}

\begin{thebibliography}{53}
\bibitem [{com()}]{com_1}%
  \BibitemOpen
  \href@noop {} {\ }\bibinfo {note} {\hspace{-0.125cm}We use here the standard
  nomenclature of SUSY-representations, see, e.g.,
  Ref.~\cite{combescure_etal_jpa_04}. Here, $n=r$ ($N=r$) is a realization with
  $r$ Hermitian (non-Hermitian) supercharge operators together with an
  involution, whereas $m=r$ ($M=r$) is a realization with $r$ Hermitian
  (non-Hermitian) supercharge operators without an involution.}\BibitemShut
  {Stop}%
\end{thebibliography}

\end{document}


\title{Second-order topology and supersymmetry in two-dimensional topological insulators: Supplemental material}

\author{Clara S. Weber}
\affiliation{Institut f\"ur Theorie der Statistischen Physik, RWTH Aachen, 
52056 Aachen, Germany and JARA - Fundamentals of Future Information Technology}
\affiliation{Department of Physics and Astronomy, University of Pennsylvania, Philadelphia, Pennsylvania 19104, USA}
\author{Mikhail Pletyukhov}
\affiliation{Institut f\"ur Theorie der Statistischen Physik, RWTH Aachen, 
52056 Aachen, Germany and JARA - Fundamentals of Future Information Technology}
\author{Zhe Hou}
\affiliation{Department of Physics, University of Basel, Klingelbergstrasse 82, 
CH-4056 Basel, Switzerland}
\author{Dante M. Kennes}
\affiliation{Institut f\"ur Theorie der Statistischen Physik, RWTH Aachen, 
52056 Aachen, Germany and JARA - Fundamentals of Future Information Technology}
\affiliation{Max Planck Institute for the Structure and Dynamics of Matter, Center for Free Electron Laser Science, 22761 Hamburg, Germany}
\author{Jelena Klinovaja}
\affiliation{Department of Physics, University of Basel, Klingelbergstrasse 82, 
CH-4056 Basel, Switzerland}
\author{Daniel Loss}
\affiliation{Department of Physics, University of Basel, Klingelbergstrasse 82, 
CH-4056 Basel, Switzerland}
\author{Herbert Schoeller}
\email[Email: ]{schoeller@physik.rwth-aachen.de}
\affiliation{Institut f\"ur Theorie der Statistischen Physik, RWTH Aachen, 
52056 Aachen, Germany and JARA - Fundamentals of Future Information Technology}


\maketitle

\section{Weyl phase for small magnetic fields}
\label{sec:weyl_weak_fields}

\begin{figure}
	 \includegraphics[width =0.95\columnwidth]{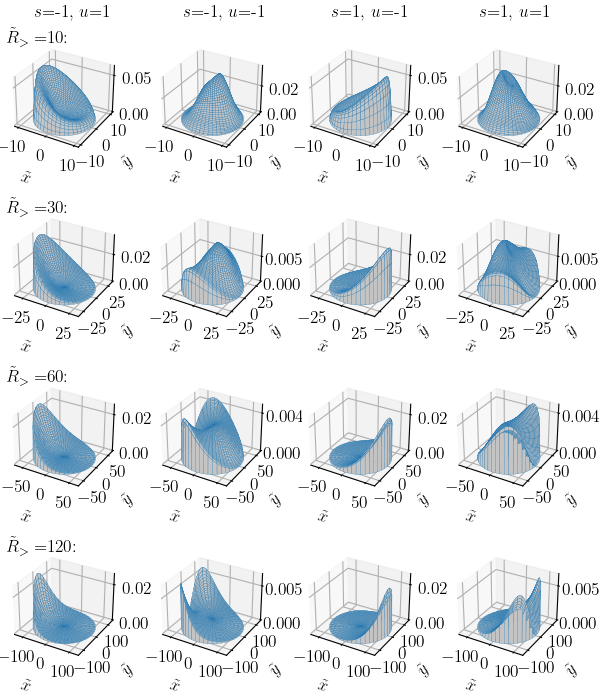}
	  \caption{The wave functions $\sum_{\sigma_z,s_z}|\bar{\psi}_{su}(x,y;\sigma_z,s_z)|^2$ as function of $\tilde{x}=x/\lambda_{\rm so}$ and $\tilde{y}=y/\lambda_{\rm so}$ of the four states with lowest absolute value of the energy for a disc with various radius $\tilde{R}_>$ and zero hole radius $\tilde{R}_< =0$ at half-integer flux $f=1/2$ in the Weyl regime $\tilde{\delta}=0$ and $\tilde{l}_B = 5$. As one can see two of the center states with $s=u=\pm 1$ move to the outer surface when $\tilde{R}_>$ is sufficiently large such that $\sqrt{\tilde{R}_>}$ exceeds significantly $\tilde{l}_B$.
	  } 
    \label{fig:weyl_weak_fields}
\end{figure}

To substantiate the analytical result that topological hole states will not exist for a Corbino disc in the Weyl phase by numerical results, we show in Fig.~\ref{fig:weyl_weak_fields} the four wave functions of lowest absolute value of the energy for a disc in the Weyl phase for increasing value of the radius of the disc. For small radius it seems as if a center state is present, but this state moves to the outer surface if the square root of the radius of the disc exceeds significantly the magnetic length. This is consistent with the analytical theory which predicts an additional topological state at the outer surface in the regime of strong localization $\tilde{l}_B \ll \sqrt{R_>}$.

\section{Topological states for Corbino disc: Comparison between numerics and analytics}
\label{sec:corbino_comparison}

Here we compare the analytical and numerical results for the topological states of a Corbino disc in the topological phase. There are four topological states $\bar{\psi}_{su}$ in the transformed basis, labelled by chiral symmetry $s=\pm 1$ and SUSY $u=\pm 1$, two at the outer surface ($s=-u=\pm 1$) and two at the inner surface ($s=u=\pm 1$). Since $\bar{\psi}_{su}$ and $\bar{\psi}_{-s,-u}$ are related by the inversion symmetry, we concentrate on the two states with chirality $s=+1$, localized around the polar angle $\varphi=0$. 

The analytical formulas for $\bar{\psi}_{1,\pm 1}(\tilde{r},\varphi;\sigma_z,s_z$ have been provided in Section IV.E of the main text as function of the radial coordinate $\tilde{r}=r/\lambda_{\rm so}$, the polar angle $\varphi$, and the spinors $\sigma_z$ and $s_z$. Omitting the normalization factor, we obtained
\begin{align}
    \nonumber
    \bar{\psi}_{1,\pm 1}(\tilde{r},\varphi;\sigma_z,s_z) &\sim \\ 
    \label{eq:TP_states_s=1}
    &\hspace{-2cm}
    \sim\left(\begin{array}{c} 0 \\ 1 \end{array}\right)_{\sigma_z}\,
    \left(\begin{array}{c} 1 \\ \mp 1 \end{array}\right)_{s_z} 
    \,\hat{\chi}_{+,n}^{(\pm 1)}(\tilde{r})\,\,e^{-\frac{1}{4}(\varphi/\Delta\varphi_\lessgtr)^2}\,,
\end{align}
with $\Delta\varphi_\gtrless = \tilde{l}_B/\sqrt{\tilde{R}_\gtrless}$,
\begin{align}
    \label{eq:chi}
    \hat{\chi}^{(\pm 1)}_{+,n}(\tilde{r}) \sim e^{i \tilde{k}^{(\pm 1)}_1 (\tilde{r}-\tilde{R}_\lessgtr)} - e^{i \tilde{k}^{(\pm 1)}_2 (\tilde{r}-\tilde{R}_\lessgtr)}\,, 
\end{align}
and 
\begin{align}
     \label{eq:k_u}
     \tilde{k}^{(u)}_{1/2} = iu \pm \sqrt{|\tilde{\delta}-1-u/\tilde{l}_B^2|}\,
    \begin{cases} 1 & {\rm for}\,\, \tilde{\delta}>1+u/\tilde{l}_B^2 \\
    iu & {\rm for}\,\, \tilde{\delta}<1+u/\tilde{l}_B^2 \end{cases}\,.
\end{align}

\begin{figure*}
	 \includegraphics[width =0.95\columnwidth]{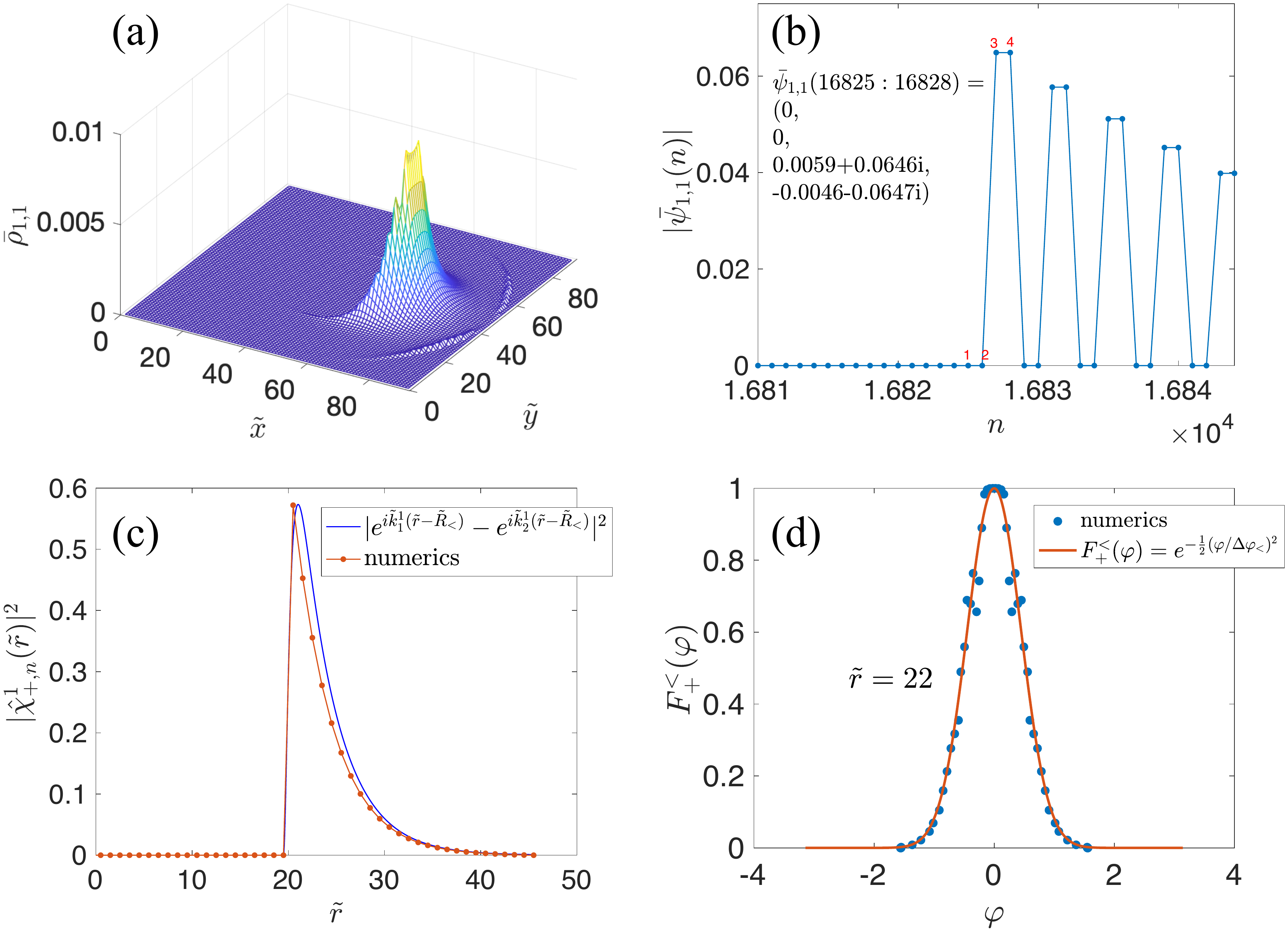}
	 \hfill\includegraphics[width =0.95\columnwidth]{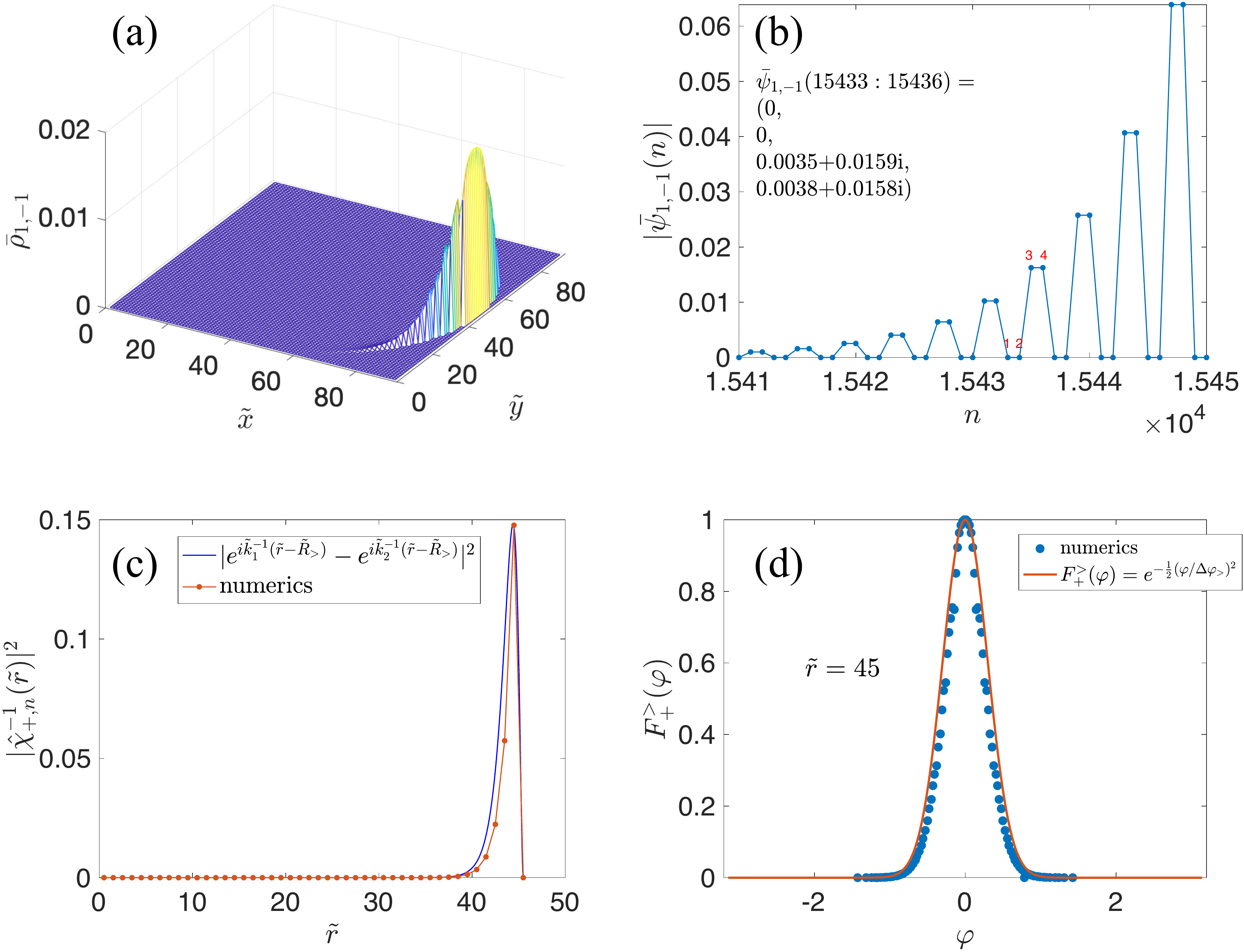}
	  \caption{Left/right part: The comparison between the analytical form (\ref{eq:TP_states_s=1}) and the tight-binding numerics for the inner/outer topological state $\bar{\psi}_{1,\pm 1}$. (a) $\bar{\rho}_{1,\pm 1}$ as defined in (\ref{eq:rho}) as function of the lattice sites labelled by $(\tilde{x},\tilde{y})$. (b) The absolute value of the wave function $|\bar{\psi}_{1,\pm 1}(n)|$ with the index $n$ accounting for both the position and spinor component. Here, the labelling $(1,2,3,4)$ corresponds to $(\sigma_z,s_z)=(1,1),(1,-1),(-1,1),(-1,-1)$. (c) The fitting for the radial dependence  $|\hat{\chi}_{+,n}^{(\pm 1)}(\tilde{r})|$ between the analytical form (\ref{eq:chi}) and the tight-binding result, where the polar angle is fixed to $\varphi=0$. (d) The fitting for the angular dependence $F_{+}^<(\varphi)$ between the analytical result (\ref{eq:F}) and the tight-binding result, where the radial coordinate is fixed to $\tilde{r}=22/45$. The other parameters in the tight-binding calculation are: $\tilde{R}_>=45$, $\tilde{R}_<=20$, $\tilde{\delta}=0.5$, $\tilde{l}_B=2$, and $f=1/2$.
	  } 
    \label{fig:comparison}
\end{figure*}

Using the tight-binding approach for a Corbino disc with $\tilde{R}_> = 45$ and $\tilde{R}_< = 20$, we show in the left/right parts of Fig.~\ref{fig:comparison}(a) the absolute value squared of the two topological states averaged over the spinor degrees of freedom
\begin{align}
    \label{eq:rho}
    \bar{\rho}_{1,\pm 1}(\tilde{x},\tilde{y}) &= \sum_{\sigma_z,s_z} |\bar{\psi}_{1,\pm 1}(\tilde{x},\tilde{y};\sigma_z,s_z)|^2 \\
    \label{eq:rho_explicit}
    &\sim  |\hat{\chi}_{+,n}^{(\pm 1)}(\tilde{r})|^2\,\,F_+^\lessgtr(\varphi) \,,
\end{align}
where $(\tilde{x},\tilde{y})$ are the coordinates of the lattice sites, and 
\begin{align}
    \label{eq:F}
    F_+^\lessgtr(\varphi) \sim e^{-\frac{1}{2}(\varphi/\Delta\varphi_\lessgtr)^2}\,. 
\end{align}
Fixing either the polar angle to $\varphi=0$ or the radial coordinate to $\tilde{r}=22/45$ for the inner/outer state, we find in Fig.~\ref{fig:comparison}(c,d) a very good agreement between the numerical and analytical results for both the radial and angular dependence of the wave functions. To check the spinor dependence we show in Fig.~\ref{fig:comparison}(b) also the four spinor components of the wave functions for each fixed lattice site. The results are consistent with the analytical predictions.

